\documentclass[aps,prd,groupedaddress,preprint,eqsecnum,nofootinbib%
]{revtex4}
\usepackage{graphicx,epsf,amssymb,amsbsy,amsfonts,amssymb,amsmath}
\usepackage[usenames, dvipsnames]{color}
\newif\ifdraft
\drafttrue
\newif\ifpreprint
\preprinttrue

\def\sect#1{section~{\ref{#1}}}
\def\fig#1{fig.~{\ref{#1}}}
\def\Fig#1{Fig.~{\ref{#1}}}
\def\figs#1#2{figs.~{\ref{#1}} and {\ref{#2}}}
\def\Figs#1#2{Figs.~{\ref{#1}} and {\ref{#2}}}
\def\app#1{appendix~{\ref{#1}}}

\def\tab#1{table~{\ref{#1}}}

\def\spa#1.#2{\left\langle#1\,#2\right\rangle}
\def\spb#1.#2{\left[#1\,#2\right]}
\def\spash#1.#2{\spa{\smash{#1}}.{\smash{#2}}}
\def\spbsh#1.#2{\spb{\smash{#1}}.{\smash{#2}}}
\def\sand#1.#2.#3{%
\left\langle\smash{#1}{\vphantom1}^{-}\right|{#2}%
\left|\smash{#3}{\vphantom1}^{-}\right\rangle}
\def\sandpp#1.#2.#3{%
\left\langle\smash{#1}{\vphantom1}^{+}\right|{#2}%
\left|\smash{#3}{\vphantom1}^{+}\right\rangle}
\def\sandpm#1.#2.#3{%
\left\langle\smash{#1}{\vphantom1}^{+}\right|{#2}%
\left|\smash{#3}{\vphantom1}^{-}\right\rangle}
\def\sandmp#1.#2.#3{%
\left\langle\smash{#1}{\vphantom1}^{-}\right|{#2}%
\left|\smash{#3}{\vphantom1}^{+}\right\rangle}

\def\ka{\kappa}

\def\tree{{\rm tree}}

\def\Tr{\, {\rm Tr}}

\def\Tra#1{\,  {\rm Tr}_{#1}}
\def\eps{\epsilon}
\def\e{\epsilon}
\def\ep{\epsilon}

\def\nn{\nonumber}

\def\sv{s_{12}}
\def\tv{s_{23}}

\def\eqn#1{eq.~(\ref{#1})}
\def\Eqn#1{Equation~(\ref{#1})}
\def\eqns#1#2{eqs.~(\ref{#1}) and~(\ref{#2})}

\def\NeqFoursYM{{${\cal N}=4$~sYM}}
\def\NeqFour{{{\cal N}=4}}

\def\NeqEight{{{\cal N}=8}}
\def\NeqOne{{{\cal N}=1}}
\def\Neqone{{{\cal N}=1}}

\def\colorf#1{\tilde f^{#1}}
\def\colorc#1{c_{(#1)}}
\def\be{\begin{equation}}
\def\ee{\end{equation}}
\def\bea{\begin{eqnarray}}
\def\eea{\end{eqnarray}}
\def\ba{\begin{eqnarray}}
\def\ea{\end{eqnarray}}

\def\MHVbar{$\overline{\hbox{MHV}}$}

\def\hrp{}
\def\hlp{\hskip -.13 cm }
\def\Bral{[}
\def\Brar{]}
\def\Tau#1{\tau_{#1}}
\def\V{{\cal V}}

\def\Frac#1#2{{\textstyle \frac{#1}{#2}}}

\def\UniversalFactor{U}

\def\tree{{\rm tree}}
\def\ra{\rightarrow}
\def\P{{\rm (P)}}
\def\NP{{\rm (NP)}}
\def\A{{\rm (A)}}
\def\B{{\rm (B)}}
\def\I{{\cal I}}
\def\onel{{(1)}}
\def\twol{{(2)}}
\def\threl{{(3)}}
\def\fourl{{(4)}}
\def\Ord{{\cal O}}

\newbox\charbox
\newbox\slabox
\def\s#1{{      
        \setbox\charbox=\hbox{$#1$}
        \setbox\slabox=\hbox{$/$}
        \dimen\charbox=\ht\slabox
        \advance\dimen\charbox by -\dp\slabox
        \advance\dimen\charbox by -\ht\charbox
        \advance\dimen\charbox by \dp\charbox
        \divide\dimen\charbox by 2
        \raise-\dimen\charbox\hbox to \wd\charbox{\hss/\hss}
        \llap{$#1$} }}

\def\subtractfour#1{\ifthenelse{#1=5}{1}{\ifthenelse{#1=6}{2}
{\ifthenelse{#1=7}{3}{\ifthenelse{#1=8}{4}{\ifthenelse{#1=9}{5}
{\ifthenelse{#1=10}{6}{\ifthenelse{#1=11}{7}{\ifthenelse{#1=12}{8}
{\ifthenelse{#1=13}{9}{\ifthenelse{#1=14}{10}{}}}}}}}}}}}

\begin{document}
\hfuzz 20pt

\ifpreprint
UCLA/10/TEP/105 $\null\hskip0.1cm\null$ \hfill 
Saclay--IPhT--T10/075 $\null\hskip0.1cm\null$ \hfill
SLAC--PUB--14137 $\null\hskip0.1cm\null$ \hfill
CERN-TH/2010-186

\fi

\title{The Complete Four-Loop Four-Point Amplitude\\
in $\NeqFour$ Super-Yang-Mills Theory}

\author{Z.~Bern${}^a$, J.~J.~M.~Carrasco${}^a$, L.~J.~Dixon${}^{b,c}$,
 H.~Johansson${}^d$ and  R.~Roiban${}^e$ }

\affiliation{
${}^a$Department of Physics and Astronomy, UCLA, Los Angeles, CA
90095-1547, USA  \\
${}^b$SLAC National Accelerator Laboratory,
              Stanford University,
             Stanford, CA 94309, USA \\
${}^c$Theory Group, Physics Department, CERN, CH--1211 Geneva 23, 
    Switzerland\\
${}^d$Institut de Physique Th\'eorique, CEA--Saclay,
          F--91191 Gif-sur-Yvette cedex, France\\
${}^e$Department of Physics, Pennsylvania State University,
           University Park, PA 16802, USA
}

\date{August, 2010}

\begin{abstract}

We present the complete four-loop four-point amplitude in $\NeqFour$
super-Yang-Mills theory, for a general gauge group and general
$D$-dimensional covariant kinematics, and including all non-planar
contributions.  We use the method of maximal cuts --- an efficient
application of the unitarity method --- to construct the result in terms
of 50 four-loop integrals.  We give graphical rules, valid in $D$-dimensions,
for obtaining various non-planar contributions from
previously-determined terms.  We examine the ultraviolet
behavior of the amplitude near $D=11/2$.  The non-planar terms are as
well-behaved in the ultraviolet as the planar terms.  However, in the
color decomposition of the three- and four-loop amplitude for
an $SU(N_c)$ gauge group, the coefficients of the double-trace terms are better
behaved in the ultraviolet than are the single-trace terms.  The
results from this paper were an important step toward obtaining the
corresponding amplitude in $\NeqEight$ supergravity, which confirmed
the existence of cancellations beyond those needed for ultraviolet
finiteness at four loops in four dimensions.  Evaluation of the
loop integrals near $D=4$ would permit tests of recent conjectures and
results concerning the
infrared behavior of four-dimensional massless gauge theory. 
\end{abstract}

\pacs{04.65.+e, 11.15.Bt, 11.30.Pb, 11.55.Bq \hspace{1cm}}

\maketitle


\section{Introduction}
\label{Introduction}

In recent years, scattering amplitudes have become an important tool
for studying fundamental issues in gauge and gravity theories.  For
example, for the maximally supersymmetric four-dimensional
super-Yang-Mills theory (\NeqFoursYM) in the planar limit, we have
recently seen an all-order resummation ansatz proposed by Smirnov and
two of the authors (BDS)~\cite{ABDK,BDS}, strong-coupling results from
Alday and Maldacena~\cite{AldayMaldacena} and the identification of of
a new symmetry --- dual (super) conformal
invariance~\cite{MagicIdentities,DualConformal,DualConformalWI}.
Together, these results have opened a new venue for studying the
holographic AdS/CFT correspondence.  Furthermore, they suggest the
remarkable possibility that planar amplitudes of \NeqFoursYM~may ultimately be
determined exactly~\cite{AMOperatorProduct}.  In the maximally
supersymmetric four-dimensional supergravity theory
($\NeqEight$)~\cite{CremmerJuliaScherk}, the remarkable ultraviolet
behavior of multi-loop graviton scattering
amplitudes~\cite{Finite,GravityThree,GravityFour,BG2010} has
challenged the widely-held belief that it is impossible to construct a
perturbatively consistent point-like theory of quantum gravity.

In this paper we compute and analyze the four-loop four-point
amplitude in \NeqFoursYM, for an arbitrary non-abelian gauge
group $G$.  The amplitude can be graphically organized into planar and
non-planar contributions.  For the case of a special unitary gauge
group, $G=SU(N_c)$, the planar contributions dominate the limit in
which the number of colors, $N_c$, tends to infinity.  The planar
terms are much simpler, and were computed previously~\cite{BCDKS}.
The non-planar contributions are the subject of this paper.

Scattering amplitudes in gauge theory and gravity have a much richer
structure than that revealed by Feynman diagrams.  One striking
example is Witten's observation that tree amplitudes are supported on
curves in twistor space~\cite{WittenTopologicalString}.  At the
multi-loop level, such structures have first been identified in theories
with maximal supersymmetry.  In \NeqFoursYM, the first hint of
powerful relations
between higher- and lower-loop amplitudes was the ``rung rule''
relation~\cite{BRY,BDDPR} for constructing particular contributions to
four-point amplitudes.  More remarkably, an iterative structure was
uncovered in the dimensionally-regularized planar two-loop four-gluon
amplitude~\cite{ABDK}, using the explicit values of the loop integrals
in an expansion around $D=4$~\cite{SmirnovTwoloop}.  This iterative
structure holds at three loops as well; it led to the all-loop BDS
ansatz~\cite{BDS} for maximally-helicity-violating (MHV) amplitudes.
For four external states, the ansatz is almost certainly correct,
given its verification at strong coupling by Alday and
Maldacena~\cite{AldayMaldacena} using the AdS/CFT correspondence.  For
four or five external legs, the assumption of dual conformal
invariance, together with the structure of the infrared divergences,
completely fixes the amplitudes' dependence on the
kinematics~\cite{DualConformalWI}, making it very likely that the BDS
ansatz is correct. In addition, a variety of explicit four- and
five-point computations confirm it through three loops in dimensional
regularization~\cite{ABDK, BDS, Iterate5pt,
SpradlinLeadingSingThreeLoop} and with a Higgs
regulator~\cite{MassiveRegulatorProgress}.  However, for six or more
external states, the BDS proposal is
incomplete~\cite{AMTrouble,BLSV,TwoLoopSixPt,TwoLoopSixPtWilson}.  To
clarify the structure of the additional ``remainder'' terms, it will
undoubtedly be important to carry out further computations at both
weak and strong coupling.  Positive steps have been taken in this
direction recently~\cite{QMWL,R6,AMNew,AMOperatorProduct}.

Much less is known about non-planar contributions to \NeqFoursYM\
amplitudes.  As a step in this direction,
the principal aim of this paper is to construct the complete
four-loop four-point amplitude in \NeqFoursYM,
including non-planar contributions, using the unitarity
method~\cite{UnitarityMethod} and various refinements of it.
We express the result in terms of 50 distinct loop
momentum integrals, each with nontrivial numerators.  (One of the
50 gives a contribution to the amplitude that vanishes after integration.)
At three loops, the analogous expression requires only
nine distinct integrals~\cite{GravityThree,CompactThree}.

Using these representations of the three- and four-loop amplitudes,
we can study their ultraviolet (UV) properties, in particular the critical
dimension $D_c(L)$ in which the first UV divergence appears, as a function
of the loop number $L$.  Based on information from some of the unitarity
cuts, a formula for $D_c(L)$ was suggested in
ref.~\cite{BDDPR} (see \eqn{SuperYangMillsPowerCount} below).
This formula corresponds to counterterms in higher dimensions
of the schematic form ${\cal D}^2 F^4$, where $F$ is the Yang-Mills
field strength, ${\cal D}$ is a covariant derivative, and the precise
color structure is not yet specified.  This form
for the counterterms was later argued for by Howe and
Stelle~\cite{HoweStelleRevisited}, based on harmonic
superspace~\cite{HarmonicSuperspace}.

Here we confirm this expected behavior for the so-called single-trace
terms in the amplitude for $G=SU(N_c)$, which correspond to
counterterms of the form $\Tr({\cal D}^2 F^4)$.  However, we also find
striking cancellations in the double-trace terms in the amplitude,
starting at three loops, such that counterterms of the form 
$\Tr({\cal D}^2 F^2) \Tr(F^2)$, {\it etc.}, are absent.
These results were first
reported in refs.~\cite{DurhamAndCopenhagenTalks}.  These
cancellations have been discussed using the pure spinor formalism in
string theory~\cite{DoubleTraceBerkovits} and field
theory~\cite{BG2010}, as well as from the point of view of
algebraic non-renormalization
theorems~\cite{DoubleTraceNonrenormalization}.  Because
of contamination from the closed-string sector, the string-based arguments
apply only to the double-trace terms with the most factors of $N_c$,
namely $N_c^{L-1}$; the field-theory arguments~\cite{BG2010} are more
general in this regard.  Here we will rearrange the color structure of the
amplitudes in order to make manifest the additional cancellations in the
double-trace terms, including all powers in $N_c$, at three
and four loops.

The present paper also provides the key input for the unitarity-based
computation of the four-loop four-graviton amplitude~\cite{GravityFour}
in $\NeqEight$ supergravity~\cite{CremmerJuliaScherk}.
On general grounds, the unitarity cuts of gravity loop amplitudes
can be obtained from those of Yang-Mills amplitudes~\cite{BDDPR}.
In a first step, generalized unitarity allows gravity loop
amplitudes to be decomposed into gravity tree amplitudes.  Then the
tree-level Kawai, Lewellen and Tye (KLT) relations~\cite{KLT} can be
used to express the gravity tree amplitudes in terms of gauge-theory
tree amplitudes.  Alternatively, one can use the new diagrammatic
numerator relations between gravity and gauge-theory tree
amplitudes developed by three of the authors~\cite{BCJ}.  Thus the gravity 
cuts are expressed as
bilinear combinations of gauge-theory cuts.  All the information
needed for the gravity computation may be found by cutting the
gauge-theory amplitude.  In the case of $\NeqEight$ supergravity, all the
sums over supersymmetric particles running in the loops are
automatically performed in the course of the \NeqFoursYM\
computation.  Because gravity does not involve color,
the non-planar contributions are just as important as the planar ones,
and the full non-planar \NeqFoursYM\ amplitude is
required.

Multi-loop $\NeqEight$ supergravity amplitudes have revealed 
UV cancellations not anticipated from earlier superspace power-counting
arguments.  The explicit UV behavior of the three- and four-loop
four-point amplitudes~\cite{GravityThree,GravityFour} exhibits strong
cancellations, beyond those needed for UV finiteness in four
dimensions.  Moreover, an analysis of certain unitarity cuts
demonstrates the existence of novel higher-loop cancellations to 
{\it all} loop orders~\cite{Finite}, based on the one-loop ``no-triangle
property''~\cite{OneloopMHVGravity,NoTriangle,NoTriangleSixPt,%
NoTriangleKallosh,BjerrumVanhove,AHCKGravity}.
These results support the proposal that $\NeqEight$ supergravity
might be a perturbatively UV finite theory of quantum gravity.
String dualities~\cite{DualityArguments} and
non-renormalization theorems~\cite{Berkovits} have also been used to
argue for both UV finiteness of $\NeqEight$ supergravity, and
for a delay in the onset of divergences, although difficulties with
decoupling towers of massive states~\cite{GOS} and technical issues
with the pure-spinor formalism~\cite{GRV2010,Vanhove2010,BG2010}
may affect 
these conclusions.

Another important reason to study the non-planar contributions to
\NeqFoursYM\ amplitudes is to investigate whether the
resummation of the four-point amplitude to all loop
orders~\cite{ABDK,BDS,AldayMaldacena} can be accomplished in some
form for non-planar amplitudes as well.
However, given the more complicated color structure of non-planar
contributions, and the probable lack of integrability for subleading terms
in the $1/N_c$ expansion, it is not clear whether such a generalization
can exist.  What is clear is that the infrared-singular behavior
would have to be understood first, before the behavior of
infrared-finite terms could be addressed.  At subleading orders in
$1/N_c$, the soft anomalous dimension matrix ${\bf \Gamma}_{S}$
controls infrared singularities due to soft-gluon
exchange~\cite{SoftMatrix,KorchemskyRadyushkin}.
The matrix structure becomes trivial in the planar, or large-$N_c$,
limit~\cite{BDS}.

Surprisingly, the two-loop soft anomalous dimension matrix in
massless gauge theory is
proportional to the one-loop one~\cite{TwoLoopSoftMatrix}.  The
proportionality constant is determined by the cusp anomalous
dimension~\cite{KorchemskyRadyushkin}.  This result was later
understood to be a consequence of an anomalous symmetry of Wilson-line
expectation values under the rescaling of their
velocities~\cite{BN,GM}.  It has been
conjectured~\cite{CompactThree,BN,GM} that the proportionality might
persist to all loop orders.  However, velocity rescaling alone, even
combined with other constraints, such as collinear factorization of
amplitudes~\cite{TwoLoopSplit,BN}, is not powerful enough to determine
the form of ${\bf \Gamma}_{S}$ beyond two loops~\cite{DGM}, except for
the matter-dependent part at three loops~\cite{Matter3l}.  In
principle, the soft anomalous dimension matrix for \NeqFoursYM\
can be determined at three and four loops, for
the case of four external massless states, using the results in
ref.~\cite{CompactThree} and in this paper.  At three loops, this 
computation would also determine ${\bf \Gamma}_{S}$ for any massless
gauge theory, because the matter-dependent part is known~\cite{Matter3l}.

To perform such a determination, the dimensionally-regularized
loop integrals entering the amplitudes, for $D=4-2\e$, would have to
be expanded in a Laurent series around $\e=0$.  If this can be
accomplished, then it is relatively straightforward to compare the results
with fixed-order formul\ae{}~\cite{STY,TwoLoopSoftMatrix} in order
to extract ${\bf \Gamma}_{S}$.
The Laurent series for the $L$-loop amplitude begins at order
$1/\e^{2L}$, and the $L$-loop soft anomalous dimension matrix
first enters at order $1/\e$ in the expansion.

At order $1/\e^2$, the $L$-loop cusp anomalous
dimension $\gamma_K^{(L)}$ appears.  While this quantity is known
in the planar limit of \NeqFoursYM\
through at least four loops~\cite{BCDKS,CSVcusp4l}, and very likely to all
loop orders~\cite{BES}, the subleading-color terms
are not known beyond three loops.  If they are nontrivial at four loops,
it would indicate the violation of ``quadratic Casimir scaling'',
or $\gamma_K^{(L)} \, \propto \, N_c^L$ for $G=SU(N_c)$, which
holds through three loops~\cite{MVV}.
Such a violation is allowed by group theory, beginning at four loops,
and hinted at by strong-coupling considerations~\cite{Armoni}.
However, it would be very useful to compute the subleading-color terms
in $\gamma_K^{(4)}$ explicitly, as it has been conjectured that
quadratic Casimir scaling will continue to hold beyond three loops~\cite{BN}.

Unfortunately, both these tests will have to await improved techniques for
the evaluation of the Laurent expansion in $\e$ of three- and higher-loop
non-planar integrals.  The non-planar integrals required for three-loop
form factors (one off-shell leg and two on-shell massless ones) have been
evaluated~\cite{ThreeLoopNonPlanarIntegrals}, but not yet those
for on-shell scattering of four massless states at three loops (let alone
four loops).

In order to determine the four-loop amplitude in terms of a set of
loop integrals, we use the unitarity method~\cite{UnitarityMethod}, in
particular the information provided by generalized unitarity
cuts~\cite{GeneralizedUnitarity,TwoLoopSplit,BCFGeneralized,FiveLoop}.
The unitarity method takes advantage of the fact that tree-level
amplitudes are much simpler than individual Feynman diagrams.  One can
exploit structures that are obscure in Feynman diagrams but visible in
on-shell amplitudes. 

For planar \NeqFoursYM, the observation that
the loop amplitudes can be expressed in terms of a restricted set of
``pseudoconformal'' integrals~\cite{MagicIdentities,BCDKS,FiveLoop,%
KorchemskyZeros} helps streamline the construction of a candidate
expression, or ansatz, that is consistent with the unitarity cuts.
For a planar graph, dual coordinates $x_i$ can be
associated with nodes of the dual graph; differences of pairs of $x_i$ 
are identified with momenta of the internal or external lines of 
the original graph.  
Loosely speaking, pseudoconformal integrals are integrals that are
invariant under conformal transformations of the dual coordinates.
It has not yet been proven that all planar loop amplitudes in
\NeqFoursYM\ are expressible
in terms of pseudoconformal integrals (especially for non-MHV
helicity configurations for six or more external states).
However, in any given loop amplitude we can directly confirm that
no other integrals appear, once the pseudoconformal ansatz satisfies all
the unitarity cuts.  This observation helped simplify recent calculations
of the planar five-loop four-point~\cite{FiveLoop} and two-loop 
six-point~\cite{TwoLoopSixPt} MHV amplitudes in \NeqFoursYM.

In contrast, for non-planar terms in the amplitude, there is
currently no useful definition of dual conformal symmetry.
In addition, non-planar contributions are inherently more intricate
than planar ones.  Nevertheless, planar and non-planar contributions
do appear to be linked via identities between diagram numerators, whose
structure is similar to the Jacobi identity obeyed by color
factors~\cite{BCJ,BCJOther,Square}.  Very recently, this structure has been
observed to hold at the three-loop level, with no cut conditions
imposed~\cite{BCJLoop}.
Indeed, for the three-loop four-point amplitude in \NeqFoursYM, one of the
planar diagrams is sufficient for determining all the non-planar ones.
It would be interesting to study these properties further, in particular to
determine the restrictions they place on the four-loop amplitude
presented here.  We leave this interesting question to the future.

A powerful way to determine large classes of both planar and non-planar
contributions is through graphical rules that capture some of the features
of generalized unitarity cuts.  The first of these rules is the rung
rule~\cite{BRY,BDDPR}.  The Jacobi-like relation between planar
and non-planar contributions can also be implemented as a graphical
manipulation~\cite{BCJ}.  There is also a ``box-substitution
rule''~\cite{FiveLoop}, which we will generalize here.  Some related
graphical identities may be found in ref.~\cite{CachazoSkinner}.  The
rules given here reduce the determination of contributions containing
four-point subdiagrams to a few simple manipulations.

The graphical rules presented in this paper capture many contributions,
but not all of them.  In order to determine the missing ones,
we used the method of maximal cuts~\cite{FiveLoop, CompactThree}.  For
four-point amplitudes in $D \geq 4$, 
{\it maximal cuts} are those generalized unitarity
cuts in which only three-point tree amplitudes appear, connected by
cut propagators, {\it i.e.,} they contain the maximum number of cut
propagators.  One first constructs a candidate expression (ansatz)
for the loop amplitude that is equal to each maximal cut, when evaluated
for the appropriate
cut kinematics.  We refer to this procedure as ``matching'' the ansatz
to the maximal cuts.  Next one adjusts the ansatz to match as well the
{\it next-to-maximal} cuts in which a cut propagator is removed from between
two of the three-point trees to form a four-point tree.  This captures
all contributions with a single four-point vertex, or ``contact term''.
Then one matches to further {\it near-maximal} cuts
with more canceled propagators.

In order to construct the complete four-loop amplitude, we used a
large number of maximal and near-maximal cuts.  To ensure that no
contributions were dropped, we then evaluated a set of 13 ``basis
cuts'' (not counting permutations of legs) which suffice to determine
any massless four-loop four-point amplitude.  We verified that our
answer matches all of these cuts.  Many of the cuts were evaluated
using four-dimensional intermediate momenta, so that powerful helicity
and supersymmetry methods could be exploited.  This leaves open the
possibility that some terms could be missed, which vanish when the cut
momenta are restricted to four dimensions.  For this reason, we
performed a large number of consistency checks, as described in
\sect{Unitaritymethodsubsection}.

The unitarity method also allows us to use an on-shell
superspace formalism to sum over states crossing cuts.  On-shell
superspaces involve only physical states and, for our purpose, 
they are generally far simpler
than their off-shell cousins.  A number of years ago Nair presented an
on-shell superspace~\cite{Nair} for MHV tree
amplitudes in \NeqFoursYM.  More recently,
this superspace has been extended to any helicity and particle
configuration.  For the computations of this paper we follow the MHV
generating-function
approach~\cite{GGK,FreedmanGenerating,FreedmanUnitarity},
as organized in ref.~\cite{SuperSum} for use in
multi-loop calculations.  We also exploit specific
super-sum results for cuts involving next-to-MHV tree
amplitudes~\cite{FreedmanUnitarity}.  A related 
procedure~\cite{AHCKGravity,RecentOnShellSuperSpace,KorchemskyOneLoop}
for covering any helicity and particle content employs the momentum 
shifts used by Britto, Cachazo, Feng and Witten to derive on-shell recursion
relations~\cite{BCFW}, extended to shifts of anti-commuting
parameters.  One-loop examples that use an on-shell superspace in
conjunction with the unitarity method may be found in
refs.~\cite{FreedmanGenerating,KorchemskyOneLoop,AHCKGravity}.
Various higher-loop examples, including four-loop ones, are given in
refs.~\cite{FreedmanUnitarity,SuperSum}.

This paper is organized as follows.  In \sect{ConstructionSection} we
describe the general structure of multi-loop amplitudes.  We also
recall the specific form of four-point amplitudes in \NeqFoursYM\ from
one to three loops, as well as the planar terms at four loops.  We
give a brief overview of the techniques used to determine the
amplitudes.  In \sect{MagicToolsSection} we describe in more detail
various tools for determining the non-planar contributions.  In
\sect{AmplitudeSection} we present the complete four-loop four-point
amplitude in terms of a set of 50 integrals.  The ultraviolet
divergence properties of the four-point amplitude through four loops,
in the critical dimension $D_c(L)$, are discussed in
\sect{UVPropertiesSection}.  In \sect{UVSubleadingColorSection}
we compute the leading UV divergence
for the double-trace terms at three loops, which appears at $D=20/3$
(in contrast to the single-trace terms, which first diverge in $D_c(3)=6$).
In \sect{ConclusionSection}, we give our conclusions and prospects for
the future.  In \app{NonTrivialCutAppendix}, we present a sample
evaluation of a nontrivial non-planar cut.  In \app{ColorAppendix} we
provide various representations of the color factors appearing in the
amplitudes.  In \app{NumeratorAppendix} we collect the numerator and
color factors for the 50 integrals entering the four-loop amplitude.


\section{Structure of multi-loop amplitudes}
\label{ConstructionSection}

Loop amplitudes in \NeqFoursYM\ exhibit
remarkable simplicity for a gauge theory.  Using the unitarity
method~\cite{UnitarityMethod}, a large variety of amplitudes have been
constructed through five loops~\cite{BRY,BDDPR,BCDKS,FiveLoop,%
Iterate5pt,TwoLoopSixPt,LeadingSingularityCalcs,SpradlinLeadingSingThreeLoop}
in terms of loop-momentum integrals.  Indeed, the structure of the
planar four-point amplitude is simple enough that an all-loop
order resummation is possible~\cite{BDS,AldayMaldacena}.
This simplicity in the planar sector has been understood in terms 
of a new symmetry dubbed ``dual conformal
symmetry''~\cite{MagicIdentities,KorchemskyZeros,DualConformal},
which is intimately connected to
integrability~\cite{AmplitudeIntegrability}. 
In this paper we focus on the non-planar contributions
to \NeqFoursYM\ amplitudes.  Although they are much
more intricate and less well understood than the planar amplitudes,
their structure is still remarkably simple, especially when compared
to amplitudes in theories with fewer supersymmetries.

In this section we begin by describing the color and
parent-graph organization of multi-loop amplitudes, 
including a review of the results for the lower-loop and planar
four-loop four-point amplitudes in \NeqFoursYM.
Then we turn to a brief review of the unitarity method.

\subsection{Color and parent-graph decomposition}
\label{ColorOrganizationSubsection}

For gauge group $G=SU(N_c)$, the leading-color (planar) terms
are particularly simple.  They have essentially
the same color structure as the corresponding tree amplitudes.
The leading-in-$N_c$ contribution to the $L$-loop $n$-point
amplitude may be written as,
\begin{equation}
{\cal A}_{n}^{(L),\,{\rm Planar}} =  g^{n-2}
 \Biggl[ { g^2 N_c \over (4\pi)^{2-\e} } \Biggr]^{L}
 \hskip -2mm
 \sum_{\sigma\in S_n/Z_n}
\hskip  -2mm 
\Tr( T^{a_{\sigma(1)}} 
   \ldots T^{a_{\sigma(n)}} )
               A_n^{(L)}(\sigma(1), \sigma(2), \ldots, \sigma(n))\,,
\label{LeadingColorDecomposition}
\end{equation}
where the $T^{a_i}$ are generators in the fundamental representation of
$SU(N_c)$, with adjoint color indices $a_i$,
and the sum runs over
non-cyclic permutations of the external legs.  In this expression we
have suppressed the (all-outgoing) momenta $k_i$, as well as
polarizations and particle types, leaving only the integer index $i$
as a collective label.  This decomposition holds for all particles in
the gauge super-multiplet, as they are all in the adjoint
representation.  The color-ordered (or color-stripped)
partial amplitudes $A_n^{(L)}$
carry no color indices; they depend only on the kinematics,
polarizations and particle type.  At leading order in the $1/N_c$
expansion they can be expressed solely in terms of planar loop integrals.

For the complete amplitude for a general gauge group $G$,
including all non-planar contributions, the parent-graph
decomposition,
\begin{equation}
{\cal A}_{n}^{(L)} =  g^{2L+n-2}
 \sum_{i\,\in\, {\rm parent}}
 a_i C_i I_i\,,
\label{ParentColorDecomposition}
\end{equation}
is more convenient than the color-trace representation.  The parent
graphs are cubic graphs --- graphs containing only three-point
vertices.
Momentum is conserved at each vertex. Every graph specifies
simultaneously a combinatorial factor $a_i$, a color dressing $C_i$
and a Feynman loop integral $I_i$.

\begin{figure}
\centerline{\epsfxsize 5 truein \epsfbox{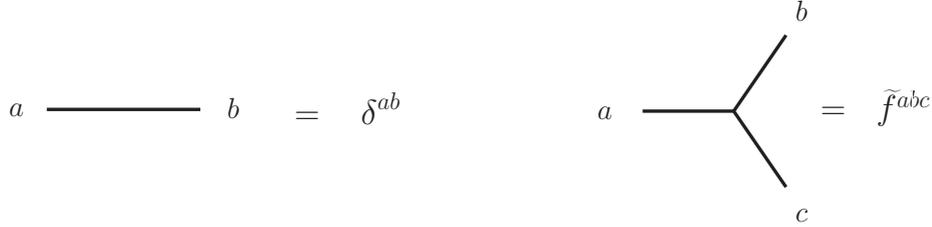}}
\caption[a]{\small
Rules for obtaining the color factors associated with cubic parent graphs. 
The roman letters $a,b,c$ are color indices.
The sign of each structure constant $\colorf{abc}$ is fixed by the
clockwise ordering of each vertex, as drawn.
}
\label{ColorDressFigure}
\end{figure}

The $C_i$ are written in terms of group structure constants. The
contractions of group indices are encoded by the graph $i$ using 
the rules given in \fig{ColorDressFigure}.  More precisely, the 
color factors $C_i$ are obtained by dressing each three-vertex 
of the parent graph with structure constants, normalized as,
\begin{equation}
\colorf{abc} = i \sqrt{2} f^{abc} = \Tr([T^a, T^b] T^c)\,,
\label{fabcdefn}
\end{equation}
where $f^{abc}$ are the standard structure constants of the gauge
group $G$, and the hermitian generators $T^a$ are normalized via
$\Tr(T^a T^b) = \delta^{ab}$. The $\colorf{abc}$ should follow the
clockwise ordering of the legs for each vertex of the parent graph.  
This clockwise ordering is important.  For example, if one redraws
a graph so that an odd number of three-point vertices have their ordering
reversed, then the signs of both $C_i$ and $I_i$ should be flipped
in tandem.

Each {\it parent integral} for the $L$-loop four-point amplitude is 
a Feynman integral with the following general structure and normalization,
\begin{equation}
I_i =  (-i)^L \int 
\biggl(\prod_{j=1}^L \frac{d^{D} \ell_j}{(2\pi)^D}\biggr) \, 
 \frac{N_i (\ell_j, k_m)}{\prod_{n=1}^{3L+1} l_n^2} \,,
\label{IntegralNormalization}
\end{equation}
where $k_m$, $m=1,2,3$, are the three independent external momenta,
$\ell_j$ are the $L$ independent loop momenta, and $l_n$ are the
momenta of the $(3L+1)$ propagators (internal lines of the graph $i$),
which are linear combinations of the $\ell_j$ and the $k_m$.
As usual, $d^{D} \ell_j$ is the $D$-dimensional measure for the
$j^{\rm th}$ loop momentum.  The numerator polynomial $N_i(\ell_j,k_m)$
is a polynomial in both internal and external momenta.

Unlike the decomposition using color traces, the parent-graph
decomposition is not unique, due to contact terms. Contact terms are 
contributions to the amplitude that
lack one or more propagators, relative to the parent graphs.  
If the contact terms were allowed to contribute as isolated integrals,
they would correspond to graphs containing quartic or higher-order
vertices.  Here we will absorb all contact terms into parent integrals, 
by multiplying and dividing by the missing propagator or propagators,
so that the corresponding term in $N_i(\ell_j,k_m)$ will contain factors
of $l_n^2$, which we refer to as {\it inverse propagators}.
However, because the associated color
factors $C_i$ can be expressed as linear combinations of other color
factors via the Jacobi identities, there is an ambiguity in choosing
the specific parent graph to absorb a given contact term.  Although
there are many valid choices, particular choices can reveal nontrivial
structures or symmetries.

For \NeqFoursYM\ amplitudes, the freedom in assigning contact terms to
parent graphs can be exploited to remove all graphs with nontrivial
two- or three-point subgraphs, as was done at three
loops~\cite{CompactThree}.  While supersymmetry and gauge invariance
may be used to show the all-order consistency of this condition on the
parent graphs, a more direct approach was taken in
ref.~\cite{CompactThree}, by verifying that an ansatz in this class is
compatible with all unitarity cuts.  Here we will take the same approach at
four loops.

\begin{figure}[th]
\centerline{\epsfxsize 1.4 truein \epsfbox{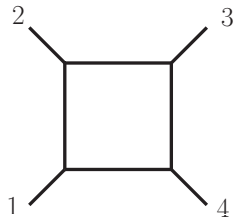}}
\caption[a]{\small The box diagram is the only parent graph at one
loop (up to permutations of the external legs).
}
\label{OneLoopIntegralsFigure}
\end{figure}

At one loop, the structure of the \NeqFoursYM\
four-point amplitude is especially simple.  We modify
\eqn{ParentColorDecomposition} slightly by extracting an overall
prefactor, and write the result as,
\begin{eqnarray}
{\cal A}_4^{(1)} \! & = & \! 
-{1 \over 8} g^4 \, {\cal K} \, \sum_{S_4}\, C^{\rm box}_{1234} 
\, I^{\rm box}(\sv,\tv) \,, \hskip .3 cm 
\label{OneLoopYMAmplitude}
\end{eqnarray}
where $g$ is the gauge coupling.  The prefactor ${\cal K}$ is defined by
\begin{equation}
{\cal K} \equiv {\cal K}(1,2,3,4)
\equiv s_{12} \, s_{23} \, A_4^{\tree}(1,2,3,4)\,,
\label{Prefactor}
\end{equation}
where $\sv = (k_1 + k_2)^2$, $\tv= (k_2 + k_3)^2$, and the $k_i$
are the external momenta.  It contains all information about the four
external states.

The unique parent graph at one loop is the box diagram shown in
\fig{OneLoopIntegralsFigure}.  For external legs ordered 1234,
the box color factor is 
\begin{equation}
C^{\rm box}_{1234} = \colorf{a_1 a_6 a_5} \colorf{a_2 a_7 a_6} \colorf{a_3 a_8 a_7}
  \colorf{a_4 a_5 a_8} \,,
\label{1loopcolortensor}
\end{equation}
where we sum over repeated indices.  This form is valid for any gauge
group.  Finally, $I^{\rm box}(\sv,\tv)$ is the one-loop box integral,
\begin{equation}
I^{\rm box}(\sv,\tv) = -i \int {d^D p \over (2\pi)^D} 
{1 \over p^2 (p-k_1)^2 (p-k_1 - k_2)^2 (p+k_4)^2 } \,.
\label{OneLoopBoxIntegral}
\end{equation}
The sum in \eqn{OneLoopYMAmplitude} runs over the 
24 permutations of external legs $\{1,2,3,4\}$, denoted by $S_4$. 
The permutations act on both the momentum and color labels.
The prefactor of $1/8$ accounts for an eightfold overcount
in the permutation sum, which we leave in to make it slightly easier
to generalize to higher loops.

In \eqn{Prefactor}, $A_4^\tree(1,2,3,4)$ stands for any
\NeqFoursYM\ tree amplitude in the canonical color order.
In four dimensions, a compact form of this
object can be written down using anti-commuting parameters $\eta$ in
the on-shell superspace formalism~\cite{Nair,GGK,FreedmanGenerating,%
RecentOnShellSuperSpace,KorchemskyOneLoop,AHCKGravity,FreedmanUnitarity},
\begin{eqnarray}
A_{4}^{\tree}(1,2,3,4)\bigr|_{D=4} &=&
{i\delta^{(8)}(\lambda_1 \eta_1+\lambda_2 \eta_2
              +\lambda_3 \eta_3+\lambda_4 \eta_4)
 \over \spa{1}.{2}\spa{2}.{3}\spa{3}.{4}\spa{4}.{1}} \,,
\label{SuperAmplitude} \\
{\cal K}(1,2,3,4)\bigr|_{D=4} &=&
-i \, {\spb1.2\spb3.4 \over \spa1.2\spa3.4}
\, \delta^{(8)}(\lambda_1 \eta_1+\lambda_2 \eta_2
              +\lambda_3 \eta_3+\lambda_4 \eta_4) \,.
\label{SuperAmplitudeII}
\end{eqnarray}
It is not difficult to verify that $(\spb1.2\spb3.4)/(\spa1.2\spa3.4)$
is symmetric under exchange of any two legs, and that ${\cal K}$
is local.  Related to
this, ${\cal K}$ represents the color-stripped four-point (linearized)
matrix elements of the local operator $\Tr F^4$, plus its
supersymmetric partners.  Therefore ${\cal K}$ is a natural prefactor
to extract from the four-point amplitude in \NeqFoursYM.

\begin{figure}[tbh]
\centerline{\epsfxsize 4 truein \epsfbox{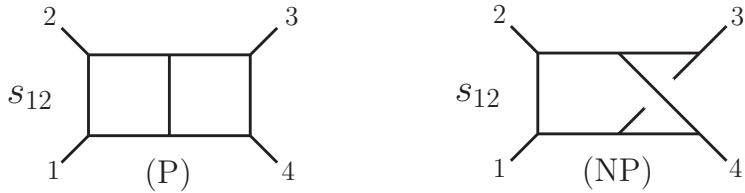}}
\caption[a]{\small The planar and non-planar parent graphs
contributing to the two-loop four-point amplitude.
The prefactors $s_{12}$ are the integral numerators
$N^{({\rm P})}$ and $N^{({\rm NP})}$, respectively.}
\label{TwoLoopIntegralsFigure}
\end{figure}

At two loops, the full \NeqFoursYM\ amplitude is given by a similar
permutation sum as for the one-loop case
(\ref{OneLoopYMAmplitude})~\cite{BRY,BDDPR},
\begin{eqnarray}
{\cal A}_4^{(2)} \! & = & \! 
{1\over 4} g^6\, {\cal K} \, \sum_{S_4}\, 
\Bigl[ C^{\rm (P)}_{1234} \, I^{\rm (\rm P)}(\sv, \tv) \, + \, 
     C^{\rm (NP)}_{1234} \, I^{\rm (NP)} (\sv, \tv) \Bigr] \,.
\hskip .3 cm 
\label{TwoLoopYMAmplitude}
\end{eqnarray}
The planar and non-planar double-box integrals, displayed in
\fig{TwoLoopIntegralsFigure}, are defined by 
\begin{eqnarray}
\hskip -.3 cm 
I^{\rm (P)}(\sv,\tv) \!&=&\!
(-i)^2 \! \int
{d^D p d^D q\over (2\pi)^{2D}} \, 
{\sv \over p^2 \, (p - k_1)^2 \,(p - k_{12})^2 \,(p + q)^2 q^2 \,
        (q-k_4)^2 \, (q - k_{34} )^2 }\,, \hskip .3 cm \nn \\
\hskip -.3 cm 
I^{\rm (NP)}(\sv,\tv) \!&=&\! 
(-i)^2 \! \int 
{d^D p d^D q \over (2\pi)^{2D}} \,
{\sv  \over p^2 (p-k_4)^2 (p+q)^2 (p+q+k_3)^2
  q^2  (q-k_1)^2 (q-k_{12})^2} \,, \nn\\
\end{eqnarray}
with $k_{12} = k_1+k_2$ and $k_{34} = k_3+k_4$. 
The permutation sum again runs over $S_4$ and acts on both momentum
and color labels. Because both graphs in \fig{TwoLoopIntegralsFigure}
have a fourfold symmetry, the
permutation sum overcounts by a factor of four.  As before, we include this
overcount and divide by an overall symmetry factor.  In a form valid
for any gauge group, the color factors of the planar and non-planar
graphs, with legs ordered 1234, are,
\begin{eqnarray}
C^{(\rm P)}_{1234} &=& 
\colorf{a_1 a_6 a_5} \colorf{a_2 a_7 a_6} 
\colorf{a_3 a_9 a_8} \colorf{a_4 a_{10} a_9}\colorf{a_7 a_8 a_{11}}
\colorf{a_5 a_{11} a_{10}}
\,, \nn \\
C^{(\rm NP)}_{1234} &=& 
\colorf{a_1 a_6 a_5} \colorf{a_2 a_7 a_6}
\colorf{a_3 a_9 a_8} \colorf{a_4 a_{11} a_{10}}\colorf{a_7 a_8 a_{10}}
\colorf{a_5 a_9 a_{11}}
\,.
\label{2loopcolortensors}
\end{eqnarray}

\begin{figure}[t] \centerline{\epsfxsize 4.6 truein 
         \epsfbox{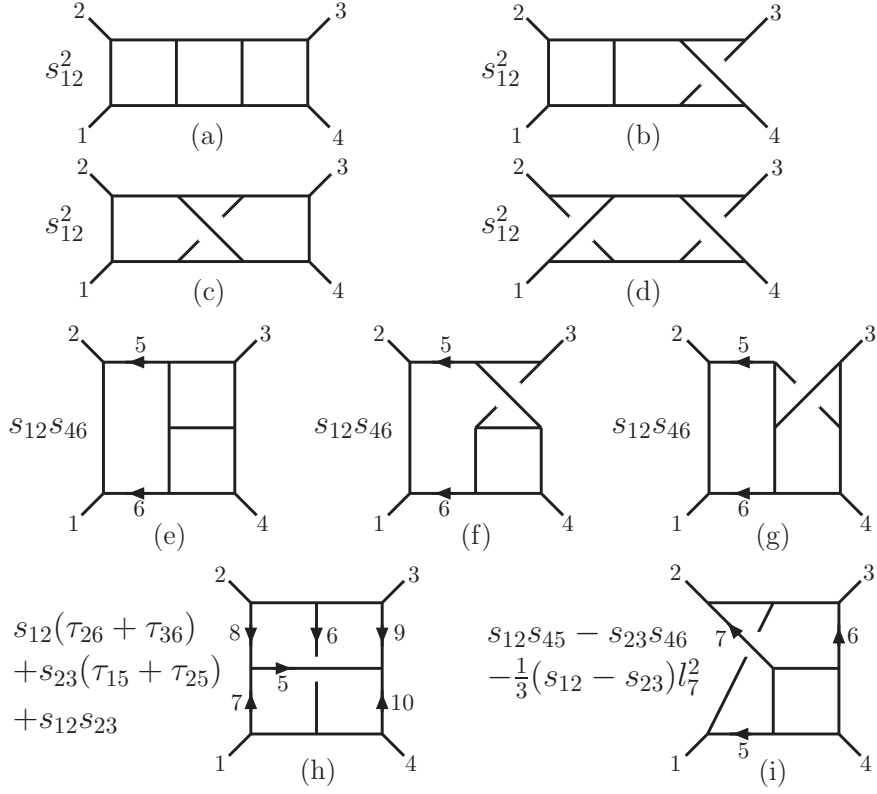}}
\caption[a]{\small The nine parent graphs for the three-loop 
four-point \NeqFoursYM\ amplitude.  
The prefactor of each diagram is the numerator polynomial $N^{(x)}$.
The kinematic invariants are defined in \eqn{InvariantsDef}.}
\label{IntegralsThreeLoopFigure}
\end{figure} 

At three loops, the fully color-dressed three-loop
four-point \NeqFoursYM\ amplitude is given
by~\cite{GravityThree, CompactThree},
\begin{eqnarray}
{\cal A}_4^{(3)} \! & = & \! -
{1\over 4}\,g^8\, {\cal K} \, \sum_{S_4}\, 
\Bigl[ C^{\rm (a)} I^{\rm (a)} 
  + C^{\rm (b)} I^{\rm (b)} 
  + {\textstyle {1\over 2}} C^{\rm (c)} I^{\rm (c)} 
  +  {\textstyle {1\over 4}} C^{\rm (d)} I^{\rm (d)}\nn \\
&& \null \hskip 2 cm 
 + 2 C^{\rm (e)} I^{\rm (e)} 
 + 2 C^{\rm (f)} I^{\rm (f)} + 4 C^{\rm (g)} I^{\rm (g)} + 
   {\textstyle {1\over 2}}  C^{\rm (h)} I^{\rm (h)} 
 + 2 C^{\rm (i)} I^{\rm (i)} 
\Bigr] \,. \hskip .3 cm 
\label{ThreeLoopYMAmplitude}
\end{eqnarray}
In this case, the integrals $I^{(x)}(\sv,\tv)$ are $D$-dimensional
loop integrals corresponding to the nine graphs shown in
\fig{IntegralsThreeLoopFigure}, using \eqn{IntegralNormalization} with
the numerator polynomials $N_i$ displayed next to the diagrams.
In \fig{IntegralsThreeLoopFigure} the
(outgoing) momenta of the external legs are denoted by $k_i$ with
$i=1,2,3,4$, while the momenta of the internal legs are denoted by
$l_i$ with $i>4$. For convenience we use the following shorthand
notation,
\begin{equation}
 s_{ij} = \left\{
\begin{array}{ll} 
(k_i +k_j)^2 &  \hskip 1cm i,j \le 4 \\
(k_i +l_j)^2 &  \hskip 1cm i \le 4<j  \\
(l_i +l_j)^2 &  \hskip 1cm  4<i,j 
\end{array}\right\}\,,
\hskip 2cm \tau_{ij} = 2 k_i\cdot l_j\,.
\label{InvariantsDef}
\end{equation}
In the three-loop case, some of the numerator polynomials contain squares
of loop momenta, which could be used to collapse propagators and generate
contact terms.  (The three-loop four-graviton amplitude
in $\NeqEight$ supergravity can be written~\cite{CompactThree} in a
form very similar to \eqn{ThreeLoopYMAmplitude}, except that there are
no color factors, and the numerator factors in the loop integrals are 
of course different.)

The color factor associated with each integral in the three-loop
amplitude is easy to write down from the parent graph,
following \fig{ColorDressFigure}.  
The expression~(\ref{ThreeLoopYMAmplitude}) is valid for any gauge
group $G$ and any dimension $D$, as verified by a direct evaluation
of the color-dressed cuts~\cite{CompactThree}.

\begin{figure}[t]
\centerline{\epsfxsize 6 truein \epsfbox{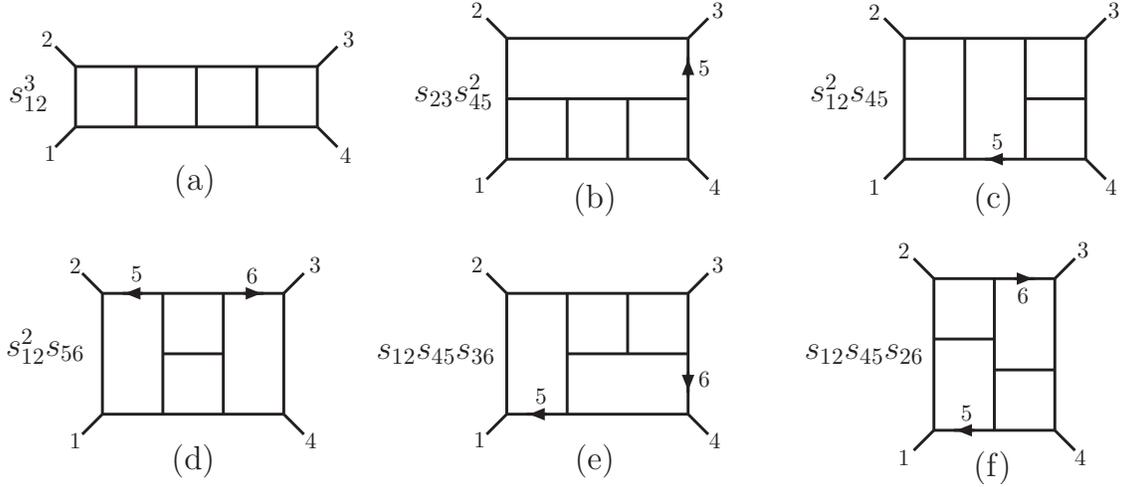}}
\caption[a]{\small Planar contributions to the four-loop
   four-point amplitude.}
\label{Planar4lFigure}
\end{figure}

\begin{figure}[t]
\centerline{\epsfxsize 4 truein \epsfbox{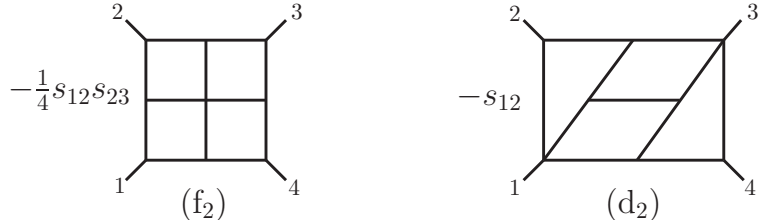}}
\caption[a]{\small Contact terms in the planar four-point amplitude,
which can be absorbed into the numerator of graph (f) of 
\fig{Planar4lFigure}. (Note that here (f$_2$) and (d$_2$) refers to the 
labeling used in ref.~\cite{BCDKS}; unlike ref.~\cite{BCDKS} we choose 
to write out explicitly the relative factors of these integrals with
respect to diagram (f).)}
\label{Planar4lNonRRFigure}
\end{figure}

In \sect{AmplitudeSection} we will present the complete four-loop
four-point amplitude of \NeqFoursYM, using the same type of parent-graph
decomposition. As we will demonstrate, this amplitude can be decomposed
into 50 distinct parent integrals, corresponding to cubic graphs with
no nontrivial two- or three-point subgraphs. As explained earlier, contact
terms are incorporated as numerator factors containing inverse propagators.
Thus we write the four-loop four-point amplitude as,
\begin{equation}
{\cal A}_4^\fourl 
= g^{10} \, {\cal K} \, \sum_{S_4} \sum_{i=1}^{50}  
a_i C_i I_i \,,
\label{FourLoopEquation}
\end{equation}
where the usual prefactor ${\cal K}$ is common to all terms, the $a_i$ are
combinatoric symmetry factors and the $C_i$ color factors. The
integrals $I_i$ are specified by the propagators associated with
the parent graph, and by the numerator polynomials $N_i$.
Each numerator polynomial is subject to various constraints.  After
accounting for the four powers of external momenta in ${\cal K}$
in \eqn{FourLoopEquation}, dimensional analysis implies that the
numerator polynomial $N_i(\ell_j,k_j)$ is of degree 6 in the momenta.
Moreover, consistency with the
known power-counting~\cite{BDDPR,HoweStelleRevisited} requires that
the numerator polynomials for \NeqFoursYM\
have a maximum degree of 4 in the loop momenta.

The planar contributions to the four-loop four-point amplitude were
presented earlier~\cite{BCDKS}, although they were given in the
color-trace decomposition, rather than the graphical decomposition used here.
Out of the eight different types of planar integrals present in the
amplitude, the six parent graphs are shown in \fig{Planar4lFigure}.
(The combinatoric factors are not included in the figure.)
The two additional contributions are contact terms and are shown in
\fig{Planar4lNonRRFigure}.  A convenient choice for absorbing these
contact terms is to assign both of them to the parent graph (f) of
\fig{Planar4lFigure}, by incorporating inverse propagators in their 
numerators.  The result, after combining two different permutations
of diagram (f), can be found in \fig{D2Figure}, diagram (28).

\subsection{Unitarity method}
\label{Unitaritymethodsubsection}

The unitarity method provides an efficient framework for
systematically constructing and verifying the expression for any massless 
multi-loop amplitude.  This
method, along with various refinements, has already been described in
some detail elsewhere~\cite{UnitarityMethod,BDDPR,%
GeneralizedUnitarity,TwoLoopSplit,BCFGeneralized,FiveLoop,CachazoSkinner}.
Here we summarize those points directly salient to our construction of
multi-loop amplitudes in \NeqFoursYM.

\begin{figure}[tbh]
\centerline{\epsfxsize 6.2 truein \epsfbox{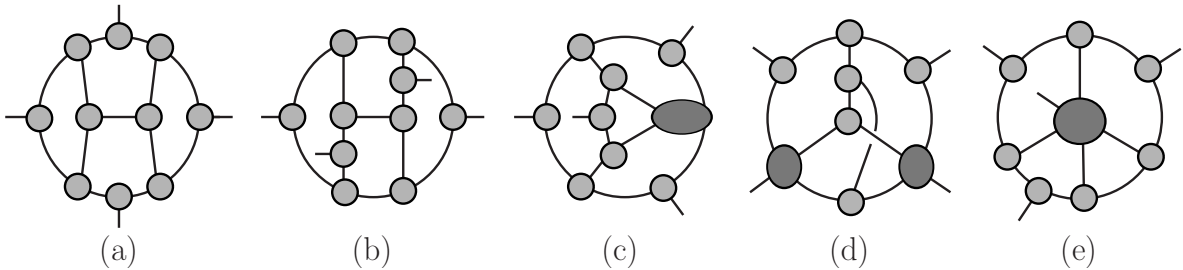}}
\caption[a]{\small Examples of maximal and near-maximal cuts used to
construct a complete ansatz for the four-loop \NeqFoursYM\
amplitude. Cuts (a) and (b) have the maximal 13 cut
conditions. Near-maximal cuts (c) have 12 conditions, and (d) and (e)
have 11 conditions. No cuts with fewer than 11 on-shell propagators are
needed to build the complete ansatz.}
\label{SampleMaximalCutsFigure}
\end{figure}

\begin{figure}[tbh]
\centerline{\epsfxsize 5.8 truein \epsfbox{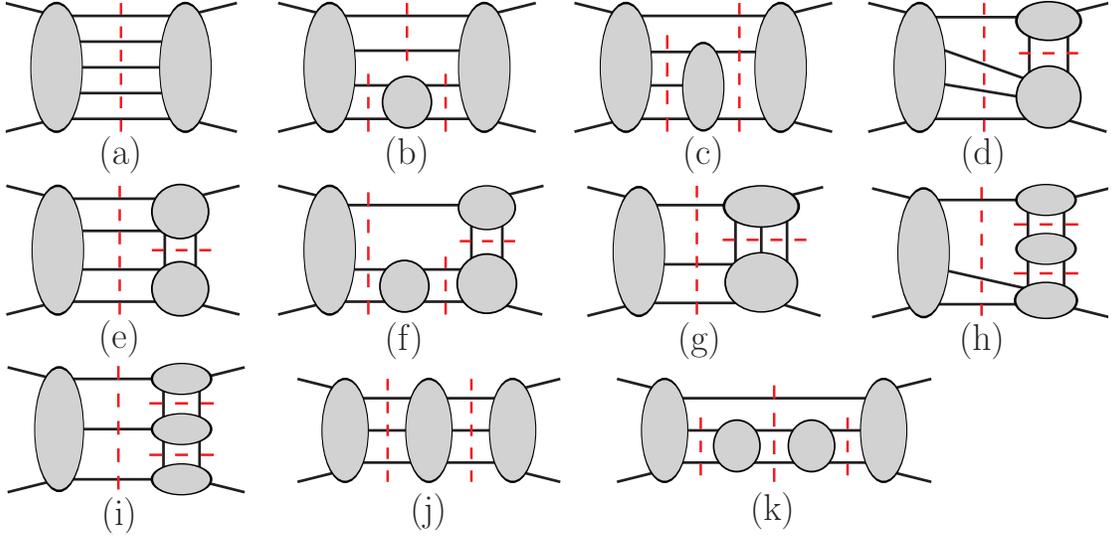}}
\caption[a]{\small These 11 cuts, along with the two-particle cuts 
in \fig{TwoParticleCutsFigure},
suffice to determine any massless four-loop four-point amplitude. As
displayed they determine any planar amplitude; to determine any
non-planar contributions one must also include cuts with the legs of
each tree sub-amplitude permuted arbitrarily.  The cuts (h) and (k)
have been evaluated in $D$ dimensions, as they are included in the box
cuts \fig{DdimensionalCutsFigure}(c) and (e), respectively.}
\label{CutBasisFigure}
\end{figure}

\begin{figure}[t]
\centerline{\epsfxsize 4 truein \epsfbox{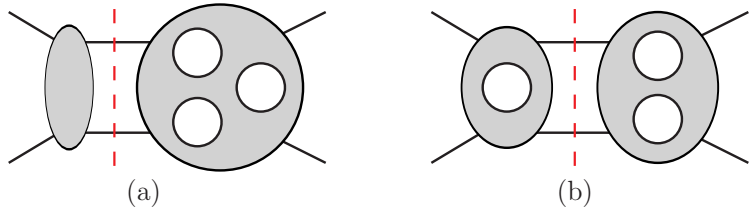}}
\caption[a]{\small The two-particle cuts.  These cuts have been
evaluated in $D$ dimensions.  Together with 
the cuts of \fig{CutBasisFigure} these form a spanning set of cuts
for any massless four-loop four-point amplitude.
}
\label{TwoParticleCutsFigure}
\end{figure}


A generalized unitarity cut is a sum over products of amplitudes,
\begin{equation}
i^c \sum_{\rm states} A_{(1)} A_{(2)} A_{(3)} \cdots 
A_{(m)} \,,
\label{GeneralizedCut}
\end{equation}
evaluated for kinematics that place all $c$ {\it cut lines} on shell.
Here we normalize the cuts to include the factor of $i$ of each cut
Feynman propagator.  Each cut-line particle appears twice in the
summand --- leaving one amplitude and entering another.
Often the cuts are chosen so that all the amplitudes are
tree amplitudes, although this is not necessary.
\Fig{SampleMaximalCutsFigure} illustrates particularly useful cuts for
determining the four-loop four-point amplitude, in which the maximal and
near-maximal number of propagators are cut.  (In order to avoid
excessive clutter, we draw maximal and near-maximal cuts without the
usual dashed lines to indicate the cuts.)  Another set of useful cuts
for confirming all contributions is depicted in
\figs{CutBasisFigure}{TwoParticleCutsFigure}.

For massless theories, the on-shell momentum condition for the
particle associated with each cut line is $l_i^2 = 0$. We sum over
all possible states of the cut-line particle.  For \NeqFoursYM\ theory
the sum is over the full supermultiplet,
including gluons, scalars and fermions. For pure-Yang-Mills theory in four
dimensions the sum is only over positive and negative helicity gluons. In
the latter case, there can also be contributions, such as loops on
massless external legs, that are not detectable in the cuts of
\figs{CutBasisFigure}{TwoParticleCutsFigure}.  While these contributions
vanish after loop integration in dimensional regularization,
they can nevertheless alter the UV behavior of the amplitude because
their vanishing is the consequence of a cancellation between UV and
IR divergences. In the case of \NeqFoursYM\ theory, for representations
of the amplitude that obey manifest power-counting~\cite{BDDPR,HoweStelleRevisited}
and do not contain three-point sub-amplitudes with one external and two 
internal legs (which is the case in this paper and in previous work), 
such massless-external-loop contributions do not appear, by virtue of 
unitarity and the vanishing of supersymmetric
three-point loop amplitudes.

The {\it cut-construction} of a multi-loop amplitude formally begins
with a generic ansatz for the amplitude in terms of multi-loop Feynman
integrals, in which the numerator polynomials of each integrand
$N_i(\ell_j,k_j)$ contain arbitrary coefficients.\footnote{The number of
independent Feynman integrals present in the four-point amplitude is
typically far smaller than the number of Feynman diagrams.} 
These coefficients are then systematically determined by comparing the
generalized cuts of the ansatz to the cuts of the amplitude. 

For a multi-loop expression to be correct for a
given theory it must satisfy all possible generalized unitarity cuts.
However, many cuts are simply special cases of other cuts.
We define a {\it spanning} set of cuts as any set whose verification
is sufficient to ensure that all other cuts are satisfied, and
thus the correctness of a multi-loop expression for the amplitude.
Below we shall describe such sets.

In the process of {\it cut-verification}, an ansatz for the desired
multi-loop amplitude --- described in terms of Feynman integrals over
loop-momenta --- is evaluated using on-shell momenta on the cut lines
at the integrand level and compared to \eqn{GeneralizedCut},
using the same cut kinematics.  This procedure requires first
identifying the subset of parent graphs in the ansatz that are nonzero on the
specified cut, and then lining up momentum labels between each such
parent graph and the generalized cut~(\ref{GeneralizedCut}).
The latter step can be performed by decomposing each of the
constituent amplitudes in \eqn{GeneralizedCut} into their own
parent graphs.  These graphs come with a labeling, which can be 
used to provide a suitable labeling for each parent graph in the
ansatz. (One may have to perform some initial relabeling
in order to avoid duplicate labels coming from different constituent
amplitudes.)  Finally, one permutes the labels in the original
representation of the ansatz so that they match this labeling.
In \app{NonTrivialCutAppendix} we illustrate this procedure in the
evaluation of a nontrivial cut at four loops.

As mentioned in the discussion of the parent-graph decomposition, the
numerators of parent graphs are not uniquely determined by cut
constraints, because of an inherent ambiguity in the assignment of
contact terms. This ambiguity corresponds to the ability to add terms
to the integrands associated with different parent graphs, such that
the complete amplitude remains unchanged.  Essentially, this ambiguity is
nothing more than the freedom to add zero to the amplitude in a nontrivial
way.  This freedom allows various representations of the amplitude,
which can expose different properties.  A particularly useful
property is for every term in every parent integral to be no 
more UV divergent than the complete amplitude; that is, all UV
cancellations are exposed. It is also rather desirable for the
numerator factors to respect the symmetries of the parent graphs.
Indeed, we will find a representation of the four-loop four-point
amplitude with both these properties.  Imposing these properties
at the beginning helps to limit the number of possible terms
in the ansatz.

In practice, as will be discussed in \sect{MaximalCuts}, it is
possible to construct a multi-loop amplitude iteratively --- building
and refining an ansatz by requiring it to be consistent with
a sequence of generalized cuts, beginning with the maximal cuts.
Note that, {\it a priori}, one can make any simplifying assumptions
that restrict the ansatz; these assumptions are justified {\it a
posteriori} by verifying the ansatz against a spanning set of cuts.
\Figs{CutBasisFigure}{TwoParticleCutsFigure} show generalized cuts used
in the verification of the four-loop four-point amplitude.  In these
figures all exposed internal lines are cut.  

Color-stripped cuts involve products of the color-ordered tree-level
partial amplitudes that appear in the trace-based color decomposition
for gauge group $SU(N_c)$.
On the other hand, color-dressed cuts are products of full tree
amplitudes, including their color factors, which are products
of structure constants $\tilde{f}^{abc}$ and are valid for any gauge group.
One can always verify a color-dressed amplitude for any gauge group
$G$ by considering a spanning set of color-dressed cuts.  In practice,
we used
color-stripped cuts in our construction of the four-loop amplitude;
in the final step we color-dressed our integrands with color factors
$C_i$ which are built from structure constants $\tilde{f}^{abc}$ and
are valid for any gauge group.

Although in principle all cuts can be evaluated analytically, some
cuts in \fig{CutBasisFigure} are rather nontrivial. It is therefore
often more practical to evaluate the cuts numerically to high
precision for a number of points in the phase space
satisfying the cut conditions.  Because the check is performed at
the level of the integrand, and does not require any numerical
integration, it can be performed at arbitrarily high precision.
  
In the rest of this section we will review technology for summing
over the $\NeqFour$ multiplets of states crossing each cut,
which is needed to evaluate the unitarity cuts.  In
\sect{MagicToolsSection} we will present two classes of unitarity cuts
--- the two-particle cut, and the box-cut --- which have been worked out
in all generality in terms of lower-loop parent-dressings.  These cuts
are particularly handy because they can be applied with very little
calculation.

\subsubsection{Supersymmetric sums over states in four dimensions}

We can simplify the evaluation of many cuts by
restricting the internal loop momenta (as well as the external momenta)
to four dimensions.  Then we can enumerate the internal states according
to their four-dimensional helicity, and apply powerful supersymmetry Ward
identities~\cite{SWI} or on-shell superspace formalisms (both of which
are valid only in four dimensions), in order to simplify the sum over 
intermediate states.  In \NeqFoursYM, this sum runs over the $\NeqFour$
super-multiplet, so we refer to it as a ``super-sum''.
For simple cuts, the sum over supersymmetric states
in~\eqn{GeneralizedCut} is easy to evaluate
component by component~\cite{BDDPR}, by making use of supersymmetry Ward
identities that relate the different tree amplitudes, and hence
relate the different terms in the state sum.  

As described in some detail in ref.~\cite{FiveLoop}, for maximal or
near-maximal cuts, it turns out that one can avoid nontrivial sums
over particles by making use of solutions to the cut conditions that
force all, or nearly all, particles propagating in the loops to be
gluons with a single helicity configuration.
Such restrictions can be arranged for cuts that contain
sufficiently many three-point tree amplitudes.
These solutions were sufficient for constructing the complete
four-point amplitude in \NeqFoursYM\ in this paper.%
\footnote{This property is special to maximally supersymmetric
amplitudes and will not hold in other theories such as QCD.}
Remarkably, this allows us to build a complete ansatz for the four-loop
amplitude avoiding all nontrivial supersymmetric sums over particles
crossing the cuts.  Even so, we must satisfy all solutions to all cut
conditions, including those that impose no restriction on the particle
content.  As discussed earlier, we verify the correctness of our construction
on a spanning set of cuts.  In such cuts we must 
sum over all allowed configurations of particles crossing
the cuts.  A good means for summing over the states is therefore
needed, especially given the nontrivial bookkeeping of states required
at four loops.

In supersymmetric theories, superspace provides an efficient way to 
track contributions from different states in the same super-multiplet.
However,
we prefer a superspace which works well with on-shell methods.  Such a
superspace is based on Nair's construction, which encodes the MHV
tree amplitudes of \NeqFoursYM~\cite{Nair}.  In
recent years, this superspace has been generalized to any
four-dimensional tree amplitude and also to loop
level~\cite{GGK,FreedmanGenerating,RecentOnShellSuperSpace,%
KorchemskyOneLoop,AHCKGravity,FreedmanUnitarity,SuperSum}.
A solution to the problem of
evaluating super-sums in generic multi-loop unitarity cuts was
given~\cite{FreedmanUnitarity}, based on an MHV-vertex
generating-function approach.  In ref.~\cite{SuperSum}, this solution
was recast into two complementary approaches for efficiently evaluating
multi-loop unitarity cuts.    In the 
first approach, the problem is recast into the calculation of the
determinant of the matrix associated with a certain system of linear
equations.  In the second approach, used in this paper, the contributions
of individual states are tracked via $SU(4)$ ``$R$-symmetry index diagrams''.

To systematically step through the many cuts we used to verify our
construction of the four-loop amplitude, it is helpful to have an
efficient and easily programmable algorithm for evaluating any cut,
with essentially no calculation.  The $R$-symmetry index
diagram method~\cite{SuperSum} is based on the observation that,
after applying the MHV-vertex expansion for tree amplitudes~\cite{CSW},
the cuts of \NeqFoursYM\ amplitudes are simply related to
those of (non-supersymmetric) pure Yang-Mills theory.  
By carrying out the super-sums in
the MHV-vertex expansion, each term contains a numerator of the form,
\begin{equation}
 (S_1 + S_2 + \cdots + S_m)^4\,,
\label{sYMNumerator}
\end{equation}
where the $S_i$'s are spinor-product monomials, such as
$\spa{i_1}.{i_2}\spb{i_3}.{i_4}\cdots\spa{i_{j-1}}.{i_j}$.  Upon
expansion of \eqn{sYMNumerator}, each quartic expression $S_iS_jS_kS_l$
corresponds to a single assignment of helicities to particles
crossing the cuts.  
Remarkably, \eqn{sYMNumerator} can be inferred instead from the much
simpler state sum for pure Yang-Mills theory, for which the
analogous numerator is
\begin{equation}
 S_1^4 + S_2^4+ \cdots + S_m^4 \,.
\label{YMNumerator}
\end{equation}
One introduces anticommuting parameters, which transform
under the $SU(4)$ $R$ symmetry of \NeqFoursYM, and
track the relative signs between $S_i$ and $S_j$ in
\eqn{sYMNumerator}.  With the aid of these parameters,
the result~(\ref{sYMNumerator}) for the
\NeqFoursYM\ cut can be read off from the pure-Yang-Mills
cut~(\ref{YMNumerator}).  A detailed description of
the algorithm, as well as the $R$-symmetry index diagrams,
may be found in ref.~\cite{SuperSum}.

Using this algorithm we have evaluated cuts
containing only MHV and \MHVbar{} vertices, 
as well as the spanning set of all 13 cuts
in \figs{CutBasisFigure}{TwoParticleCutsFigure} and their
permutations.  These evaluations confirm that the ansatz
we constructed for the amplitude, using graphical rules and information
provided by the maximal and near-maximal cuts,
captures all contributions that are nonzero when loop momenta
are restricted to four dimensions.

Elvang, Freedman and Kiermaier~\cite{FreedmanUnitarity} have used the
MHV-vertex expansion to provide very compact expressions for the
super-sums for cuts (a) and (j) in \fig{CutBasisFigure}, which contain
next-to-MHV (and next-to-\MHVbar) amplitudes.
We also compared the cuts of our ansatz to their results,
and found agreement.\footnote{%
We thank H.~Elvang, D.~Freedman and M.~Kiermaier for
  assistance in implementing their expressions.}  
Cut (a) is particularly powerful
because it checks most of the terms in all 50 of the parent graphs
in \eqn{FourLoopEquation}, providing an important independent check.


\subsubsection{$D$-dimensional cuts}

\begin{figure}[t]
\centerline{\epsfxsize 5.4 truein \epsfbox{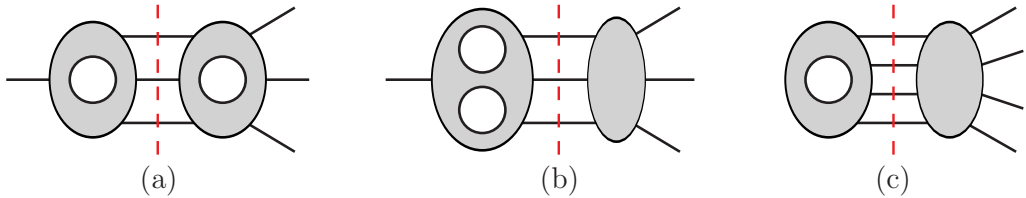}}
\caption[a]{\small The three box cuts.  These cuts have been
  evaluated in $D$ dimensions.  }
\label{DdimensionalCutsFigure}
\end{figure}

The method outlined above efficiently solves the problem of
evaluating unitarity cuts in four dimensions in \NeqFoursYM.  However,
because we are interested in computing the amplitudes in
$(D>4)$-dimensions, this is not sufficient; the loop momenta
are $D$-dimensional.  Even if we are interested in amplitudes in
four dimensions, we need to use a 
(supersymmetry-preserving~\cite{DimRed}) form  of dimensional
regularization to regulate infrared singularities.
If the cuts are evaluated in four dimensions, as described in the previous
subsection, then terms that vanish when the loop momenta are
restricted to four dimensions, but are non-vanishing in $D$ dimensions,
could be missed.  Indeed, amplitudes in theories with fewer
supersymmetries~\cite{DDimUnitarity,DimShift}, and \NeqFoursYM\
amplitudes with more than four external legs~\cite{TwoLoopSixPt},
do contain such terms.  Unfortunately, generic
$D$-dimensional cuts are significantly more complicated
than their four-dimensional counterparts.  For continuous values of $D$,
there is no helicity formalism, nor is there a useful on-shell superspace.
Some of the additional complexity can avoided by
working in $D$ dimensions, but with the $\NeqFour$ states organized
according to the $\NeqOne$ super-Yang-Mills theory in ten
dimensions~\cite{BCDKS}.
Nevertheless, evaluating general cuts in this way at four loops is
still difficult.

It is important to note that we do not need the full power of
$D$-dimensional cuts to construct the amplitude, but only to identify
any potential terms dropped by the four-dimensional cuts.  Such terms
are rare or even nonexistent for \NeqFoursYM\ at low orders and
low multiplicity.  Specifically, for four-point amplitudes through three
loops, and for the planar contributions through four loops, explicit
computation has revealed~\cite{GSB,BRY,BDDPR,CompactThree,BCDKS} 
that the $D$-dimensional versions of the amplitudes are obtained
simply by replacing the four-dimensional loop integration measure with
the $D$-dimensional one,
\begin{equation}
\int {d^4 p\over (2\pi)^4}  \rightarrow \int {d^D p\over (2\pi)^D} \,,
\label{SimpleDimensionContinuation}
\end{equation}
and reinterpreting all Lorentz products of momenta as $D$-dimensional
ones. Based on this evidence, we have every reason to believe that
\eqn{SimpleDimensionContinuation} holds as well for the non-planar
contributions at four loops. Although we have not checked all cuts in
$D$ dimensions in this paper, we have performed a set of strong
consistency checks to make it extremely unlikely that any
$D$-dimensional contributions have been missed. Such checks include
all two-particle cuts, and all generalized cuts that isolate a
four-particle sub-amplitude (box cuts), as shown
in~\fig{DdimensionalCutsFigure}.  As noted some time
ago~\cite{BRY,BDDPR}, for four-point amplitudes, the 
iterated two-particle cuts automatically
give the same result in $D$ dimensions as in
four dimensions.  In the next section we will explain why the
box cuts have the same property. Another powerful check comes from the
new diagrammatic numerator identities~\cite{BCJ,BCJLoop}, which hold
in any dimension.  They allow us to obtain many non-planar terms
directly from planar ones.  At four loops, the latter are known to be
valid in $D$ dimensions~\cite{BCDKS}, at least for external gluon
states.  Because of these checks it is rather unlikely that any terms
were dropped in extending the four-dimensional loop-momentum integrand
to $D$ dimensions.  Nevertheless, it would still be useful to evaluate
a complete set of unitarity cuts in $D$ dimensions.  As a step in this
direction, the unitarity cuts have been confirmed for six-dimensional
external and cut momenta~\cite{FutureD6}, using the helicity formalism
of Cheung and O'Connell~\cite{OConnell} and the on-shell superspace of
Dennen, Huang and Siegel~\cite{SixDimSusy}.


\section{Constructing a compact ansatz}
\label{MagicToolsSection}

In the process of constructing amplitudes, it is helpful to have
a toolkit that allows one to write down large classes of terms with
essentially no
computation.  Even heuristic rules motivated by observed structures,
or tools that capture only a subset of terms, can be rather
useful. Typically, such rules allow one to quickly fix the
structurally simplest terms in the amplitude, allowing the remaining
effort to be focused on the more intricate ones. This strategy is
especially potent when combined with the method of maximal
cuts~\cite{FiveLoop}, which (as discussed below) allows a relatively
small set of contributions to be considered in isolation.

For planar \NeqFoursYM~there are a set of powerful graphical tools.
The oldest of these tools is the ``rung insertion rule''~\cite{BRY},
which generates certain higher-loop contributions from lower-loop
ones. More recently, the observed dual conformal properties of planar
\NeqFoursYM~\cite{MagicIdentities,BCDKS} amplitudes have led to a
powerful method for determining them, up to
prefactors~\cite{FiveLoop,TwoLoopSixPt,SpradlinLeadingSingThreeLoop,Vergu}
that can be determined straightforwardly from cuts.  Heuristic rules
for determining the prefactors in the planar four-point case have been
given as well~\cite{FiveLoop,KorchemskyZeros,CachazoSkinner}.
Unfortunately, it is not clear how to extend the notion of dual
conformal invariance to non-planar contributions.  
We also remark that the planar terms in the four-loop
four-point amplitude were determined previously (without using
dual conformal invariance)~\cite{BCDKS}.  The non-planar contributions
are much more intricate. In this section we will discuss tools that
are useful for identifying both planar and (more importantly)
non-planar contributions.

We begin by reviewing and extending some particularly useful cuts
that can be expressed very simply, and in all generality, in terms
of lower-loop expressions.  We also discuss a tree-level identity
that allows many multi-loop contributions to be constructed, 
up to potential contact terms.  We will close this section
by discussing the method of maximal cuts, which provides a systematic
tool for constructing all terms in any amplitude, including contact terms.


\subsection{Two-particle cuts}
\label{TwoParticleCutSection}

\begin{figure}[tbh]
\centerline{\epsfxsize 1.9 truein \epsfbox{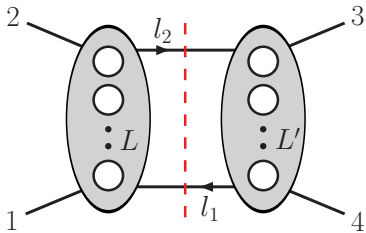}}
\caption[a]{\small A two-particle cut that may be used to construct
contributions to the $(L+L'+1)$-loop amplitude from those at
$L$ and $L'$ loop orders.}
\label{TwoParticleCutFigure}
\end{figure}

A {\it two-particle cut} of a multi-loop four-point amplitude
has the form shown in \fig{TwoParticleCutFigure} --- it divides
the amplitude into two lower-loop four-point amplitudes.
Four-point amplitudes in \NeqFoursYM\ have an especially simple dependence
on the external states.  This fact makes it possible to
immediately write down the numerator factors for parent graphs 
that have two-particle cuts, in terms of lower-loop numerator factors
(up to potential contact term ambiguities).  This method can be applied
to non-planar parent graphs as well, making it especially powerful.

This simplicity relies on the observation that all \NeqFoursYM\ four-point
amplitudes can be expressed in a common factorized form,
\begin{equation}
{\cal A}_4^{(L)}(1,2,3,4)
= g^{2L + 2} {\cal K}(1,2,3,4) \, \UniversalFactor^{(L)}(1,2,3,4)\,,
\label{FourPointFactorization}
\end{equation}
where ${\cal A}_4$ represents the full color-dressed amplitude (as
distinguished from the color-stripped $A_4$).
All of the state dependence in ${\cal A}_4$ is carried by
the kinematic prefactor ${\cal K}=s_{12}s_{23} A_4^\tree$, also
defined in \eqn{Prefactor}.
All of the color dependence is carried by the state-independent universal
factor $\UniversalFactor^{(L)}$.  We will use this factorization of
color and state dependence to determine the terms in 
$\UniversalFactor^{(L)}$ that are visible in two-particle cuts,
iteratively in terms of lower-loop universal factors.  This result 
will be valid in $D$ dimensions, whenever the lower-loop universal factors
are valid in $D$ dimensions.

For four-dimensional external momenta and states,
\eqn{FourPointFactorization} follows from the Ward identities for
maximal supersymmetry.  These identities relate all four-point
amplitudes to each other at any loop
order~\cite{SWI,BDDPR,FreedmanSWI}.  From explicit computations, we
know that this equation holds in $D$ dimensions for all states at one
and two loops and at least for gluon amplitudes at three
loops~\cite{GSB,BRY,BDDPR,GravityThree}.  Using the observation that,
in theories with 16 supercharges, the number of states ($2^8$) in a
massive representation of the supersymmetry algebra is the same as the
number of states in a product of two short (massless) representations
($2^4\times 2^4$), Alday and Maldacena \cite{AldayMaldacena} argued
that the intermediate states in a $2\rightarrow 2$ scattering process
form a single supermultiplet in any dimension. This argument suggests
that \eqn{FourPointFactorization} also holds in any dimension, with
the loop factor $U^{(L)}$ capturing the $L$-loop correction to this
multi-particle intermediate state.  We therefore assume that
\eqn{FourPointFactorization} is valid in $D$ dimensions, for any
two-particle cut of the four-loop four-point $\NeqFour$ amplitude.

In order to treat the tree-level case, $L=0$, on an equal footing
with loop level, we note that the color-dressed tree amplitude
in \NeqFoursYM\ can be written as 
\begin{equation}
{\cal A}_4^{(0)}(1,2,3,4)=  g^2 A_4^{(0)}(1,2,3,4) \times \\
\biggl (
 \colorf{a_4a_1b}\colorf{ba_2a_3} 
+\colorf{a_3a_1b}\colorf{ba_2a_4}  \frac{s_{23}}{s_{13}}
\biggr )\, ,
\label{TreeAllColors}
\end{equation}
where $A_4^{(0)} \equiv A_4^\tree$, and we have used the color-Jacobi
identity to eliminate the color factor $\colorf{a_1a_2b}\colorf{ba_3a_4}$ in 
favor of the other two.  We also used the fact that ${\cal K}$ is
crossing symmetric (see \eqn{SuperAmplitudeII}), which implies 
that all the orderings of the color-ordered tree amplitude
$A_4^{(0)}(1,2,3,4)$ are related simply to each other, up to
ratios of kinematic invariants.  Dividing \eqn{TreeAllColors}
by ${\cal K}$, we see that the state-independent color-dressed
universal factor at tree level, $\UniversalFactor^{(0)}$, defined
by \eqn{FourPointFactorization}, is given by,
\begin{equation}
{\UniversalFactor}^{(0)}(1,2,3,4) = 
 \biggl(
 \frac{\colorf{a_4a_1b}\colorf{ba_2a_3}} {s_{12} s_{23}}
+\frac{\colorf{a_3a_1b}\colorf{ba_2a_4}} {s_{12} s_{13}}
 \biggr)\,.
\label{TreeUniversalFactor1}
\end{equation}

In general, the universal factor $\UniversalFactor^{(L)}$ is a 
sum of $L$-loop integrals.  The
integrands entering $\UniversalFactor^{(L)}$ are rational functions of
momentum invariants involving the loop and external momenta. 
Explicit formul\ae~for the universal factors for
$L=1,2,3$, including planar and non-planar contributions, may be found by
matching \eqn{FourPointFactorization} with the known amplitudes
already presented in
\sect{ColorOrganizationSubsection}:~(\ref{OneLoopYMAmplitude}),
(\ref{TwoLoopYMAmplitude}) and (\ref{ThreeLoopYMAmplitude}).

Next we evaluate the generic two-particle color-dressed cut
depicted in \fig{TwoParticleCutFigure}.  It cuts the $(L+L'+1)$-loop
amplitude ${\cal A}_4^{(L+L'+1)}(1,2,3,4)$ into the two four-point
amplitudes ${\cal A}_4^{(L)}(-l_1,1,2,l_2)$ and
${\cal A}_4^{(L')}(-l_2,3,4,l_1)$ of loop orders $L$ and $L'$,
respectively.  The cut has the form,
\begin{equation}
{\cal A}_4^{(L+L'+1)}(1,2,3,4)\Big|_{\rm 2\hbox{-}cut}
= i^2 \, \sum_{\NeqFour \atop {\rm states}}
{\cal A}_4^{(L)}(-l_1,1,2,l_2) \, {\cal A}_4^{(L')}(-l_2,3,4,l_1) \,,
\label{GenLoopTwoParticleCut}
\end{equation}
where the state sum is over the particles with momenta $l_1$ and $l_2$.

Using the factorization~(\ref{FourPointFactorization}) and the 
state-independence of $\UniversalFactor^{(L)}$, we can immediately
rewrite the cut as follows:
\begin{eqnarray}
\UniversalFactor^{(L+L'+1)}(1,2,3,4)\Big|_{\rm 2\hbox{-}cut}
 \times {\cal K}(1,2,3,4) 
&=& i^2 
\, \UniversalFactor^{(L)}(-l_1,1,2,l_2) \, 
  \UniversalFactor^{(L')}(-l_2,3,4,l_1) \nonumber\\
&& \hskip0.3cm \times
  \sum_{\NeqFour \atop {\rm states}}
  {\cal K}(-l_1,1,2,l_2) \, {\cal K}(-l_2,3,4,l_1) \,.
\label{GenLoopTwoParticleCutRewrite}
\end{eqnarray}
Substituting in the definition of ${\cal K}$ given above, we find:
\begin{eqnarray}
\UniversalFactor^{(L+L'+1)}(1,2,3,4)\Big|_{\rm 2\hbox{-}cut}
\times s_{12} s_{23} A_4^{(0)}(1,2,3,4)
&=& i^2 
\, \UniversalFactor^{(L)}(-l_1,1,2,l_2) \, 
  \UniversalFactor^{(L')}(-l_2,3,4,l_1) \nonumber \\
&& \hskip-2.0cm \times
 s_{12}^2 \, s_{2 l_2} s_{4 l_1}  \sum_{\NeqFour \atop {\rm states}} 
  \, A_4^{(0)}(-l_1,1,2,l_2) \,  
    A_4^{(0)}(-l_2,3,4,l_1) \,. \nonumber \\
\label{GenLoopTwoParticleCutRewrite2}
\end{eqnarray}
To evaluate this, we use the sewing relation between two four-point
color-ordered \NeqFoursYM\ trees~\cite{BRY,BDDPR},
\begin{eqnarray}
\sum_{\NeqFour \atop \rm states}
 A_4^{(0)}(-l_1,1,2,l_2) A_4^{(0)}(-l_2,3,4,l_1)
= - i \sv \tv \, A_4^{(0)}(1,2,3,4) {1\over s_{2l_2} s_{4l_1}} \,.
\label{SewingRelation}
\end{eqnarray}
This sewing relation is valid in any dimension $D$ and for any
external states in the $\NeqFour$ multiplet.  A straightforward way to
confirm \eqn{SewingRelation} is to work in $D=10$ and evaluate the sum
over states in components, using the fact that in $D=10$ \NeqFoursYM\
is equivalent to an $\Neqone$ theory composed of a gluon and a gluino.
By dimensional reduction the sewing relation~(\ref{SewingRelation})
then holds in any dimension $D\le 10$.  Recently, this equation has also
been verified directly in six dimensions using an on-shell
superspace~\cite{SixDimSusy}. 

\begin{figure}[tb]
\centerline{\epsfxsize 5.8 truein \epsfbox{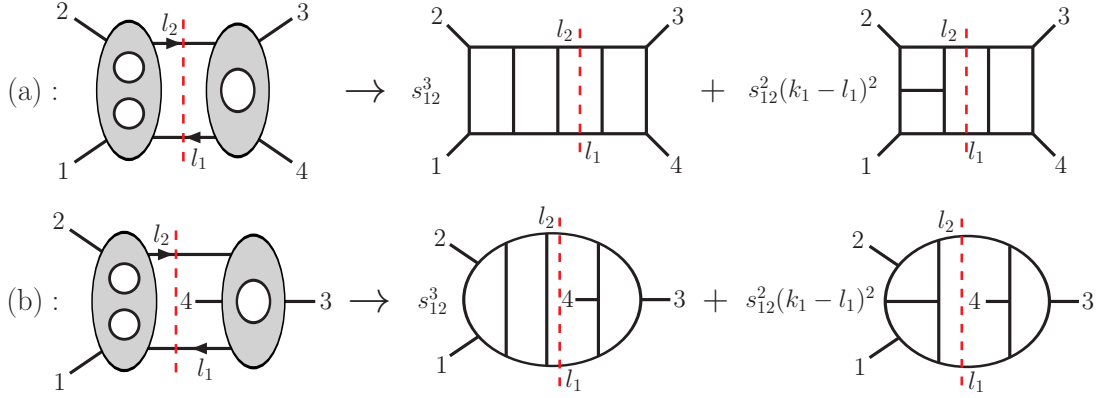}}
\caption[a]{\small Sample contributions to the full color-dressed
  two-particle cut for $L=2$ and $L'=1$.  The diagrams on the
  right-hand side show some of the terms in
  $\UniversalFactor^{(4)}(1,2,3,4)$ that are constructed from these
  cuts, using \fig{TwoParticleCutConstructPiecesFigure}.  The explicit
  color factors, as well as factors of $i$, have been omitted.}
\label{TwoParticleCutConstructFigure}
\end{figure}

\begin{figure}[tb]
\centerline{\epsfxsize 5 truein
\epsfbox{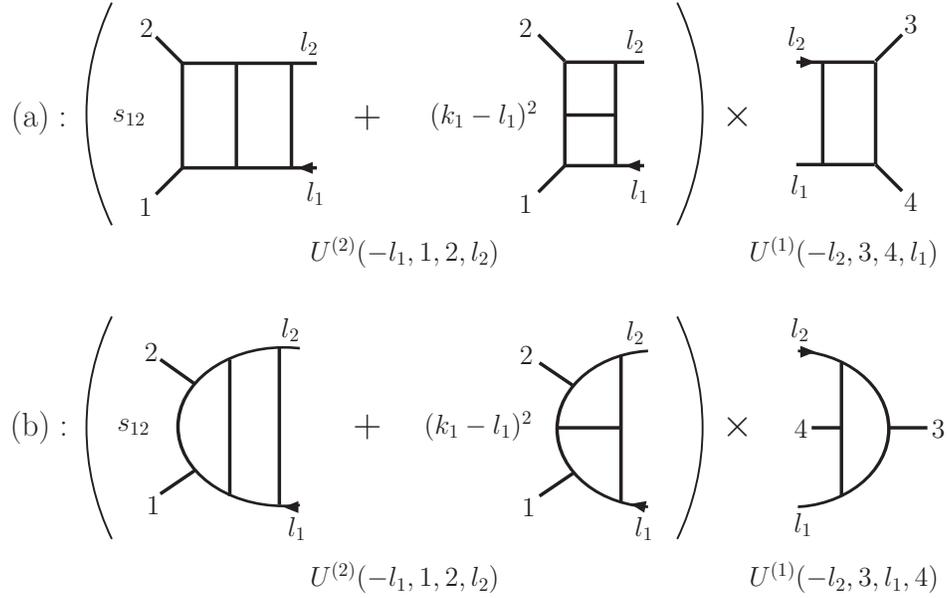}}
\caption[a]{\small  The lower-loop integral functions entering the cuts
on the left-hand side of \fig{TwoParticleCutConstructFigure}.  This figure
displays in detail how the prefactor of the planar double-box
integral appears in $U^{(2)}(-l_1,1,2,l_2)$ as either $s_{12}$
or $(k_1-l_1)^2$, depending on the permutation.}
\label{TwoParticleCutConstructPiecesFigure}
\end{figure}

Applying \eqn{SewingRelation} to  \eqn{GenLoopTwoParticleCutRewrite2},
we find the key equation for building all contributions
from two-particle cuts directly in terms of the $\UniversalFactor$s: 
\begin{equation}
\UniversalFactor^{(L+L'+1)}(1,2,3,4)\Big|_{\rm 2\hbox{-}cut}
= i \, s^2_{12}\, \UniversalFactor^{(L)}(-l_1,1,2,l_2) 
             \, \UniversalFactor^{(L')}(-l_2,3,4,l_1) \,.
\label{TwoParticleSigmaSewing}
\end{equation}
\Eqn{TwoParticleSigmaSewing} is rather powerful.  No complicated
calculations remain in order to obtain all contributions visible in
two-particle cuts; they are given simply by taking the product of
lower-loop results.  The color-dressed $\UniversalFactor^{(L+L'+1)}$
is given immediately as a sum over products of individual integrals
residing inside the $\UniversalFactor^{(L)}$ and
$\UniversalFactor^{(L')}$ factors, up to terms that vanish because of
the on-shell conditions, $l_1^2 = l_2^2 = 0$.  (As a straightforward
exercise, one can verify that the one-loop universal
factor $\UniversalFactor^{(1)}$ --- which can be extracted from
\eqn{OneLoopYMAmplitude} --- satisfies this equation, using the
tree-level universal factor given in \eqn{TreeUniversalFactor1}.)

\Figs{TwoParticleCutConstructFigure}{TwoParticleCutConstructPiecesFigure}
illustrate diagrammatically some of the terms generated by
\eqn{TwoParticleSigmaSewing} for the case $L=2$ and $L'=1$. For
simplicity, we draw only the planar contributions of
$\UniversalFactor^{(2)}(-l_1,1,2,l_2)$, encoding the $\tilde{f}^{abc}$
visually in the diagrams, and we omit all factors of $i$.  The
denominator factors in $\UniversalFactor^{(2)}$ and
$\UniversalFactor^{(1)}$ correspond to propagators that are visible
on the left- and right-hand sides of
\fig{TwoParticleCutConstructPiecesFigure}, respectively.
Therefore they are accounted for graphically in
$\UniversalFactor^{(3)}|_{\rm 2\hbox{-}cut}$
simply by connecting the $l_1$ and $l_2$ legs of the corresponding 
diagrams.  Similarly, the numerator factor for each parent graph on the
right-hand side of \fig{TwoParticleCutConstructFigure} is given by
forming the product of the numerator factors for the two sewn
subdiagrams in \fig{TwoParticleCutConstructPiecesFigure} (taking into
account the proper permutation of legs), and then multiplying by two
powers of $s_{12}$.

This diagrammatic interpretation of the two-particle cuts provides a
rather simple tool for generating many higher-loop contributions from
known lower-loop ones.  It is the mechanism behind the rung
rule~\cite{BRY,BDDPR}.  For the planar contributions at four loops,
the two-particle cuts have either $L=2$, $L'=1$, as in
\fig{TwoParticleCutConstructFigure}, or else $L=3$, $L'=0$.  Together,
they capture diagrams (a)-(e) in \fig{Planar4lFigure}, but not diagram
(f).  (Diagram (f) can still be guessed from the rung rule,
or constructed using a box cut, as described in the next subsection.)
These cuts also do not guarantee the absence of contact terms that have
no two-particle cuts, such as diagrams (f$_2$) and (d$_2$) in
\fig{Planar4lNonRRFigure}.  For the full four-loop amplitude described
in \sect{AmplitudeSection}, 33 of the 50 parent graphs contain
two-particle cuts (graphs 1 through 27, and graphs 40 through 45).
The two-particle cuts capture the majority of the terms contributing
to these graphs. Because the two-particle cut sewing algebra is
valid in $D$ dimensions, all contributions obtained by iterating
two-particle cuts are automatically valid in $D$ dimensions.
Surprisingly, the two-particle cuts capture the majority of terms in
the 33 parent graphs containing them.  The fact that so many
potential contact terms are absent hints at further structures
to be uncovered.

We note that in $\NeqEight$ supergravity, the two-particle cuts have
an equally simple structure~\cite{BDDPR}, which can be exploited
analogously.


\subsection{Box cuts}
\label{BoxCutSection}

\begin{figure}[t]
\centerline{\epsfxsize 5.5 truein \epsfbox{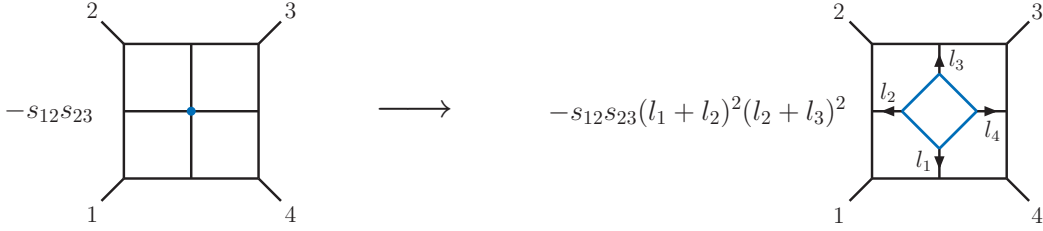}}
\caption[a]{\small The box-substitution rule~\cite{FiveLoop}
  for generating a higher-loop contribution by inserting a one-loop
  four-point box subintegral into a four-point vertex. In
  this example, we substitute a box into the central four-point vertex
  in the four-loop ``window'' diagram. The result is a five-loop
  integral that cannot be obtained from two-particle cuts. 
  (Note that an overall normalization factor of $s_{12}s_{23}$ has been
  absorbed into ${\cal K}$, relative to ref.~\cite{FiveLoop}.)}
\label{BoxSubRuleFigure}
\end{figure}

The simple structure of the four-point amplitude in \NeqFoursYM\
can also be applied to (generalized) four-particle cuts that isolate
a four-point sub-amplitude.
A simple version of this generalization appeared already~\cite{FiveLoop}
as a ``box-substitution rule''.  It allowed the
construction of $L$-loop contributions with a box subgraph,
starting from $(L-1)$-loop contributions with a contact interaction, as
illustrated in \fig{BoxSubRuleFigure}. Related rules were
discussed in conjunction with
leading singularities~\cite{CachazoSkinner}.  Here we promote the
box-substitution rule into a more general cut for \NeqFoursYM\
amplitudes in $D$ dimensions, which we call the ``box cut''.

\begin{figure}[t]
\centerline{\epsfxsize 4.5 truein \epsfbox{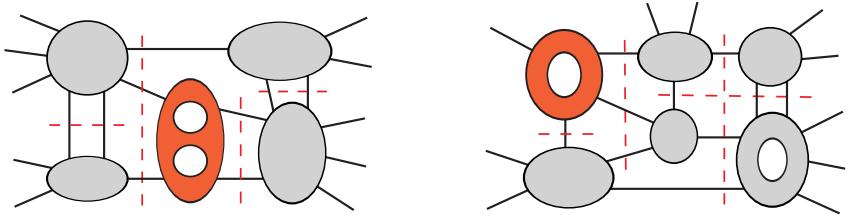}}
\caption[a]{\small Two examples of multi-particle multi-loop ``box
  cuts''. They reduce to lower-loop cuts by replacing the darker (red)
  four-point sub-amplitudes by four-point color-ordered trees,
  multiplied by known numerator and denominator factors. 
  This property allows these cuts to be computed easily in $D$
  dimensions, once the amplitudes with fewer loops are
  known. White holes represent loops and the darker
  sub-amplitudes mark four-point amplitudes amenable to reduction.}
\label{GeneralBoxCutFigure}
\end{figure}

Consider the generalized cut of an $L$-loop $n$-point amplitude,
\begin{equation}
{\cal A}_n^{(L)}\Big|_{\rm box~cut} \equiv 
\sum_{\NeqFour \atop \rm states}
 {\cal A}_{(1)}\cdots {\cal A}_{4, (i)}^{(L')} \cdots {\cal A}_{(m)} \,,
\label{BoxCut}
\end{equation}
that is composed of a generic set of color-dressed amplitude factors,
except for the $i^{\rm th}$ such factor, which we take to be a
color-dressed $L'$-loop four-point sub-amplitude,
${\cal A}_{4, (i)}^{(L')}$.  (There may be additional cut conditions
imposed on this sub-amplitude; its internal kinematics are irrelevant
for the subsequent discussion.)  Example of such box cuts are given
in \fig{GeneralBoxCutFigure}.

The $L'$-loop four-point sub-amplitude of \NeqFoursYM\
is special because of the factorization property
(\ref{FourPointFactorization}).  Labeling the cut legs by $l_1,l_2,l_3,l_4$,
we have,
\begin{equation}
 {\cal A}_{4,(i)}^{(L')}(l_1,l_2,l_3,l_4) 
= A_{4,(i)}^{(0)}(l_1,l_2,l_3,l_4)
\,  (l_1+l_2)^2 (l_2 + l_3)^2 
\, \UniversalFactor^{(L')}(l_1,l_2,l_3,l_4) \,,
\label{FourPointFactorizationBoxCut}
\end{equation}
where, as in the previous section, we use ${\cal A}_4$ to represent
the color-dressed amplitude, and only the color-ordered tree
amplitude factor $A_{4,(i)}^{(0)}$ depends on the states crossing
the cuts.  Therefore we can pull the factor $\UniversalFactor^{(L')}$ 
out of the sum over states in \eqn{BoxCut}, leading to a simpler 
expression in the summand,
\begin{equation}
{\cal A}_n^{(L)}\Big|_{\rm box~cut}=(l_1+l_2)^2 (l_2 + l_3)^2
\UniversalFactor^{(L')}(l_1,l_2,l_3,l_4)
 \, \sum_{\NeqFour \atop \rm states} 
 {\cal A}_{(1)}\cdots  A_{4, (i)}^{(0)} \cdots  {\cal A}_{(m)} \,.
\label{BoxCutTrue}
\end{equation}
The state-sum is identical to a lower-loop cut, that of the
$(L-L')$-loop amplitude, but utilizing the color-ordered contribution
to the $i^{\rm th}$ tree.  This fact immediately gives a simple relation
between the $L$-loop box cut and contributions to the reduced
$(L-L')$-loop cut under the same cut conditions.

We can formally write down an equation relating the cut of an $L$-loop
amplitude to a cut of a lower-loop one as,
\begin{equation}
{\cal A}_n^{(L)}\Big|_{\rm box~cut}
= (l_1+l_2)^2 (l_2+l_3)^2
\,\UniversalFactor^{(L')}(l_1,l_2,l_3,l_4) \, 
\tilde{{\cal A}}_n^{(L-L')}\Big|_{\rm cut} \,.
\label{BoxCutReduction}
\end{equation}
We introduced the reduced cut $\tilde{{\cal A}}$ notation to emphasize that
the state-sum in \eqn{BoxCutTrue} is exactly a $(L-L')$ loop unitarity
cut which is color-dressed with $\colorf{abc}$ everywhere, except for the
four-point color-ordered tree amplitude whose associated color factors are
accounted for in the $L'$-loop universal factor $\UniversalFactor^{(L')}$.

Given a generalized cut that isolates an $L'$-loop four-point sub-amplitude
with legs $l_1,l_2, l_3, l_4$, we can re-express the box cut
as a recipe that can be applied easily to individual diagrammatic
(integral) contributions:
\begin{itemize}
\item Split up the cut into three parts as in \eqn{BoxCutReduction}:
  The reduced cut, $\tilde{{\cal A}}_n^{(L-L')}\big|_{\rm cut}$, 
  the kinematic factor $(l_1+l_2)^2 (l_2+l_3)^2$, and the loop integrals
  $\UniversalFactor^{(L')}$ of the four-point sub-amplitude.
   (This latter part generalizes to $L'$ loops the one-loop box
   integral of the box substitution rule.)
\item Express the reduced cut of the known lower-loop amplitude in a
   diagrammatic form that corresponds to a covariant integral
   representation.
\item The diagrams of the reduced cut may contain spurious propagators
  in the $(l_1+l_2)^2$ or $(l_2 + l_3)^2$ channels, which upon
  multiplication cancel against the $(l_1+l_2)^2 (l_2+l_3)^2$
  prefactor in \eqn{BoxCutReduction}.  The result is always a diagram
  with an internal four-point contact vertex.
\item To recover the integrals of the original box cut, insert the
  four-point integrals of $\UniversalFactor^{(L')}$ ({\it e.g.,} the box
  integral for $L'=1$) into the obtained four-point contact vertex of
  each diagram.
  \end{itemize}
  
\begin{figure}[t]
\centerline{\epsfxsize 5.5 truein \epsfbox{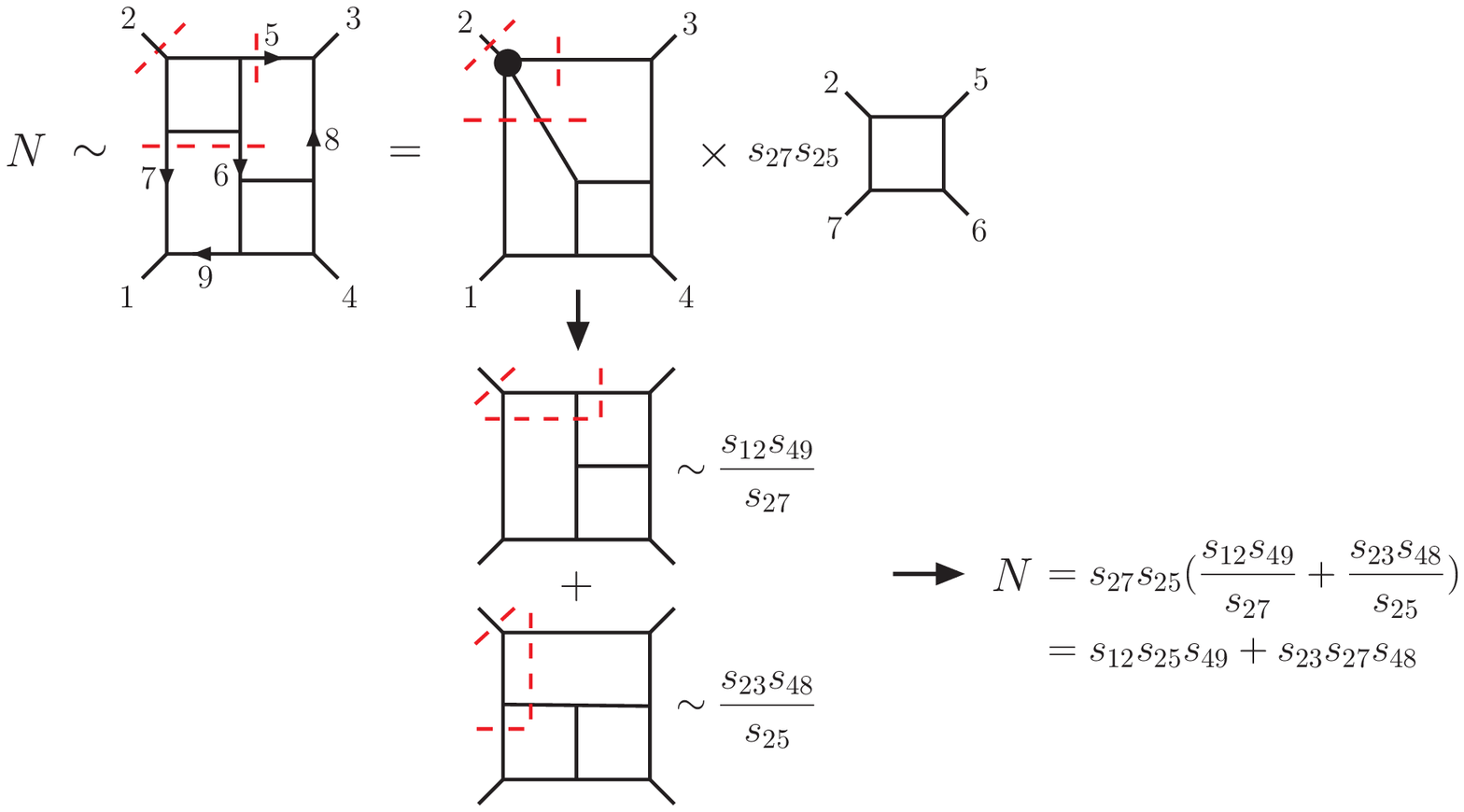}}
\caption[a]{\small Application of the box cut to determine the
  numerator $N$ for the four-loop parent graph in
  \fig{Planar4lFigure}(f).  The (red) dashed cut conditions around the
  upper left box in this diagram allow us to replace it by the product
  of a reduced cut diagram, some kinematical factors and a box
  integral.  The reduced cut diagram is then expanded into two
  three-loop ``tennis-court'' diagrams, corresponding to the two
  allowed channels of the marked four-point vertex.  The relevant
  kinematical pieces of the tennis-court diagrams, {\it i.e.} the
  numerators and the spurious propagators, are extracted from the
  known three-loop contribution in
  \fig{IntegralsThreeLoopFigure}(e). Assembling all the kinematical
  factors gives the result for $N$, which is free of spurious
  propagators.  The result is consistent with \fig{Planar4lFigure}(f);
  here the numerator is symmetrized with respect to the 
  $(1 \leftrightarrow 3)$ symmetry of this parent graph.}
\label{boxcutFigure}
\end{figure}

\Fig{boxcutFigure} shows how the box cut can be used to determine
the numerator polynomial for \fig{Planar4lFigure}(f), using
the three-loop information in \fig{IntegralsThreeLoopFigure}(e).
Although this example is planar (and is presented in a color-ordered
way in the figure), it is just as simple to use the box cut
for non-planar contributions.  For example, inserting a box into the
four-point vertex in \fig{boxcutFigure} in a non-planar fashion generates
contributions to parent graph 29 in the full four-loop amplitude.

The box cut is an extremely efficient way to obtain contributions
to parent graphs that contain a lower-loop four-point subgraph.
As mentioned earlier, the $L'=1$ box cut is closely
related to the box-substitution rule.  The box cut also generates
contributions that are consistent with the rung rule~\cite{BRY}.

Box cuts capture a majority of those terms in the complete
four-loop four-point amplitude in \sect{AmplitudeSection} that are
not determined by two-particle cuts.  Of the 17 parent graphs
that do not have two-particle cuts, 13 of them have box cuts.
The only four that have neither two-particle cuts nor box cuts
are graphs 39, 48, 49 and 50.
In fact, most of the parent integrals have multiple box cuts,
allowing us to constrain their numerators under complementary cut
conditions, and to fix many of the contact terms.

From the above covariant derivation it follows that box cuts are
valid in any dimension, if both the reduced cut $A_n^{(L-L')}\big|_{\rm
cut}$ and the four-point universal factor $\UniversalFactor^{(L')}$ are
known in $D$ dimensions.  As a practical matter, the universal factors
entering the lower-loop amplitudes should already be known
in $D$ dimensions, prior to attempting the higher-loop
calculation in $D$ dimensions.  In the case relevant to this paper,
$L=4$ and $n=4$, all we need as input are the $L'=1,2,3$ four-point
amplitudes, which are indeed known in $D$
dimensions~\cite{BRY,BDDPR,GravityThree}.

The effectiveness of the box cut suggests that one should investigate
analogous ``pentagon cuts'', {\it etc.}, which isolate sub-amplitudes
with five or more legs.  Both the color and kinematic structure of the
five- and higher-point loop amplitudes is, however, more
intricate, and there is no simple factorization property
similar to \eqn{FourPointFactorizationBoxCut}.  (See for example,
the five- and six-point loop amplitudes described in
refs.~\cite{DimShift,TwoLoopSixPt,Vergu}.)

The box cut also easily generalizes to $\NeqEight$ supergravity, because
its four-point amplitude has a factorized form similar to
\eqn{FourPointFactorization}, which is related to the existence
of analogous supersymmetric Ward identities~\cite{SWI}.


\subsection{Color-kinematic duality}

In the early 1980s, radiation zeroes appearing in certain 
gauge-theory cross sections were traced back to a curious identity
obeyed by tree-level four-point amplitudes~\cite{Zhu,GHL}. 
This curiousity turns out to be the simplest of a set of relations
arising from a general tree-level duality between color factors
and kinematic numerators~\cite{BCJ}.  If one assumes that the duality
holds for an arbitrary number of external states, one can
derive~\cite{BCJ} new relations among color-ordered tree amplitudes,
which have since been proven~\cite{BoFengBCJProof}.
Similar relations among string theory amplitudes have also
been proven recently~\cite{BjerrumBohrJacobi,StiebergerJacobi}.
In the low-energy limit, the string-theory relations become identical to 
two types of field-theory relations:  the Kleiss-Kuijf
relations~\cite{KleissKuijf} (which follow from color considerations
alone~\cite{DDDM}) and the amplitude relations which 
follow from the color-kinematic duality.

In this subsection, we discuss how the four-point tree-level
color-kinematic identity may be combined with generalized
unitarity at the loop level~\cite{BCJ}, particularly
to the construction of the four-loop \NeqFoursYM\ amplitude.  In short,
the identity relates sets of three parent graphs that only differ in how
a four-point cubic tree graph is glued into the rest of the graph.
Evidence that the color-kinematic duality also holds directly
at the loop level, {\it without} the need to impose on-shell conditions, 
was presented recently for the three-loop four-point
amplitude~\cite{BCJLoop}.

\begin{figure}[tbh]
\centerline{\epsfxsize 5. truein \epsfbox{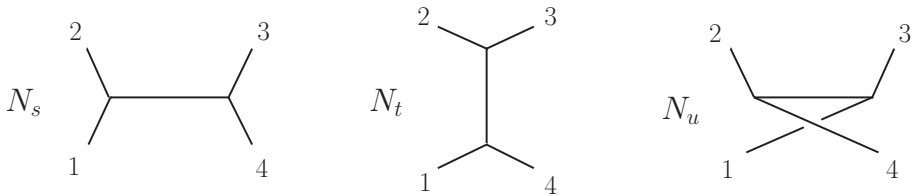}}
\caption[a]{\small Graphs for the four-point tree amplitude.
Contact terms are absorbed into the diagrams as inverse 
propagators.  Each diagram is associated with a color factor obtained
by dressing the vertices with an $\colorf{abc}$, as in
\sect{ColorOrganizationSubsection}.}
\label{ColorJacobiFigure}
\end{figure}

Consider the color-dressed four-point tree amplitude.  
Just as for the loop amplitudes discussed in
\sect{ColorOrganizationSubsection}, it can be written as a sum
of color factors $C_i$ multiplied by kinematic factors.  The
kinematic factors can be further divided into denominators, which are
propagators associated with (tree-level) parent graphs,
and numerators $N_i$.  As in \sect{ColorOrganizationSubsection},
contact terms can be absorbed into the $N_i$, so that we require
only the three cubic graphs shown in \fig{ColorJacobiFigure}.
In this representation the amplitude is,
\begin{equation}
{\cal A}^\tree_4 = g^2 \biggl( \frac{N_s C_s}{s} + \frac{N_t C_t}{t} + 
              \frac{N_u C_u}{u} \biggr) \,,
\label{FullTree}
\end{equation}
where $s = (k_1 + k_2)^2$, $t = (k_2 + k_3)^2$ and $u=(k_1 + k_3)^2$
correspond to the three channels, and 
\begin{eqnarray}
&& 
C_s \equiv \colorf{a_1 a_2 b} \colorf{b a_3 a_4} \,, \hskip 1 cm 
C_t \equiv \colorf{a_2 a_3 b} \colorf{b a_4 a_1} \,,  \hskip 1 cm 
C_u \equiv \colorf{a_4 a_2 b} \colorf{b a_3 a_1 } \,,
\end{eqnarray}
are color factors corresponding to the three graphs in
\fig{ColorJacobiFigure}.  The color factors of
the graphs satisfy the Jacobi identity,
\begin{equation}
C_u = C_s - C_t \,.
\label{JacobiIdentity}
\end{equation}

The $N_i$ in \eqn{FullTree} contain momentum invariants, polarization
vectors, spinors and superspace Grassmann parameters.  The only real
restriction on them is that \eqn{FullTree} gives the correct color-dressed
tree amplitude.  Hence there is a tremendous amount of freedom in the 
definition of the numerator factors. (Non-local $N_i$ could even be
allowed.)   This freedom is just the tree-level
analog of the inherent ambiguity in the multi-loop parent-graph
decomposition mentioned in \sect{Unitaritymethodsubsection}.
We refer to the invariance of \eqn{FullTree} under this 
freedom as a ``generalized gauge invariance.''  
For every such generalized gauge choice for the four-point \NeqFoursYM\
amplitude, the numerator factors must satisfy the identity~\cite{BCJ},
\begin{equation}
N_u = N_s - N_t \,,
\label{TwistIdentity}
\end{equation}
in concordance with the color Jacobi identity (\ref{JacobiIdentity}).
We emphasize that the identity (\ref{TwistIdentity}) is only between
the numerator factors; it does {\it not} involve the propagators
associated with the $s$, $t$, and $u$ channel graphs.  It is fairly
straightforward to check that these identities hold in $D$ dimensions
by direct computation~\cite{Zhu}.  Although it is not relevant to this
section, it should be noted that for higher-point tree amplitudes
the color-kinematic duality is only manifest for certain special
generalized gauge choices~\cite{BCJ,Square}.

In conjunction with the unitarity method, the tree-level four-point
numerator identity~(\ref{TwistIdentity}) becomes quite powerful.  In
every multi-loop parent graph that contains four on-shell propagators
arrayed around a four-point tree sub-amplitude, it relates the
numerator factor to those of two other parent graphs satisfying those
conditions~\cite{BCJ}.  The three multi-loop parent graphs correspond
to gluing in the four-point cubic tree graph in its $s$, $t$, or $u$
channel configuration.
For every line of each cubic graph this identity will always relate the
numerators of three graphs.  However, the relations do not have to be
manifest in a given amplitude representation, because of the freedom 
to move contact terms\footnote{One can automatically disregard such
contact terms by considering near-maximal cuts where only the central
propagator in the four-point tree graph is off shell.}
associated with other propagators between different graphs.

These relations allow one to take kinematic numerator
information, obtained using dual-conformal symmetry (for planar
graphs), two-particle cuts and box cuts, and export that information
to other parent graphs or contributions for which such methods are
{\it not} applicable.

\begin{figure}[t]
\centerline{\epsfxsize 4.5 truein \epsfbox{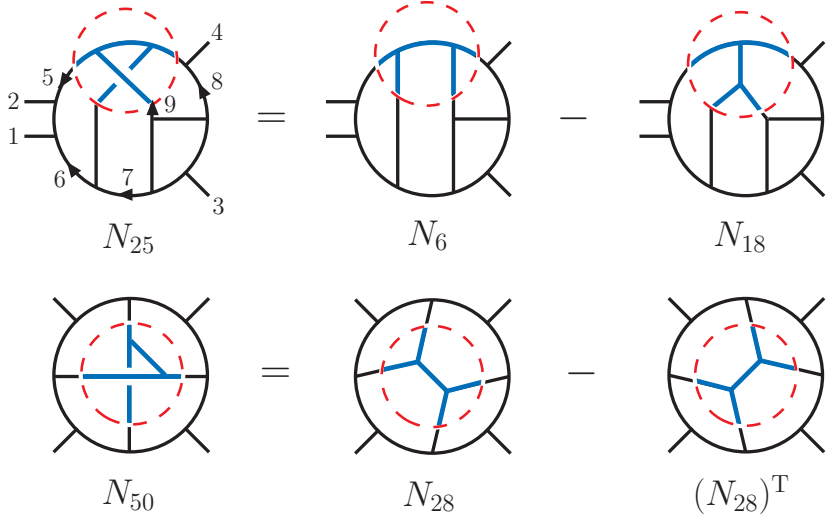}}
\caption[a]{\small Two nontrivial relations between numerators of planar
  and non-planar contributions at four loops, which follow from the
  tree-level numerator relations in \fig{ColorJacobiFigure}.  For the three
  graphs in each relation, the configuration of lines outside the region
  marked by a dashed circle is identical; the only difference
  is for the lines fully inside this region.
  The relation on the second line involves the same
  graph numerator $N_{28}$ with two different labelings of momenta.
}
\label{TwistExamplesFigure}
\end{figure}

To see how this works, consider the two four-loop examples illustrated
in \fig{TwistExamplesFigure}.  In each case, the numerators on the
right-hand side are planar, and are relatively simple to obtain using
two-particle cuts ($N_{6}$ and $N_{18}$) or box cuts ($N_{28}$, at least
up to contact terms).  We then use the four-point tree color-kinematic
duality to obtain the bulk of the non-planar graph numerators on the
left-hand side of each equation.  The numerators of the four-point
tree amplitudes entering the cut satisfy the
relation~(\ref{TwistIdentity}) (see also \fig{ColorJacobiFigure}).
The remaining contributions from the outside of the dashed circle
are identical in all three contributions.  Hence we
obtain a numerator identity for the loop integrands,
\begin{equation}
N_a\bigr|_{\rm cut} = (N_b - N_c) \bigr|_{\rm cut} \,,
\label{NumeratorIntegralIdentity}
\end{equation}
where the $N_x$ are the kinematic numerators of the integrals
corresponding to the graphs in \fig{TwistExamplesFigure}.  This
relation is valid in $D$ dimensions.  However, it holds only for the
numerator terms that are nonvanishing under the imposed on-shell
conditions, in which the four legs crossing the dashed circle are put 
on shell. Also, it should be realized that some contact terms may be
distributed for convenience into other graphs.  That is, there are a large
number of coupled equations obtained from these constraints, and it is
not necessary to satisfy each one simultaneously for all contact
terms.

To illustrate these ideas in more detail, consider the identity on the
first row of \fig{TwistExamplesFigure}.  It involves the numerators
$N_{25}, N_{6}$ and $N_{18}$ which are presented in the next section,
in figs.~\ref{BC1Figure}, \ref{D1Figure} and \ref{D2Figure}, and also
in \app{NumeratorAppendix}. We relabel the lines to match the labels in
\fig{TwistExamplesFigure}, obtaining,
\begin{eqnarray}
N^{\rm fig.\,\ref{TwistExamplesFigure}}_{25}
&=&-s_{12}(s_{45}s_{38}-s_{12}s_{37})
+ l_9^2 s_{12}s_{45} + l_6^2 l_8^2 s_{12}
\,,\nn \\
N^{\rm fig.\, \ref{TwistExamplesFigure}}_{6}
&=&s^2_{12}s_{45}\bigr|_{\{3\leftrightarrow 4, 5\rightarrow 7\}}
= s^2_{12}s_{37} \,,\nn \\
N^{\rm fig.\, \ref{TwistExamplesFigure}}_{18}
&=&s_{12}s_{35}s_{46}\bigr|_{\{3\leftrightarrow 4, 6\rightarrow 8\}}
= s_{12}s_{45}s_{38}\,.
\label{RelabeledNumerators}
\end{eqnarray}
In addition, numerator $N^{\rm fig.\,\ref{TwistExamplesFigure}}_{25}$
picks up a sign relative to $N_{25}$ in \fig{D2Figure}. That is because
the deformation of graph 25 in \fig{D2Figure} into the graph in
\fig{TwistExamplesFigure} requires an odd number of three-vertex
reorderings (three).  Each reordering results in a minus sign (from
the structure constants) for the color factor $C_{25}$, and a
corresponding minus sign for $N_{25}$.

Using \eqn{RelabeledNumerators}, it is easy to see that the numerator
relation (almost) holds on the cut:
\begin{equation}
N^{\rm fig.\,\ref{TwistExamplesFigure}}_{25}\bigr|_{\rm cut} =
 (N^{\rm fig.\,\ref{TwistExamplesFigure}}_{6} - 
N^{\rm fig.\,\ref{TwistExamplesFigure}}_{18}) \bigr|_{\rm cut} \,.
\end{equation}
The on-shell conditions on the legs crossing the dashed circle include
$l_9^2=0$, so the term $l_9^2 s_{12}s_{45}$ in \eqn{RelabeledNumerators}
should be set to zero.  What about the term $l_6^2 l_8^2 s_{12}$?
It is not zero on the cut, so it should be accounted for.  The alert
reader will notice that canceling propagators 6 and 8 in graph 25
in \fig{D2Figure} gives a graph that is topologically identical to
that obtained by canceling propagators 5 and 8 (or 6 and 7) in graph 28.
Also, terms containing $l_5^2 l_8^2$ and $l_6^2 l_7^2$ are present in
$N_{28}$.  These features allow the $l_6^2 l_8^2 s_{12}$ contact term
in $N_{25}$ to be moved elsewhere to be consistent with the identity.
However, the presence of overlapping identities can complicate
their application, when all contact terms are retained.

The second relation in \fig{TwistExamplesFigure} works similarly.  The
same graph 28 appears twice on the right-hand side, with two
different labelings.  It is worth noting that for our choice of
numerators $N_{28}$ and $N_{50}$, as given in
\figs{D2Figure}{E2Figure} and \app{NumeratorAppendix}, this particular
relation holds even including all contact terms.

It has been conjectured recently~\cite{BCJLoop} that a representation
exists for all multi-loop amplitudes in which all color-kinematic
duality relations are manifest for all graphs, and with no internal
on-shell conditions imposed.  This conjecture has been confirmed
for the three-loop four-point amplitude of \NeqFoursYM, as well as for
certain lower-loop cases~\cite{BCJLoop}, but it remains to be tested
more generally.  Strong evidence in favor of the conjecture would be
provided if the four-loop amplitude presented here can be rearranged
into such a duality-satisfying form.  We leave this exercise to
future work.

At four loops, the three rules just presented can be used to generate
all non-contact-term contributions to the \NeqFoursYM{} amplitude,
as well as many of the contact-term contributions.  To ensure that all
contact terms are captured correctly, we turn to the method of maximal
cuts.


\subsection{Method of Maximal Cuts}
\label{MaximalCuts}

The method of maximal cuts~\cite{FiveLoop,CompactThree} offers a
particularly efficient means for determining the numerator polynomials
for each parent integral.  In this method we start from generalized
cuts with the maximum number of cut propagators (maximal cuts)
and match these cuts against an initial ansatz.
If an ansatz has been constructed that covers all non-contact-term
contributions (for example, by using the three rules just presented),
then this step is merely one of cut-verification.  Next we systematically
reduce the number of cut propagators (by one at each step) and match 
these (near-maximal) cuts --- capturing in the process all potential
contact contributions. 

It is important that massless on-shell three-point amplitudes are
non-vanishing and non-singular~\cite{BCFGeneralized}, for 
appropriate choices of complex cut loop
momenta~\cite{GoroffSagnotti,WittenTopologicalString}.  The maximal
cuts of four-point amplitudes involve products of only three-point
tree amplitudes, and are the simplest cuts to evaluate.  Near-maximal
cuts, in which one or two of the maximal-cut propagators have been
allowed to go off shell, are the next simplest to evaluate, and so on.

The advantage of the maximal-cut method is that
it allows one to focus on a small number of terms at a time,
namely those that become nonvanishing when a particular propagator
is allowed to go off shell.  This feature reduces the computational
complexity at each stage, allowing us to efficiently find compact
representations of amplitudes with the
desired properties.  We note that the ``leading-singularity''
technique, which is applicable to maximally supersymmetric amplitudes,
is also based on cutting a maximal or near-maximal number of
propagators~\cite{CachazoSkinner,CachazoLeading,LeadingSingularityCalcs},
but in addition it makes use of further
conditions from hidden singularities that are special to four dimensions.

In practice the method of maximal cuts allows the sequential improvement
of an ansatz for the numerator factor of each parent graph.
Every new cut identifies the presence of {\it missing pieces}, which
were left undetermined by the previous cuts, and which can be assigned
to one of the parent graphs contributing to the cut.
Because these pieces vanish on the previous set of cuts, they will
contain an inverse propagator factor associated with the last propagator
to be allowed to go off shell.
Once new cuts cease to reveal any more missing pieces, the ansatz is
generally complete and is ready for systematic cut-verification.

Although the maximal-cut method can be applied to $D$-dimensional
cuts, in order to simplify their evaluation we restrict many of the
cuts to have four-dimensional momenta for both internal and external
lines.  As we often evaluate these cuts numerically, it is useful
to build an ansatz for any missing pieces, which consists
of a Lorentz-covariant numerator polynomial containing unknown
constant coefficients.  We reduce the number of unknowns in the
ansatz by assuming that no individual term in it violates the
expected ultraviolet power-counting bound
(\eqn{SuperYangMillsPowerCount} below)~\cite{BDDPR,HoweStelleRevisited}.
These assumptions are, of course, validated by comparing against a
spanning set of cuts after the amplitude has been constructed.

At four loops, the bound~(\ref{SuperYangMillsPowerCount})
predicts that at most four powers of loop
momenta (or at most two inverse propagator factors $l_n^2$) can 
appear in any numerator polynomial.  This restriction allow us to
focus our attention on the maximal and near-maximal cuts that have at
least 11 cut conditions, $l^2_i=0$, out of the maximal 13 (corresponding
to the 13 propagators of the parent graphs).
Examples of such cuts are shown in \fig{SampleMaximalCutsFigure}.
At the level of 11 cut conditions,
there are always some quartic monomials of the form $l_n^2l_m^2$ that
are non-vanishing. As one cycles through all cuts at this
level, all such quartic terms will be detected, and their coefficients
will be fixed. Similarly, one can show that these cuts will
detect all quartic monomials of the form $p^2q^2$ and more generally
$(p\cdot p')(q\cdot q')$, where $p,p',q$ and $q'$ are linear
combinations of the loop momenta and external momenta.  We can
continue the procedure of removing on-shell conditions, one
by one, until we end up with a spanning set of cuts.  However,
in practice, it is much simpler to stop the construction phase as
soon as we suspect that the ansatz is complete.  

The ansatz is then confirmed by checking that it matches
the minimal spanning set of 11 cuts in \fig{CutBasisFigure},
plus the two two-particle cuts in \fig{TwoParticleCutsFigure}.
We refer to this set as a spanning set because the information
it provides is equivalent to that contained in all possible cuts, and
minimal because any further reduction could only involve tadpole-like
contributions.  To show that it is a spanning set, we show that
it includes all the information in the ordinary two-, three-, four- and
five-particle cuts.  First of all, \Fig{CutBasisFigure}(a) is just the
ordinary five-particle cut.  The information from the ordinary two-particle
cuts is given by \fig{TwoParticleCutsFigure}(a) and (b).  Ordinary
four-particle cuts consist of a tree-level six-point amplitude
multiplied by a one-loop six-point amplitude.  We can reproduce the
information in these cuts by studying those generalized cuts in which
we further cut the one-loop six-point amplitude in all inequivalent
ways (omitting three-point trees).  This procedure leads to
\fig{CutBasisFigure}(b), (c), (d) and (e).  Finally, ordinary
three-particle cuts leave either the product of a tree-level
five-point amplitude and a two-loop five-point amplitude (with further
cuts leading to \fig{CutBasisFigure}(f), (g), (h), (i), (j) and (k)),
or the product of two one-loop five-point amplitudes (which does not
lead to any new cut).  In this classification, we can omit a cut if another
cut already appears with a subset of the cut propagators.

\begin{figure}[t]
\centerline{\epsfxsize 6.5 truein \epsfbox{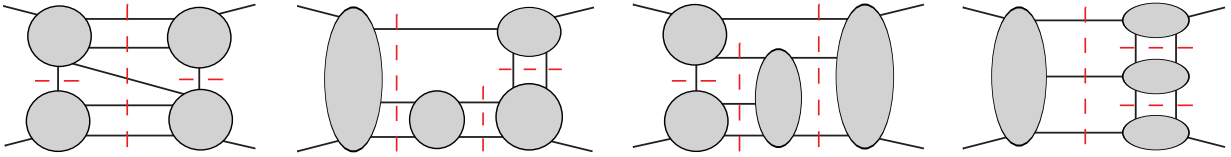}}
\caption[a]{\small Four-loop examples of ``MHV/\MHVbar-amplitude cuts'',
  which are composed entirely of four- and five-point tree amplitudes.}
\label{MHVCutsFigure}
\end{figure}

Suppose one assumes the existence of a representation of the four-loop
amplitude in which each term in the numerator polynomial for each
parent graph has no more than two inverse propagators, consistent with
the known \NeqFoursYM\ power-counting~\cite{BDDPR,HoweStelleRevisited}.
In this case, one only needs to check near-maximal cuts with
at most two canceled propagators.  Because of this, the spanning set of cuts
can be restricted to products of four- and five-point tree amplitudes,
as illustrated in \fig{MHVCutsFigure}. (A six-point tree amplitude requires
three propagators to be canceled from a maximal cut.)  We refer to these as
``MHV/\MHVbar-amplitude cuts'', because all tree amplitudes appearing in
the cuts are either MHV or conjugate \MHVbar{} amplitudes.
The MHV/\MHVbar-amplitude cuts are useful because they are simpler to
evaluate than the spanning set in
\figs{CutBasisFigure}{TwoParticleCutsFigure};
the super-sums are particularly easy to evaluate~\cite{SuperSum}.

We note that when the cuts are verified using color-stripped amplitudes,
in order to capture all non-planar contributions we must include cuts
where the legs of each tree entering the cuts are permuted in all
possible inequivalent ways.


\section{The complete four-loop amplitude}
\label{AmplitudeSection}

\begin{figure}[t]
\centerline{\epsfxsize 6. truein \epsfbox{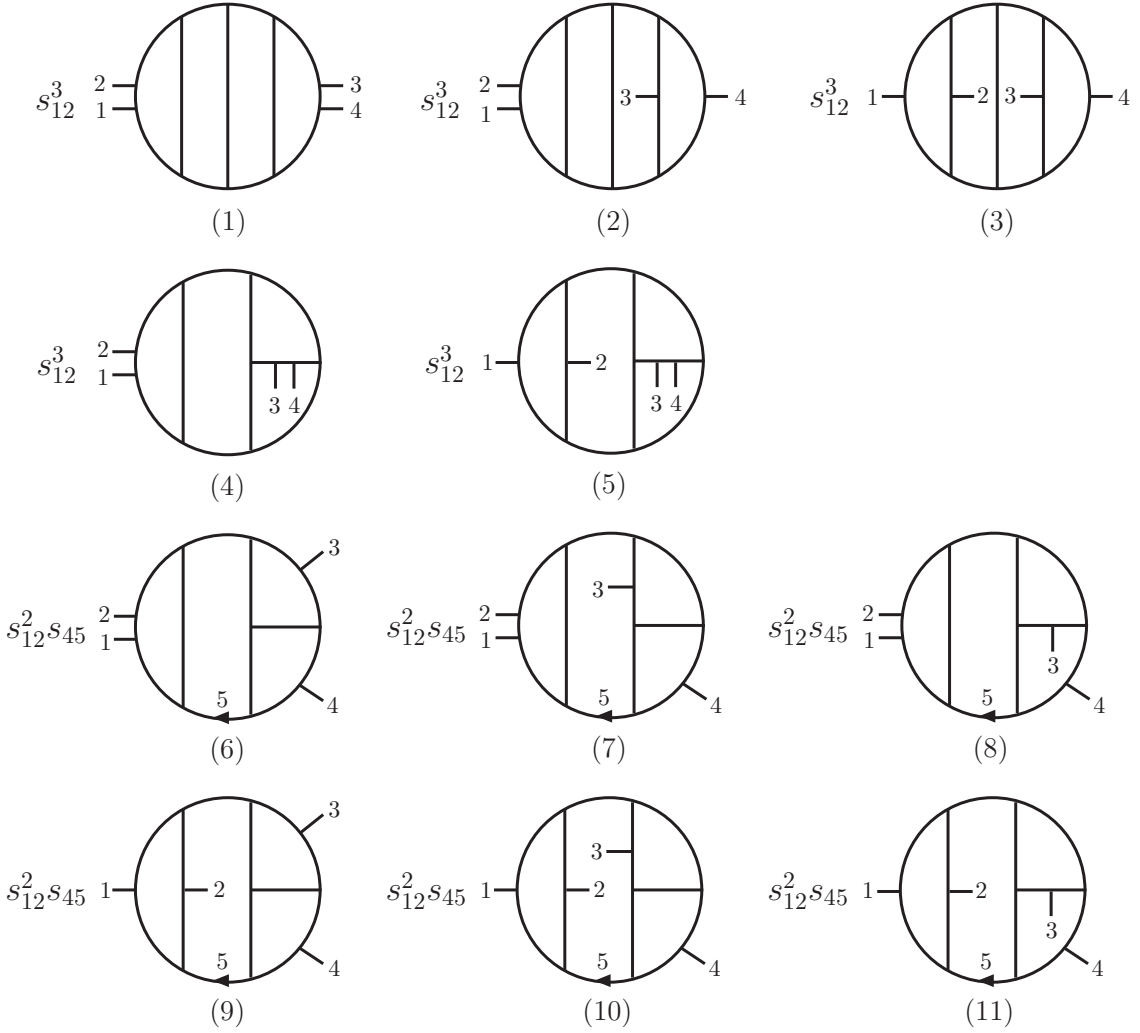}}
\caption[a]{\small Integrals (1)-(11) appearing in the four-loop
amplitude. The graphs encode denominator factors as 
the Lorentz square of the momenta flowing across every internal line, 
and color factors following the rules defined in \fig{ColorDressFigure}. 
The momentum-dependent factor in front of each graph represents the 
numerator factor
$N_{i}$ that resides inside the integral $I_{i}$, where $(i)$
is the label below each graph. Numbers $1,2,3,4$ label the external
(outgoing) momenta. The internal legs carry momenta as signified by
the arrows (here only leg 5 is labeled). The kinematic
variables are defined as $s_{ij}=s_{i,j}=(l_i+l_j)^2$ and in the
following figures $s_{i,\overline{j}}=(l_i-l_j)^2$ and
$\tau_{ij}=2l_i \cdot l_j$, where $l_i$ is the momentum of leg $i$.
A specific (clockwise) orientation of each cubic vertex (in the plane of
the figure) is implied here.  Due to the antisymmetry of
the structure constants, any noncyclic reordering of a vertex should be
accompanied by a sign flip of the numerator factor.}
\label{BC1Figure}
\end{figure}

\begin{figure}[tbh]
\centerline{\epsfxsize 6.2 truein \epsfbox{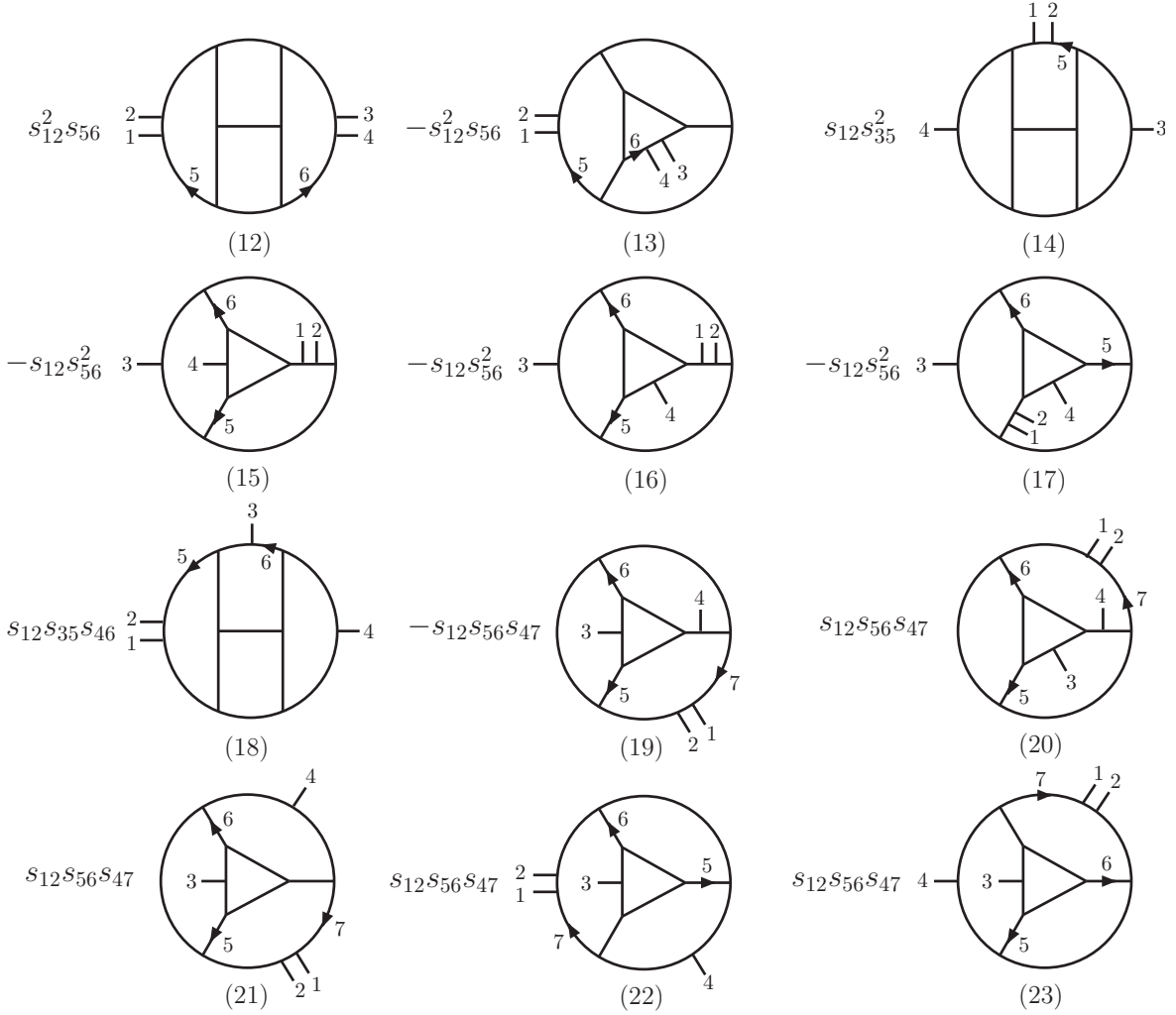}}
\caption[a]{\small  Integrals (12)-(23) appearing in the four-loop amplitude.
The notation follows that of \fig{BC1Figure}.
}
\label{D1Figure}
\end{figure}

\begin{figure}[tbh]
\centerline{\epsfxsize 5.8 truein \epsfbox{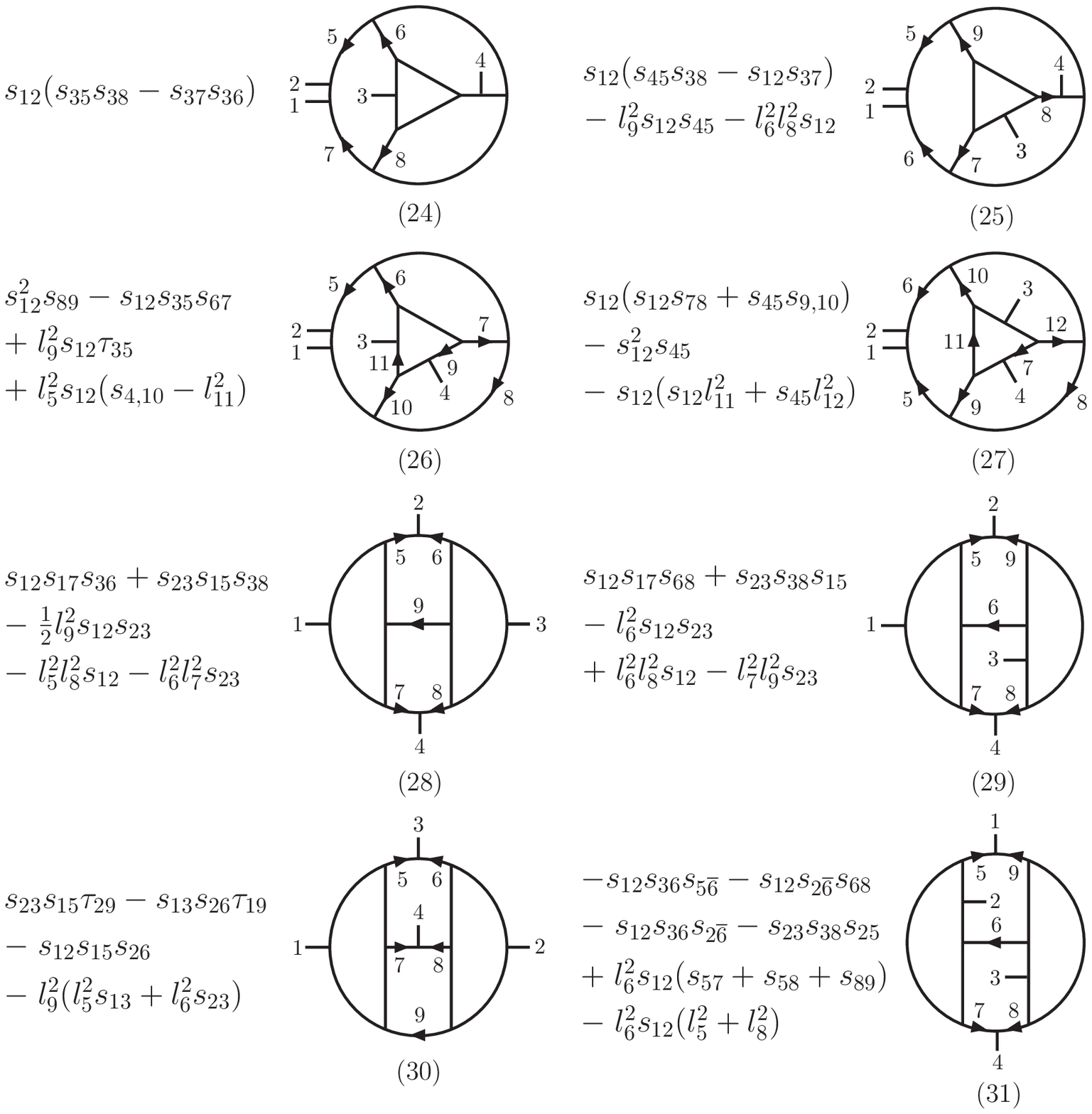}}
\caption[a]{\small  Integrals (24)-(31) appearing in the four-loop amplitude. 
The notation follows that of \fig{BC1Figure}.}
\label{D2Figure}
\end{figure}

\begin{figure}[tbh]
\centerline{\epsfxsize 6.4 truein \epsfbox{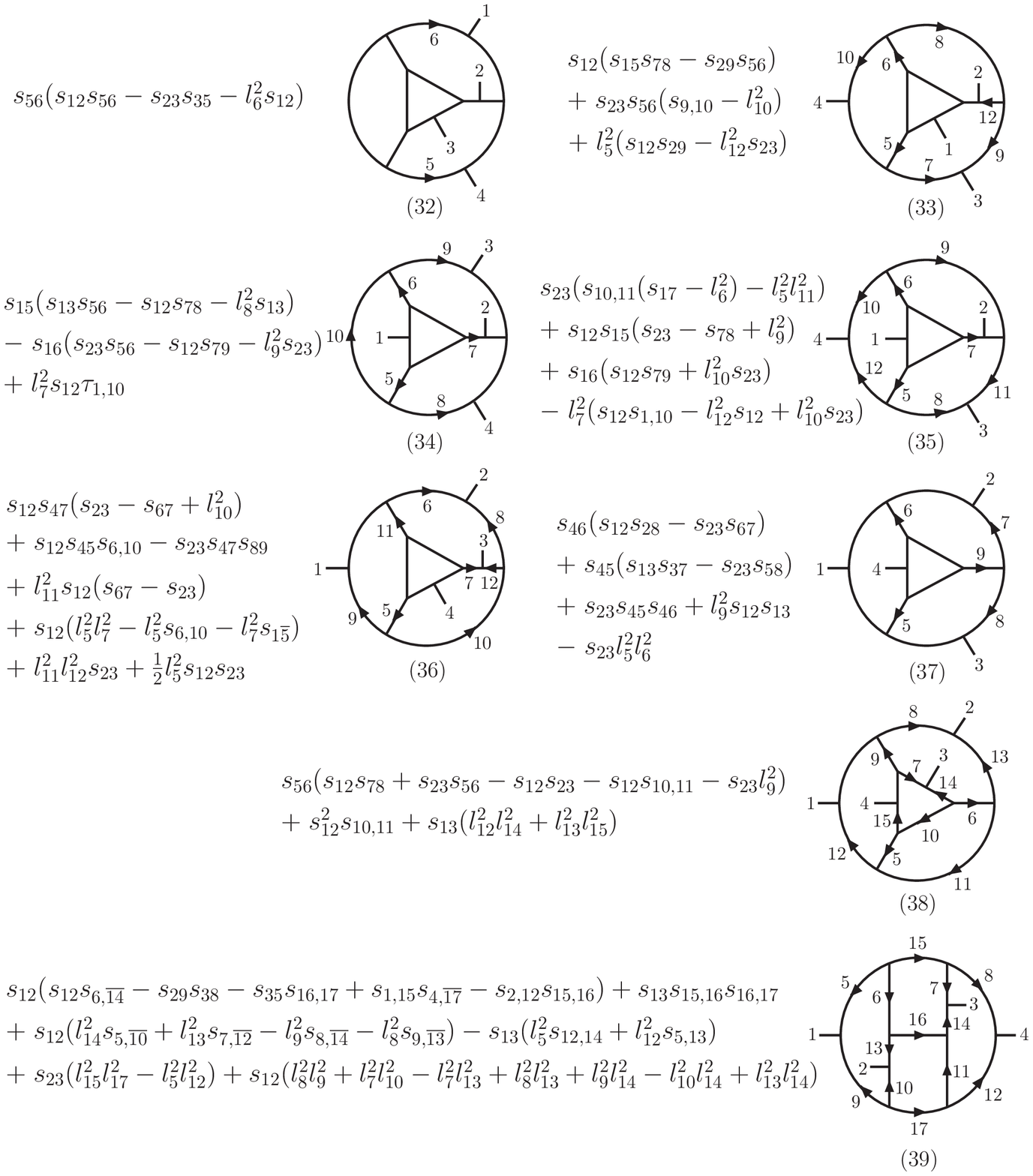}}
\caption[a]{\small  Integrals (32)-(39) appearing in the four-loop amplitude. 
The notation follows that of \fig{BC1Figure}.}
\label{D3Figure}
\end{figure}

\begin{figure}[tbh]
\centerline{\epsfxsize 6 truein \epsfbox{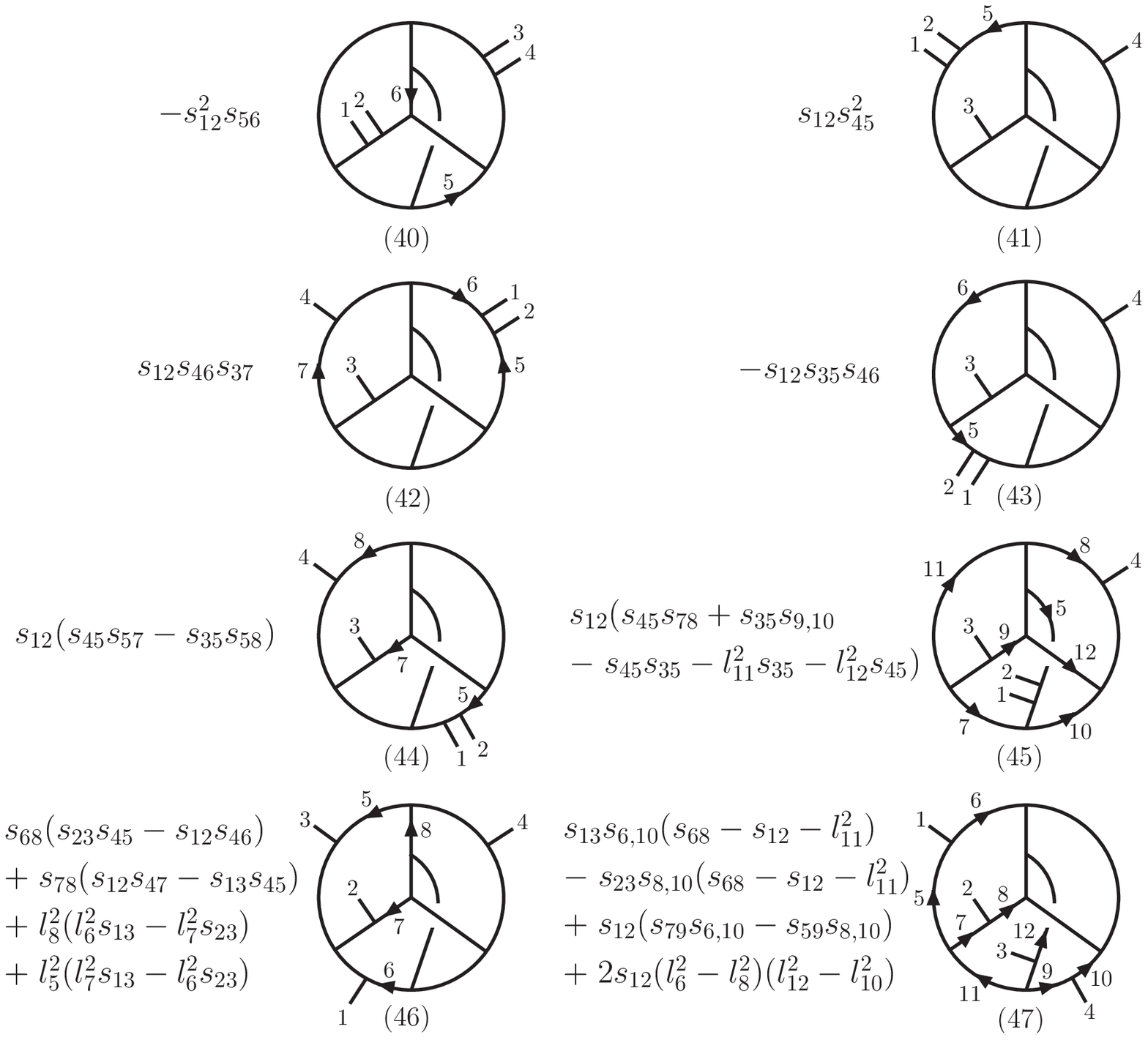}}
\caption[a]{\small  Integrals (40)-(47) appearing in the four-loop amplitude. 
The notation follows that of \fig{BC1Figure}. }
\label{E1Figure}
\end{figure}

\begin{figure}[tbh]
\centerline{\epsfxsize 6.4 truein \epsfbox{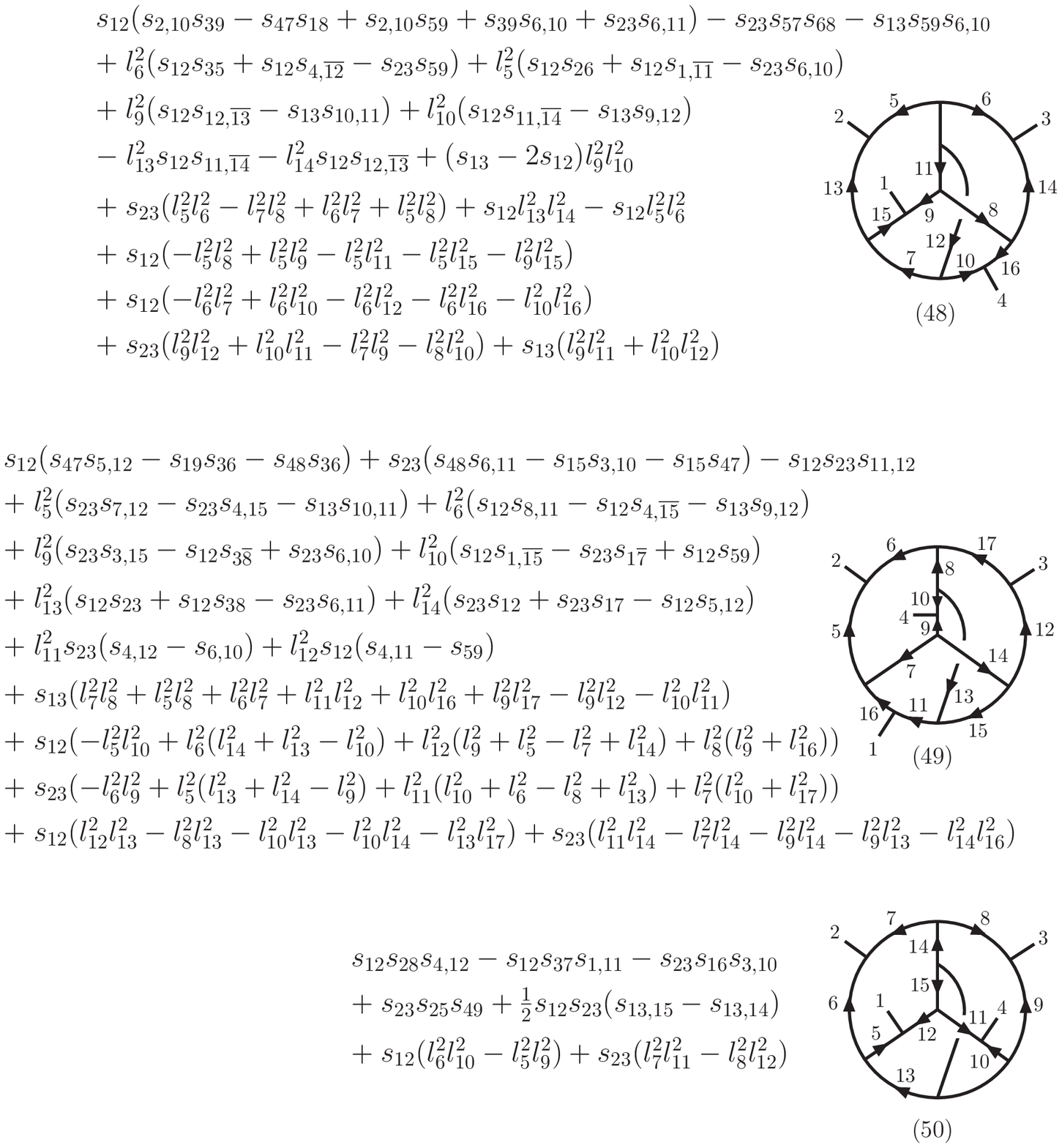}}
\caption[a]{\small Integrals (48)-(49) appearing in the four-loop
  amplitude.  Integral (50) is required for the ansatz to match the
  color-stripped cut at the level of the integrand. However, it
  does not contribute to the color-dressed amplitude.
   The notation follows that of \fig{BC1Figure}.
}
\label{E2Figure}
\end{figure}

We applied the construction methods outlined in the previous section
to the four-loop four-point \NeqFoursYM\ amplitude.  The
resulting amplitude is given by,
\begin{eqnarray}
{\cal A}_4^{\fourl} & = &
g^{10}\, {\cal K} \, \sum_{S_4} \, 
\Bigl[ 
{\Frac{1}{4}}\I_{1}+{\Frac{1}{4}}\I_{2}+{\Frac{1}{16}}\I_{3}
+{\Frac{1}{4}}\I_{4}+{\Frac{1}{8}}\I_{5}+{\Frac{1}{2}}\I_{6}
+{\Frac{1}{2}}\I_{7}+\I_{8}+{\Frac{1}{4}}\I_{9} \nn \\ && 
+{\Frac{1}{4}}\I_{10}+{\Frac{1}{2}}\I_{11}+{\Frac{1}{4}}\I_{12}
+{\Frac{1}{2}}\I_{13}+{\Frac{1}{2}}\I_{14}+{\Frac{1}{4}}\I_{15}
+\I_{16}+{\Frac{1}{2}}\I_{17}+\I_{18}+\I_{19}\nn \\ && 
+\I_{20}+\I_{21}+\I_{22}+\I_{23}+{\Frac{1}{2}}\I_{24}+\I_{25}
+\I_{26}+{\Frac{1}{2}}\I_{27}+{\Frac{1}{4}}\I_{28}+\I_{29}
+{\Frac{1}{2}}\I_{30}\nn \\ && 
+{\Frac{1}{2}}\I_{31}+\I_{32}+\I_{33}+{\Frac{1}{2}}\I_{34}
+\I_{35}+\I_{36}+{\Frac{1}{2}}\I_{37}+{\Frac{1}{4}}\I_{38}
+{\Frac{1}{2}}\I_{39}+{\Frac{1}{4}}\I_{40}\nn \\ && 
+{\Frac{1}{2}}\I_{41}+\I_{42}+\I_{43}+{\Frac{1}{2}}\I_{44}
+{\Frac{1}{4}}\I_{45}+{\Frac{1}{2}}\I_{46}+{\Frac{1}{8}}\I_{47}
+{\Frac{1}{2}}\I_{48}+{\Frac{1}{2}}\I_{49} + {\Frac{1}{8}} \I_{50}
\Bigr]
 \,,  \hskip .3 cm 
\label{FourLoopYMAmplitude}
\end{eqnarray}
where the prefactor ${\cal K}$, defined in
\eqn{Prefactor}, encodes the full external-state dependence, and
$\I_{i}=C_{i}I_{i}$ are the color-dressed four-loop integrals. The
$C_{i}$ are color factors obtained by dressing the parent graphs
with structure constants $\colorf{abc}$, and are given explicitly in
\app{ColorAppendix}.  The $I_{i}(\sv,\tv)$ are
$D$-dimensional loop integrals, defined in terms of numerator factors
$N_i$ in \eqn{IntegralNormalization}, and corresponding to the 50
four-loop cubic parent graphs in
figs.~\ref{BC1Figure}-\ref{E2Figure}.  The 50$^{\rm th}$ graph,
which appears in \eqn{FourLoopYMAmplitude} and in \fig{E2Figure}, is
needed to match all cuts; however, it integrates to zero, and its
associated color factor $C_{50}$ also vanishes. Thus its contribution
to the integrated color-dressed amplitude is doubly vanishing.
(Another reason we list this \NeqFoursYM\ contribution is because it
gives a nonvanishing input into the construction of the corresponding
$\NeqEight$ supergravity amplitude~\cite{GravityFour}.)

As in the lower-loop amplitudes in \sect{ColorOrganizationSubsection},
the sum runs over the
24 independent permutations of legs $\{1,2,3,4\}$, denoted by $S_4$,
which act on both kinematic and color labels.  The numerical
coefficients in front of the integrals in \eqn{FourLoopYMAmplitude}
are symmetry factors $1/S$, where $S$ is the number of elements in
the discrete automorphism group of the corresponding unlabeled
graph.  As before, these factors compensate for overcounting.

As mentioned in \sect{MagicToolsSection},
the parent graphs containing two-particle cuts, namely 
graphs 1--27 and 40--45, are the simplest to obtain; the bulk of the
terms in their numerator polynomials are constructed using
\eqn{TwoParticleSigmaSewing}.  The remaining terms in these graphs,
and all of the terms in the remaining 17 parent graphs, are obtained
using box cuts (\eqn{BoxCutReduction}), 
plus the color-kinematic duality
relation~(\ref{NumeratorIntegralIdentity}),
as well as an evaluation of the near-maximal cuts.
The amplitude's construction was followed by a confirmation of the
complete set of cuts in \figs{CutBasisFigure}{TwoParticleCutsFigure}.

\section{Ultraviolet properties in higher dimensions}
\label{UVPropertiesSection}

In this section we examine the UV behavior of the four-point
\NeqFoursYM\ amplitude in its critical dimension, {\it i.e.,} the 
lowest dimension in which it diverges.  This issue is of some
importance because it provides a simpler venue than $\NeqEight$
supergravity for studying UV cancellations in theories with maximal
supersymmetry.  For \NeqFoursYM, an analysis of supersymmetric
cancellations in two-particle cuts (and more generally, in ordinary
$m$-particle cuts that have only MHV amplitudes on either side of the
cut), suggested the UV finiteness bound~\cite{BDDPR},
\begin{equation}
D < D_c = 4+\frac{6}{L}   \hskip 1 cm (L > 1) \,.
\label{SuperYangMillsPowerCount}
\end{equation}
One-loop amplitudes, corresponding to $L=1$, are special as they are
UV finite for $D<8$, not $D<10$.  The
bound~(\ref{SuperYangMillsPowerCount}) is somewhat stronger than
earlier superspace power counting bounds~\cite{HoweStelleYangMills},
although all bounds agree that the theory is UV finite in $D=4$.
The bound (\ref{SuperYangMillsPowerCount}) is
consistent with a formulation of the theory having $3/4$ of the
supercharges manifestly realized, and it has been
confirmed~\cite{HoweStelleRevisited} using ${\cal N} = \nobreak 3$
harmonic superspace~\cite{HarmonicSuperspace}.  Explicit computations
(including the ones discussed in this paper) demonstrate that this
bound is saturated through at least four
loops~\cite{BRY,BDDPR,ABDK,Finite}.  It is straightforward to verify from
the planar amplitude in ref.~\cite{FiveLoop} that the same result holds
in the large-$N_c$ limit at five loops.

An interesting question concerns the UV behavior of the different color
terms.  For an $SU(N_c)$ gauge group it is convenient to expand the
amplitude in the trace basis --- {\it i.e.} traces of products of
generators in the fundamental representation, as in
\eqn{LeadingColorDecomposition}, but including also terms that are
subleading in the $1/N_c$ expansion.  For a four-particle
amplitude there are only single- and double-trace terms.  This
decomposition follows quite closely that of the terms in the 
effective action that have four or fewer field strength tensors.
The single-trace terms satisfy the finiteness bound
(\ref{SuperYangMillsPowerCount}).  As noted already in
refs.~\cite{DurhamAndCopenhagenTalks}, double-trace terms at three and
four loops exhibit additional cancellations which increase their
critical dimension.  These cancellations have been discussed in
refs.~\cite{DoubleTraceBerkovits,BG2010},
which suggest double-trace terms should instead satisfy the
finiteness bound,
\begin{equation}
D < 4 + \frac{8}{L}   \hskip 1 cm (L > 2) \,.
\label{SuperYangMillsDoubleTracePowerCount}
\end{equation}
This formula is equivalent to the statement that an additional momentum
invariant ($s$, $t$ or $u$) can be extracted from the double-trace terms
than the single-trace terms, at each loop order.  As a statement
about counterterms, it implies that the leading double-trace
counterterms for $L>2$ would have four covariant derivatives,
for example $\Tr({\cal D}^4 F^2) \Tr(F^2)$, in contrast to the two
derivatives characteristic of the single-trace counterterms
$\Tr({\cal D}^2 F^4)$.  In a superspace description, the dimension of
the double-trace counterterms would then be consistent, curiously,
with the existence of a superspace formalism for these terms
that preserves manifestly all sixteen supercharges.

Here we provide details of the cancellations observed
in refs.~\cite{DurhamAndCopenhagenTalks} and affirm the double-trace
power-count bound (\ref{SuperYangMillsDoubleTracePowerCount}) for
three and four loops. Moreover, by direct integration of the results
we show that at three loops the double-trace bound is saturated.
At four loops, we have not evaluated the required integrals
in $D=6$, so it is possible (though perhaps unlikely) that the
behavior of the double-trace terms is better than this bound.


\subsection{One-loop ultraviolet divergence}
\label{OneLoopDivergenceSubsection}

Before turning to the four-loop amplitude, it is useful to review and
expand on the lower-loop results.  Let us consider first the
color-dressed one-loop amplitude in \eqn{OneLoopYMAmplitude}.  For each
of the contributing integrals it is necessary to specify the
corresponding color factor. The trace basis, in which the color
factors are expressed in terms of traces of products of group
generators $T^{a_i}$ in the fundamental representation, is convenient for
discussing the UV divergences for an $SU(N_c)$ gauge group. The
basis elements are:
\begin{equation}
\Tr_{ijkl}\ \equiv\ \Tr(T^{a_i}T^{a_j}T^{a_k}T^{a_l}), \qquad
\Tr_{ij}\ \equiv\ \Tr(T^{a_i}T^{a_j}) = \delta^{a_ia_j} \,.
\label{Trdef}
\end{equation}
(In general one may also add to the basis $\Tr(T^{a_i})$,
$\Tr(T^{a_i}T^{a_j}T^{a_k})$, {\it etc.}\  For four-point amplitudes
the traces of length one and three must appear together; for $SU(N_c)$
they are not necessary, because $\Tr(T^a) = 0$.)

The color factor of the box integral, given in
\eqn{1loopcolortensor}, may be expressed in the trace basis as
\begin{equation}
C^{\rm Box}_{1234} =N_c \, ( \Tr_{1234} + \Tr_{1432} )
 + 2 \, ( \Tr_{12} \Tr_{34} + \Tr_{14} \Tr_{23}
        + \Tr_{13} \Tr_{24} ) \,.
\label{C1234Tr}
\end{equation}
The lowest dimension in which the one-loop box integral in
\eqn{OneLoopBoxIntegral} develops an ultraviolet divergence
is $D=8$.  Near $D=8$, we have 
\begin{equation}
I^{{\rm box},\,D=8-2\ep}(s_{12},s_{23})\Bigr|_{\rm pole}
= {1\over 6 \, (4\pi)^4 \, \e} \,.
\label{I41pole}
\end{equation}
From \eqns{OneLoopYMAmplitude}{C1234Tr} it then follows that 
the divergence of the  one-loop amplitude in the critical dimension
is given by
\begin{eqnarray}
{\cal A}_4^{\onel}(1,2,3,4) \Bigr|^{SU(N_c)}_{\rm pole}
&=&  - { g^4 \,  {\cal K} \over 6 \, (4\pi)^4 \, \e } 
\Bigl( N_c \, ( 
  \Tr_{1324} + \Tr_{1423} + \Tr_{1243} + \Tr_{1342}
 + \Tr_{1234} + \Tr_{1432} )
\nonumber\\ &&\hskip1.8cm\null
+ 6 \, ( \Tr_{12} \Tr_{34} + \Tr_{14} \Tr_{23}
       + \Tr_{13} \Tr_{24} ) \Bigr) \,.
\label{OneLoopPole}
\end{eqnarray}
At this loop order, the coefficients of the double-trace terms, 
relative to the single-trace ones, are fixed by a $U(1)$
decoupling identity, or dual Ward identity~\cite{BKColor}.

For a general gauge group $G$ the UV divergence is expressed in terms of
three independent color tensors: the two tree-level tensors and the
irreducible one-loop tensor $C^{\rm Box}_{1234}$ in
\eqn{1loopcolortensor}:
\begin{equation}
{\cal A}_4^{\onel}(1,2,3,4) \Bigr|^{G}_{\rm pole}= 
- { g^4 \,  {\cal K} \over 6 \, (4\pi)^4 \, \e } 
\Bigl( -\frac{1}{2} 
C_A (\colorf{a_1 a_2 b}\colorf{b a_3 a_4} + \colorf{a_2 a_3 b}\colorf{b a_4 a_1} )
 + 3  C^{\rm Box}_{1234}  \Bigr) \,,
\label{OneLoopPoleGeneral}
\end{equation}
where $C_A$ is the quadratic Casimir in the adjoint representation,
normalized as in \eqn{Casimirs} of \app{ColorAppendix}.
Note that the Bose symmetry of the divergence is not manifest in this
form.  As discussed in \app{ColorAppendix}, any one-loop
four-point quantity can be expressed in terms of these three color
tensors. Thus at one loop, in the critical dimension $D_c=8$, 
the coefficients of all listed color structures diverge and
there are no additional hidden cancellations.


\subsection{Two-loop ultraviolet divergence}
\label{TwoLoopDivergenceSubsection}

A similar analysis may be carried out for 
the two-loop four-point \NeqFoursYM\ amplitude, which is given 
in \eqn{TwoLoopYMAmplitude} in terms of planar and 
non-planar double-box integrals $I^{\P}$ and $I^{\NP}$. 
These integrals first diverge in $D=7$ and their poles in 
$D=7-2\e$ are~\cite{BDDPR},
\begin{eqnarray}
 V^{\P} &\equiv& { I^{\P,\, D=7-2\e} \bigr|_{\rm pole} \over s_{12} }
 = - {\pi\over20 \, (4\pi)^7 \, \e}
\,,  \label{planarseven}\\
 V^{\NP} &\equiv& { I^{\NP,\, D=7-2\e}\bigr|_{\rm pole} \over s_{12} }
 = - {\pi\over30 \, (4\pi)^7 \, \e}
\,.  
\label{nonplanarseven}
\end{eqnarray}

For an $SU(N_c)$ gauge group, the planar and non-planar two-loop color
tensors $C^{\P}_{1234}$ and $C^{\NP}_{1234}$ defined in
\eqn{2loopcolortensors} become, in the trace basis, 
\begin{eqnarray}
C^{\P}_{1234} &=& 
(N_c^2 + 2) \, ( \Tr_{1234} + \Tr_{1432} )
+ 2 \,  ( \Tr_{1243} + \Tr_{1342} )
- 4 \, ( \Tr_{1423} + \Tr_{1324} )
\nonumber\\ &&\hskip0.0cm \null
+ 6 N_c \Tr_{12} \Tr_{34}
 \,,
\label{CP1234Tr}\\
C^{\NP}_{1234} &=& 
  2 \, ( \Tr_{1234} + \Tr_{1432} )
+ 2 \,  ( \Tr_{1243} + \Tr_{1342} )
- 4 \, ( \Tr_{1423} + \Tr_{1324} )
\nonumber\\ &&\hskip0.0cm \null
+ 2 N_c {}\, ( 2 \Tr_{12} \Tr_{34} 
            - \Tr_{13} \Tr_{24} - \Tr_{14} \Tr_{23} )
\,.
\label{CNP1234Tr}
\end{eqnarray}
The full amplitude was originally presented in the trace
basis~\cite{BRY}.  In terms of $V^{\P}$ and $ V^{\NP}$,
the UV divergence of the amplitude~(\ref{TwoLoopYMAmplitude}) is,
\begin{eqnarray} 
\label{TwoLoopPoleVForm}
{\cal A}_4^{\twol}(1,2,3,4) \Bigr|^{SU(N_c)}_{\rm pole} 
&=&  - \, g^6 \, {\cal K} \,
\biggl[ \Bigl( N_c^2 \, V^{\P} + 12 ( V^{\P} + V^{\NP} ) \Bigr)
\\ && \hskip-1.5cm\null
 \times \Bigl( s_{12} \, ( \Tr_{1324} + \Tr_{1423} ) 
       + s_{23} \, ( \Tr_{1243} + \Tr_{1342} )
       + s_{13} \, ( \Tr_{1234} + \Tr_{1432} ) \Bigr)
\nonumber\\ &&\hskip-1.5cm \null
- 12 \, N_c \, ( V^{\P} + V^{\NP} )
  \Bigl( s_{12} \Tr_{12} \Tr_{34}
       + s_{23} \Tr_{14} \Tr_{23}
       + s_{13} \Tr_{13} \Tr_{24} \Bigr) \biggr] \,.
\nonumber
\end{eqnarray}
Inserting the planar and non-planar integral
poles~(\ref{planarseven}) and~(\ref{nonplanarseven}) into
\eqn{TwoLoopPoleVForm}, the UV divergence becomes,
\begin{eqnarray} 
\label{TwoLoopPole}
{\cal A}_4^{\twol}(1,2,3,4) \Bigr|^{SU(N_c)}_{\rm pole} 
&=&  { g^6 \, \pi \, {\cal K} \over 20 \, (4\pi)^7 \, \ep}
\Bigl[ (N_c^2+20) 
  \Bigl( s_{12} \, ( \Tr_{1324} + \Tr_{1423} )\\ 
                       &&\hskip3.5cm \null 
       + s_{23} \, ( \Tr_{1243} + \Tr_{1342} )
       + s_{13} \, ( \Tr_{1234} + \Tr_{1432} ) \Bigr)
\nonumber\\ &&\hskip2.0cm \null
- 20 N_c \, \Bigl( s_{12} \Tr_{12} \Tr_{34}
       + s_{23} \Tr_{14} \Tr_{23}
       + s_{13} \Tr_{13} \Tr_{24} \Bigr) \Bigr] \,.
\nonumber
\end{eqnarray}

As was the case at one loop, $U(1)$ decoupling identities
at two loops~\cite{BDFD} relate the coefficients of the single- 
and double-trace structures.  These identities provide an {\it a priori}
justification that the double-trace terms diverge whenever the single-trace 
terms diverge, at this loop order.  In slightly more detail, the
double-trace coefficient is equal to the negative of a subleading-color 
($N_c^0$) single-trace coefficient, plus the sum
of all three leading-color ($N_c^2$) single-trace coefficients.
(See eqs.~(4.48)--(4.50) of ref.~\cite{BDFD}.)
However, in the case of \eqn{TwoLoopPole}, the
relevant leading-color sum vanishes by the identity
$s_{12}+s_{23}+s_{13}=0$.
Thus group theory enforces the equality of the two ``20''s
appearing in \eqn{TwoLoopPole}.  

We may also analyze the color structure for a general gauge group $G$.
In this case, the UV divergence of the two-loop amplitude depends on
five independent color tensors, which we take to be the tree-level
and one-loop tensors that already appear in the one-loop
divergence~(\ref{OneLoopPoleGeneral}), plus two new independent
(irreducible) two-loop tensors.
We take the latter to be two independent permutations of the
tensor $C^{\P}_{1234}$.  (All other tensors are related to these by
repeated application of the Jacobi identity.)  In terms of the pole
parts~(\ref{planarseven}) and (\ref{nonplanarseven}) of the planar and
non-planar double-box integrals, the two-loop divergence is:
\begin{eqnarray}
\label{TwoLoopPoleVFormGeneral}
{\cal A}_4^{\twol}(1,2,3,4) \Bigr|^{G}_{\rm pole} 
&=&  g^6 \, {\cal K} \biggl[ 
\frac{1}{4} C_A^2  V^{\P} (s_{12}\colorf{a_2 a_3 b}\colorf{b a_4 a_1}
 + s_{23}\colorf{a_1 a_2 b}\colorf{b a_3 a_4}) \\ && \null\hskip-0.4cm
+ (V^{\P} +V^{\NP}) (3 C_A C^{\rm Box}_{1234} s_{13} + 
2 C^{\P}_{1234} (s_{12} - s_{13}) + 2 C^{\P}_{2341}(s_{23} - s_{13}))
\biggr] .\nn 
\end{eqnarray}
The five color tensors appearing in \eqn{TwoLoopPoleVFormGeneral}
form a basis in the space of two-loop four-point color tensors
(see \app{ColorAppendix}).  Each of their coefficients diverges in
the critical dimension $D_c=7$, and we see no natural combination
of coefficients for which the $D=7$ divergence cancels.


\subsection{Three-loop ultraviolet divergence}
\label{ThreeLoopDivergenceSubsection}

\begin{figure}[tbh]
\centerline{\epsfxsize 2.7 truein \epsfbox{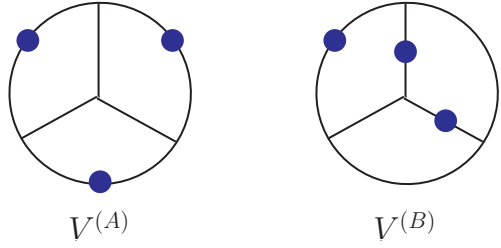}}
\caption[a]{\small The vacuum-like graphs describing the
three-loop UV divergences in $D=6$. The large (blue) dots indicate
that a propagator appears squared, or doubled.}
\label{Vacuum3loopsFigure}
\end{figure}
 
Let us proceed with a similar discussion of the UV divergences of the
three-loop four-point amplitude given in \eqn{ThreeLoopYMAmplitude}
and \fig{IntegralsThreeLoopFigure}. In this case the critical
dimension is $D_c=6$.  In the
representation of the three-loop amplitude given in
\fig{IntegralsThreeLoopFigure}, integrals (a)-(d) and (h) are
all finite in the critical dimension.  Only integrals (e), (f), (g)
and (i) contribute to the UV divergence, because they have numerator
polynomials that are quadratic in the loop momentum.
To extract the divergence we follow the procedure discussed in
refs.~\cite{Vladimirov,MarcusSagnotti,CompactThree,GravityFour}, {\it i.e.}
we expand in small external momenta and keep only the leading terms.
In this limit, the contributing integrals reduce to,
\begin{eqnarray}
&&I^{\rm (e)} \rightarrow s_{12} \, V^{\A}\,, \hskip .4 cm
I^{\rm (f)} \rightarrow s_{12} \, V^{\B}\,, \hskip .4 cm
I^{\rm (g)} \rightarrow s_{12} \, V^{\B}\,, \hskip .4 cm  \nn\\
&&
I^{\rm (i)} \rightarrow (s_{12} - s_{23})
 \Bigl( V^{\B} - {1 \over 3} V^{\A} \Bigr)\,, 
\label{YM3integralpoles}
\end{eqnarray}
where $V^{\A}$ and $V^{\B}$ are the vacuum-like
graphs in \fig{Vacuum3loopsFigure}. These integrals are both UV and
infrared divergent. As discussed in ref.~\cite{CompactThree}, to extract the
UV divergence we inject off-shell momenta at two vertices, thus
removing the infrared divergence.  The values of their UV divergences are:
\begin{eqnarray}
V^{\A}\Bigr|_{\rm pole} &=& -\frac{1}{6 \, (4\pi)^9 \, \e} \,, 
\label{VA} \\
V^{\B}\Bigr|_{\rm pole} &=& -\frac{1}{6 \, (4\pi)^9 \, \e}
\biggl( \zeta_3 - {1\over3} \biggr) \,.
\label{VB}
\end{eqnarray}

The relevant three-loop color factors $C^{\rm (e)},\, C^{\rm
(f)},\,C^{\rm (g)}$ and $C^{\rm (i)}$ are easy to express in the
trace basis,
\begin{eqnarray}
C^{\rm (e)} &=& N_c^3 \, ( \Tr_{1234} + \Tr_{1432} )
       + 2 N_c^2 \Tr_{12} \Tr_{34}
\nonumber\\ && \null \hskip 0 cm 
       + 2 N_c \, ( 4 \, ( \Tr_{1234} + \Tr_{1432} )
                  - 2 \, ( \Tr_{1243} + \Tr_{1342} )
                  - 3 \, ( \Tr_{1423} + \Tr_{1324} ) )
\nonumber\\ && \null \hskip 0 cm 
       - 4 \, ( \Tr_{12} \Tr_{34} + \Tr_{14} \Tr_{23}
             + \Tr_{13} \Tr_{24} ) \,,
\label{CeTr}\\
C^{\rm (f)} &=& 
  - 2 N_c^2 \, ( \Tr_{14} \Tr_{23} + \Tr_{13} \Tr_{24} )
\nonumber\\ && \null \hskip 0 cm 
       + 2 N_c \, ( 4 \, ( \Tr_{1234} + \Tr_{1432} )
                  - 2 \, ( \Tr_{1243} + \Tr_{1342} )
                  - 3 \, ( \Tr_{1423} + \Tr_{1324} ) )
\nonumber\\ && \null \hskip 0 cm 
       - 4 \, ( \Tr_{12} \Tr_{34} + \Tr_{14} \Tr_{23}
             + \Tr_{13} \Tr_{24} ) \,,
\label{CfTr}
\\
C^{\rm (g)} &=& 
       - 4 N_c^2 \Tr_{12} \Tr_{34}
\nonumber\\ && \null \hskip 0 cm 
       + 2 N_c \, ( 3 \, ( \Tr_{1234} + \Tr_{1432} )
                  - 3 \, ( \Tr_{1243} + \Tr_{1342} )
                      -  \Tr_{1423} - \Tr_{1324}  ) \hskip .6 cm 
\nonumber\\ && \null \hskip 0 cm 
       - 4 \, ( \Tr_{12} \Tr_{34} + \Tr_{14} \Tr_{23}
             + \Tr_{13} \Tr_{24} ) \,,  
\label{CgTr}\\
C^{\rm (i)} &=& 
     2 N_c^2 ( \Tr_{12} \Tr_{34} - \Tr_{14} \Tr_{23} )
\nonumber\\ && \null \hskip 0 cm 
       + 2 N_c \, ( \Tr_{1243} + \Tr_{1342} 
                 -  \Tr_{1423} - \Tr_{1324} ) \,.
\label{CiTr}
\end{eqnarray}

Using these expressions, as well as the
relation~(\ref{YM3integralpoles}) between the leading poles of the
integrals $I^{\rm (e)},\,I^{\rm (f)},\,I^{\rm (g)},\,$ and $I^{\rm
(i)}$ and the vacuum integrals $V^{\A}$ and $V^{\B}$, we find
that the leading UV divergence of the three-loop amplitude in
dimension $D=6-2\epsilon$ is
\begin{eqnarray} 
{\cal A}_4^{\threl}(1,2,3,4) \Bigr|^{SU(N_c)}_{\rm pole} 
&=&  2 \, g^8 \, {\cal K} \, 
   \Bigl( N_c^3 \, V^{\A}
       + 12 \, N_c \, ( V^{\A} + 3 \, V^{\B} ) \Bigr)
\label{ThreeLoopPoleVForm}\\ 
&&\hskip-1.2cm \null
\times \Bigl( s_{12} \, ( \Tr_{1324} + \Tr_{1423} )
           + s_{23} \, ( \Tr_{1243} + \Tr_{1342} )
           + s_{13} \, ( \Tr_{1234} + \Tr_{1432} ) \Bigr) \,.
\nonumber
\end{eqnarray}
Inserting the UV pole parts~(\ref{VA}) and (\ref{VB}) of $V^{\A}$
and $V^{\B}$ then gives,
\begin{eqnarray} 
{\cal A}_4^{\threl}(1,2,3,4) \Bigr|^{SU(N_c)}_{\rm pole} 
&=&  - { g^8 \, {\cal K} \over 3 \, (4\pi)^9 \, \e}
   ( N_c^3 + 36 \, \zeta(3) \, N_c )
\label{ThreeLoopPole}\\ &&\hskip-1.2cm \null
\times \Bigl( s_{12} \, ( \Tr_{1324} + \Tr_{1423} )
           + s_{23} \, ( \Tr_{1243} + \Tr_{1342} )
           + s_{13} \, ( \Tr_{1234} + \Tr_{1432} ) \Bigr) \,.
\nonumber
\end{eqnarray}

\begin{figure}[tb]
\centerline{\epsfxsize 3.5 truein \epsfbox{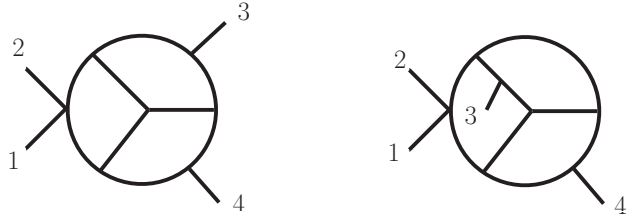}}
\caption[a]{\small The leading contributions with numerators quadratic in 
loop momentum in \fig{IntegralsThreeLoopFigure} can be absorbed 
into the two contact diagrams displayed here, up to relabeling of 
external legs.  Further rearrangements push the leading terms into
the diagrams of \fig{ThreeLoopNoDoubleTraceFigure}
}
\label{ThreeLoopContactFigure}
\end{figure}

\begin{figure}[tb]
\centerline{\epsfxsize 5.5 truein \epsfbox{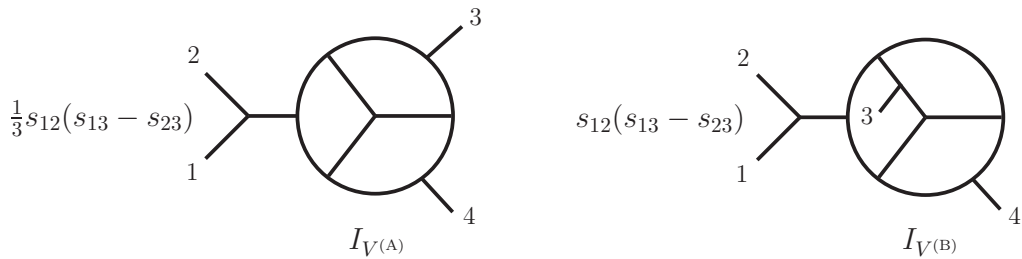}}
\caption[a]{\small At three loops the leading divergence can be
  rearranged so that it comes from parent graphs which are
  one-particle reducible, depicted here in the $s_{12}$ channel.  The
  color factors of these graphs have no double-trace
  contributions. The complete contribution comes from summing over all
  24 permutations of external legs and multiplication by a symmetry
  factor of $1/4$ to remove double counts in both graphs.}
\label{ThreeLoopNoDoubleTraceFigure}
\end{figure}

A remarkable feature of the three-loop UV
divergence~(\ref{ThreeLoopPoleVForm}) in the critical dimension is the
absence of double-trace terms.  From eqs.~(\ref{CeTr})-(\ref{CiTr})
it is clear that such terms exist separately
for each integral and cancel only in the complete amplitude. This
cancellation was first noted~\cite{DurhamAndCopenhagenTalks} as a
consequence of the calculation described in this paper.  Two rather
different discussions of this property have been presented recently.
One approach is based on
the pure-spinor formalism, both in string theory in the low-energy
limit~\cite{DoubleTraceBerkovits} and more recently in field
theory~\cite{BG2010}.  The other approach is based on algebraic
non-renormalization theorems~\cite{DoubleTraceNonrenormalization},
following up on earlier work~\cite{BHS}.

Motivated by these results, especially the structure of the leading
contributions in the string theory analysis of
ref.~\cite{DoubleTraceBerkovits}, we rearrange the leading terms in
the small-momentum expansion of the integrals in
\fig{IntegralsThreeLoopFigure} such that double-trace terms are
manifestly absent. This representation is found by noting that all
terms quadratic in loop momentum, can be placed into contact-term
diagrams, of the form in \fig{ThreeLoopContactFigure}.  This fact
suggests that the leading terms can be absorbed into the 
graphs with external propagators 
displayed in \fig{ThreeLoopNoDoubleTraceFigure}, if we
include a factor of $s_{12}$ to cancel the external propagator.
After dressing the graphs with color, and a factor of
$(s_{13} - s_{23})$ so that the graphs have the proper antisymmetry, 
it is not difficult to verify that with the numerical coefficients in
\fig{ThreeLoopNoDoubleTraceFigure}, the expression
\begin{eqnarray} 
{\cal A}_4^{\threl}(1,2,3,4) 
&=&  {1\over4} \, g^8 \, {\cal K} \sum_{S_4}
      \Bigl[ C^{V^{\A}}_{1234} I_{V^{\A}}
           + C^{V^{\B}}_{1234} I_{V^{\B}} \Bigr]
\; +\; \hbox{subleading},
\label{LeadingPiecesThreeLoop}
\end{eqnarray}
has the proper leading behavior.  Here $I_{V^{\A}}$, $I_{V^{\B}}$
and $C^{V^{{\A}}}_{1234}$, $C^{V^{{\B}}}_{1234}$ correspond to the
integrals and color factors indicated by the graphs in
\fig{ThreeLoopNoDoubleTraceFigure}.  The dropped subleading terms are
better behaved in the ultraviolet than the displayed leading terms and
are not relevant for our discussion.  The integrals in
\eqn{LeadingPiecesThreeLoop} are essentially just the two vacuum graphs
$V^{\A}$ and $V^{\B}$, promoted to one-particle-reducible four-point
integrals, and dressed with appropriate numerator factors. The color
factors $C^{V^{{\A}}}_{1234}$ and $C^{V^{{\B}}}_{1234}$ are given by
\begin{eqnarray}
C^{V^{\A}}_{1234} 
& = & \colorf{a_1 a_2 a_5} \colorf{a_5 a_6 a_7} \colorf{a_6 a_{10} a_8} 
         \colorf{a_8 a_{11} a_9} \colorf{a_9 a_{12} a_7} \colorf{a_3 a_{13} a_{10}}
    \colorf{a_{13} a_{14} a_{11}} \colorf{a_4 a_{12} a_{14}} \nn \\
&=& N_c (N_c^2 + 12) \Bigl( \Tr_{1 2 3 4} +  \Tr_{1 4 3 2}
       - \Tr_{1 2 4 3} - \Tr_{1 3 4 2} \Bigr) \,,  \nn\\
C^{V^{\B}}_{1234}& = &  \colorf{a_1 a_2 a_5} \colorf{a_5 a_6 a_7} 
   \colorf{a_6 a_{13} a_{10}} \colorf{a_3 a_{10} a_8} \colorf{a_8 a_{11} a_9}
   \colorf{a_{13} a_{14} a_{11}} \colorf{a_7 a_9 a_{12}} \colorf{a_4 a_{12} a_{14}} \nn\\
& = & 12 N_c \Bigl(\Tr_{1 2 3 4} + \Tr_{1 4 3 2} - \Tr_{1 2 4 3} -
                  \Tr_{1 3 4 2} \Bigr)\,, 
\label{CVACVBTraceBasis}
\end{eqnarray}
exposing the manifest absence of double-trace terms in the leading
divergence.  Because the color factors are one-particle reducible, they
are proportional to the tree-level color factors; they are given by
\begin{equation}
 C^{V^{\A}}_{1234}\,=\,
C_{V^{\A}} \, \colorf{a_1 a_2 b} \colorf{b a_3 a_4} \,,
 \qquad C^{V^{\B}}_{1234}
\,=\, C_{V^{\B}} \, \colorf{a_1 a_2 b} \colorf{b a_3 a_4}
\,.
 \label{generalGGcolor3loop}
\end{equation}
These equations may also be taken as the definition of the 
three-loop scalar invariants $C_{V^{\A}}$ and $C_{V^{\B}}$.

The graphs in \fig{ThreeLoopNoDoubleTraceFigure}, dressed with
the two types of one-particle-reducible color structure,
reproduce the original representation of the divergence.
Thus they account for the absence of double-trace terms for
the terms with two powers of loop momentum in the numerator.
Double-trace terms may only arise from subleading terms
in the small-external-momentum expansion of the integrals in
\fig{IntegralsThreeLoopFigure}. In \sect{3loopSubDivSection}
we will show that the three-loop double-trace terms develop UV
divergences starting at $D=20/3$; therefore the double-trace bound
suggested in \eqn{SuperYangMillsDoubleTracePowerCount} is saturated at
$L=3$.  We note that the leading divergences at one and two loops are
{\it not} associated with powers of loop momentum in the numerator
of the parent integral (which can be considered contact terms in the UV
limit).  Thus a rearrangement of the leading-divergence
contributions similar to \eqn{generalGGcolor3loop}
is not possible at one and two loops.  Therefore the
single- and double-trace terms have the same critical dimension
at these loop orders.

We now turn to the independent question of the divergence structure
for a general gauge groups $G$, starting from the representation in
\eqn{ThreeLoopYMAmplitude}.  Following a similar procedure as at one
and two loops we write the three-loop four-point color tensors in
terms of six independent tensors, which we may take to be the five
lower-loop ones used in \eqn{TwoLoopPoleVFormGeneral} plus one
irreducible three-loop tensor (see \app{ColorAppendix}).
Summing over the various color permutations, the divergence of
the four-particle three-loop amplitude in $D=6-2\epsilon$ dimensions
is
\begin{equation} 
{\cal A}_4^{\threl}(1,2,3,4) \Bigr|^{G}_{\rm pole}
 =  g^{8} \, {\cal K}\,
 {\cal V}^{(3)}\, \Bigl( s_{12} \, \colorf{a_2 a_3 b}\colorf{b a_4 a_1} + 
                   s_{23} \,\colorf{a_1 a_2 b}\colorf{b a_3 a_4} \Bigr) \, .
\label{ThreeLoopPoleG}
\end{equation}
(Alternatively, one can use \eqn{LeadingPiecesThreeLoop}
to arrive at \eqn{ThreeLoopPoleG}.)
The color tensor of the three-loop divergence is proportional to
the tree-level color tensor.  Only two out of the six independent
color tensors for a general gauge group $G$ are present in the
divergence.  However, the scalar coefficient ${\cal V}^{(3)}$ involves
additional group invariants $C_{V^{\A}}$ and $C_{V^{\B}}$,
defined in \eqn{generalGGcolor3loop},
which do not appear below three loops,
\begin{equation} 
{\cal V}^{(3)} =
- 2 \, ( C_{V^{\A}} V^{\A} + 3 C_{V^{\B}} V^{\B} ) \, .
\label{ThreeLoopCasimirs}
\end{equation} 
They are nontrivial group invariants constructed out of structure 
constants, with the index contraction following the topology of the 
vacuum diagrams $V^{\A}$ and $V^{\B}$;
their explicit definitions are given in \eqn{Casimirs}, and
their values for $SU(N_c)$ are provided in \eqn{CasimirsForSUNc}.
Although they are not reducible to $C_A$ alone, they are related to
the standard group invariants $C_A$ and $(d_A^{abcd})^2$ by,
\begin{eqnarray} 
C_{V^{\A}} - C_{V^{\B}} &=& \frac{C_A^3}{8} \,,
\label{CVACVBdiff}\\
\frac{1}{3} C_{V^{\A}}+ \frac{2}{3} C_{V^{\B}}
&=& \frac{d_A^{abcd}d_A^{abcd}}{N_A C_A} \,,
\label{CVACVBsum}
\end{eqnarray} 
where $d_A^{abcd}$ is the totally symmetric rank four tensor
in the adjoint 
representation\footnote{The invariant tensor $d_A^{a_1a_2a_2a_4}$
is defined as
$
d_A^{a_1a_2a_2a_4}=\frac{1}{4!}\sum_{\pi\in S_4}
\Tr_A(T^{a_{\pi_1}}T^{a_{\pi_2}}T^{a_{\pi_3}}T^{a_{\pi_4}}) \,,
$
where $S_4$ is the set of permutations of four objects.}, and $N_A$ is
the dimension of the adjoint representation. However, the invariants
$C_{V^{\A}}$ and $C_{V^{\B}}$ are more natural in our context, because
they correspond directly to the color factors of vacuum-like diagrams.

We can make a few observations related to the one-particle-reducible
form of the UV-divergent terms displayed in
\fig{ThreeLoopNoDoubleTraceFigure}.  We note that the group invariants
in \eqn{ThreeLoopCasimirs} can be promoted naturally to rank-three
invariant tensors, as suggested by the marked points on the internal
lines of the vacuum diagrams $V^{\A}$ and $V^{\B}$ in
\fig{Vacuum3loopsFigure}, which denote doubled propagators.
This connotation suggests that
\begin{equation} 
C^{abc}_{V^{\A}} \equiv C_{V^{\A}}\colorf{abc}\,,\qquad
C^{abc}_{V^{\B}} \equiv C_{V^{\B}}\colorf{abc}\,,
\end{equation} 
are the most primitive yet nontrivial three-loop rank-three 
color tensors that can be constructed only out of structure constants.
All other three-loop rank-three tensors built from a single string
of structure constants reduce to a multiple of $C_A^3 \colorf{abc}$.  
It is remarkable that the UV divergence depends only on the three-loop
invariants in \eqn{ThreeLoopCasimirs}. Moreover, they follow a simple
pattern: the color invariants and vacuum integrals are in one-to-one
correspondence, with no mixing between the terms, and their relative
numerical coefficient can be interpreted
({\it via} \eqn{LeadingPiecesThreeLoop}) as having a combinatorial
origin. We shall see below that the four-loop divergences follow a
similar pattern.

Finally, we comment on the degree of
transcendentality\footnote{Riemann $\zeta$ values $\zeta_n$ are
assigned degree of transcendentality $n$, logarithms are assigned
degree 1, polylogarithms ${\rm Li}_n$ degree $n$ and rational numbers
are assigned degree $0$.}  of the divergences.  In four dimensions,
infrared-divergent terms of \NeqFoursYM\ amplitudes, expanded in Laurent
series around $D=4$, exhibit a uniform degree of transcendentality
through at least three and four loops in the planar
case~\cite{BDS,BCDKS}, and through two loops for the full color
dependence~\cite{NNS}. This property is related to the uniform degree
of transcendentality observed for the anomalous dimension of twist-two
operators~\cite{KLOV}.
Information on UV-divergent terms in $D>4$ is more limited.  Through
two loops, the simple structure of the vacuum integrals
(\ref{I41pole}), (\ref{planarseven}) and (\ref{nonplanarseven}) enforces
a uniform degree of transcendentality.
As can be seen from \eqns{VA}{VB}, this is no longer the case at three
loops: a non-uniform degree of transcendentality can and does occur
({\it cf.} \eqn{ThreeLoopPole}).
It is interesting to note, however, that for gauge group $SU(N_c)$
the coefficient of each power of $N_c$ does have a uniform degree of 
transcendentality, due to the
appearance of the particular linear combination $V^{\A} + 3 \,
V^{\B}$ in the subleading-color terms in
\eqn{ThreeLoopPoleVForm}.  The same combination of vacuum integrals
appears in the UV pole for the three-loop $\NeqEight$ supergravity
amplitude in $D=6-2\e$, ensuring that it also has a uniform degree of
transcendentality~\cite{CompactThree}.
On the other hand, for a general gauge group, \eqn{ThreeLoopCasimirs}
does not display any interesting behavior with respect to
transcendentality.  The significance and
generality of these facts may be clarified further by evaluating the
UV singular terms of the four-loop amplitude, as we do below.


\subsection{Four-loop ultraviolet divergence}

The four-loop four-point amplitude is given in
\eqn{FourLoopYMAmplitude} and figs.~\ref{BC1Figure}--\ref{E2Figure}.
According to \eqn{SuperYangMillsPowerCount}, we expect the four-loop
planar amplitudes to start diverging in the critical dimension
$D_c=11/2$.  Indeed, an inspection of the parent integrals in
figs.~\ref{BC1Figure}--\ref{E2Figure} reveals that 
many of them have numerator polynomials $N_i$ that are
quartic in the loop momentum, leading to logarithmic divergences
in $D_c=11/2$.  For $G=SU(N_c)$ we have no reason to expect color
single-trace terms to exhibit further cancellations compared to
\eqn{SuperYangMillsPowerCount}.  However, because all divergences 
in the critical dimension can be interpreted as arising from contact
terms, we expect that, as for the three-loop case, divergences
from double-trace terms may cancel, increasing the dimension in 
which such divergences first appear; below we show in detail how this
occurs.

\begin{figure}[tbh]
\centerline{\epsfxsize 6 truein \epsfbox{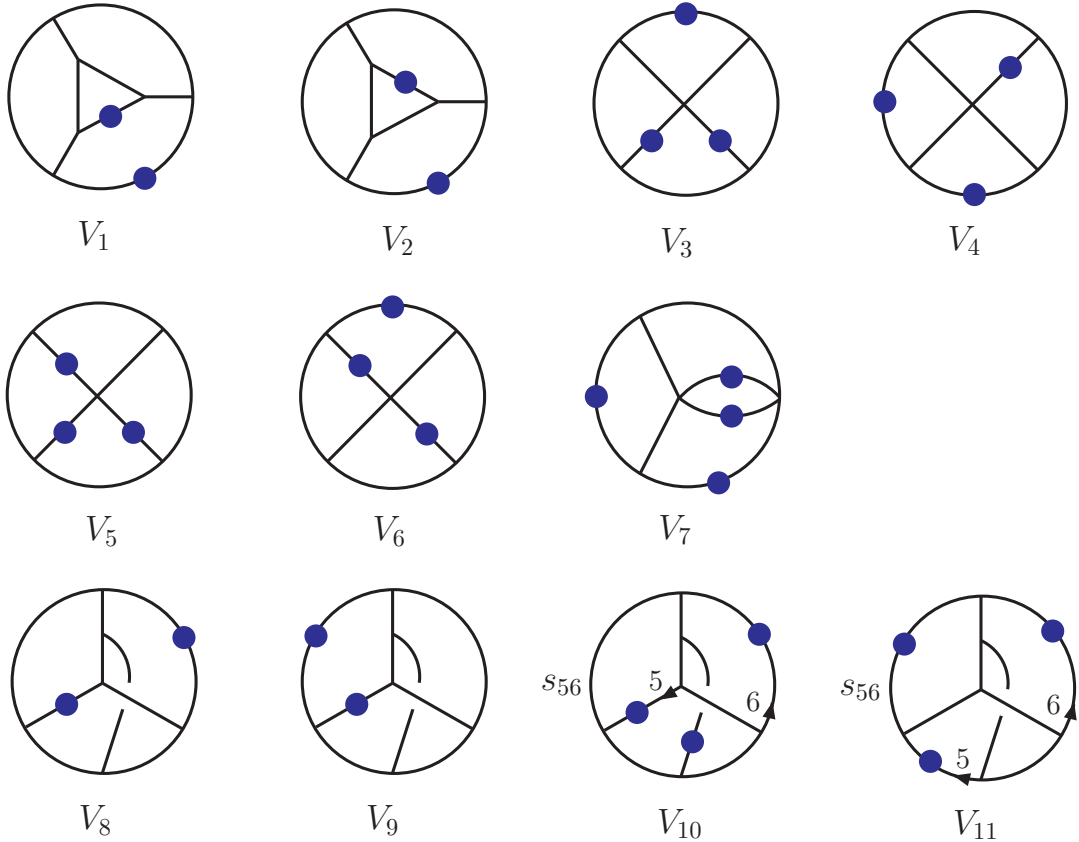}}
\caption[a]{\small Four-loop vacuum diagrams describing the UV
divergences of individual graphs in the four-loop \NeqFoursYM\
amplitude.  A dot indicates that a given propagator appears squared
in the integral. For $V_{10}$ and $V_{11}$, the factor of $s_{56}$
indicates the insertion of $s_{56} = (l_5+l_6)^2$ into the numerator 
of the integral, where lines 5 and 6 are marked in the figure.}
\label{Vacuum4loopsFigure}
\end{figure}

Of the 49 non-vanishing integrals in \eqn{FourLoopYMAmplitude}, 29
diverge in $D_c=11/2$.  Following a similar
analysis as at three loops, their leading divergences may be expressed in
terms of 11 vacuum integrals, $V_1$ through $V_{11}$, shown in
\fig{Vacuum4loopsFigure}:
\begin{eqnarray}
&&
I_{14} \ra 
 s_{12} V_{1}
\, , \quad 
I_{15} \ra 
 -s_{12} V_{1}
\, , \quad 
I_{16} \ra 
 -s_{12} V_{2}
\, , \quad 
I_{17} \ra 
 -s_{12} V_{2}
\, , \nn \\&&
I_{18} \ra 
 s_{12} V_{1}
\, , \quad 
I_{19} \ra 
 -s_{12} V_{2}
\, , \quad 
I_{20} \ra 
 s_{12} V_{2}
\, , \quad  
I_{21} \ra 
 s_{12} V_{3}
\, , \nn \\&&
I_{22} \ra 
 s_{12} V_{5}
\, , \quad 
I_{23} \ra 
 s_{12} V_{5}
\, , \quad 
I_{25} \ra 
 -s_{12} V_{4}
\, , \quad 
I_{26} \ra 
 -s_{12} V_{2}
\, , \nn \\&&
I_{29} \ra 
 s_{12} (V_{2} + V_{4})
\, , \quad 
I_{31} \ra 
 -s_{23} V_{2}
\, , \nn \\&&
I_{32} \ra 
 -s_{12} (V_{1} - V_{5}) - s_{23} V_{2} 
\, , \quad 
I_{33} \ra 
 - s_{12} (V_{2} - V_{5} - V_{6}) - s_{23} V_{1} 
\, , \nn \\&&
I_{34} \ra 
 (s_{23}-s_{13})(V_{4} - V_{6})
\, , \quad 
I_{35} \ra  
 s_{12} V_{3}-s_{23} (V_{2} + V_{3} -V_{4} - V_{5} + V_{7})
\, , \nn \\&&
I_{36} \ra 
 s_{12} V_{7}
\, , \quad 
I_{37} \ra 
 -s_{23} (V_{5} + 2 V_{6})
\, , \quad 
I_{38} \ra 
 2 s_{13} V_{2}
\, , \nn \\&&
I_{39} \ra 
 2 s_{12} (V_{2} - V_{3} + V_{5}) - s_{23} (V_{2} - 2 V_{3})
\, , \quad  
I_{41} \ra 
 s_{12} V_{8}
\, , \nn \\&&
I_{42} \ra 
 s_{12} V_{8}
\, , \quad
I_{43} \ra 
 -s_{12} V_{5}
\, , \quad
I_{45} \ra 
 s_{12} (3 V_{8} - 2 V_{10})
\, , \nn  \\&&
I_{46} \ra 
 {1\over 2} (s_{23} -s_{13}) (V_{4} -  V_{5} - 2V_{6}  + V_{9})
\, , \nn \\&&
I_{48} \ra 
 - s_{12} (V_{4} + V_{5})
 - s_{23} (2 V_{6} - V_{8} + 3 V_{9} - 2 V_{11})
 - s_{13} (2 V_{8} - V_{9})
\, , \nn \\&&
I_{49} \ra 
 s_{13} \Bigl(V_{3}- {1\over 2} V_{4} - {1\over 2} V_{5} + V_{7} +V_{8} 
   - {5\over 2} V_{9}-V_{10} +V_{11} \Bigr)
\,,
\label{YM4integralpoles}
\end{eqnarray}
and all other integrals are finite in the critical dimension.

The color factors of the 50 parent integrals in
figs.~\ref{BC1Figure}--\ref{E2Figure} are collected in
\app{NumeratorAppendix}. For gauge group $SU(N_c)$ it is straightforward
to evaluate these color factors in the trace basis.  We refrain from
including these expressions directly, due to their length. (In
\app{ColorAppendix} we decompose the $C_i$ in a basis of color
tensors for a general gauge group, and list the basis elements
in trace basis.)
However, the structure of the UV divergence is substantially
simpler.  Upon using the explicit color factors and the reduction to
vacuum integrals~(\ref{YM4integralpoles}), we find that the UV
divergence of the color-dressed amplitude depends only on the three
integrals $V_1$, $V_2$ and $V_8$:
\begin{eqnarray} 
{\cal A}_4^{\fourl}(1,2,3,4) \Bigr|^{SU(N_c)}_{\rm pole} 
&=& 
 - 6 \, g^{10} \, {\cal K} \, N_c^2
   \Bigl( N_c^2 \, V_1
       + 12 \, ( V_1 + 2 \, V_2 + V_8 ) \Bigr)
\label{FourLoopPoleVForm}\\ &&\hskip-1.2cm \null
\times \Bigl( s_{12} \, ( \Tr_{1324} + \Tr_{1423} )
           + s_{23} \, ( \Tr_{1243} + \Tr_{1342} )
           + s_{13} \, ( \Tr_{1234} + \Tr_{1432} ) \Bigr)
 \,.
\nonumber
\end{eqnarray}
Thus, we find that double-trace terms are absent from the
divergence in the critical dimension $D_c$, as was the case at three
loops.  Another interesting feature is that terms independent
of $N_c$ are also absent from the divergence. Only the leading and
next-to-leading powers of $N_c$ are present in the single-trace
divergence.  Finally, using eqs.~(\ref{TwoLoopPoleVForm}),
(\ref{ThreeLoopPoleVForm}) and (\ref{FourLoopPoleVForm}),
the divergences for $L=2,3,4$ all have the form,
\begin{eqnarray} 
{\cal A}_4^{(L)}(1,2,3,4) \Bigr|^{SU(N_c)}_{\rm pole} 
&=& 
 (-1)^{L-1} \, (L-1)! \, g^{2L+2} \, {\cal K} \, N_c^L
   \biggl( \Bigl( 1 + \frac{12}{N_c^2} \Bigr) \, V^{\rm planar}
           + \frac{12}{N_c^2} \, V^{\rm non-planar} \biggr)
\nonumber\\ &&\hskip-1.0cm \null
\times \Bigl( s_{12} \, ( \Tr_{1324} + \Tr_{1423} )
           + s_{23} \, ( \Tr_{1243} + \Tr_{1342} )
           + s_{13} \, ( \Tr_{1234} + \Tr_{1432} ) \Bigr)
\nonumber\\ &&\hskip0.0cm\null
+ \delta_{L,2} \times (\hbox{double-trace-terms})
 \,,
\label{LLoopPoleVForm}
\end{eqnarray}
where $V^{\rm planar}$ and $V^{\rm non-planar}$ come from
planar and non-planar four-point integrals, respectively.

\begin{figure}[tb]
\centerline{\epsfxsize 6.3 truein \epsfbox{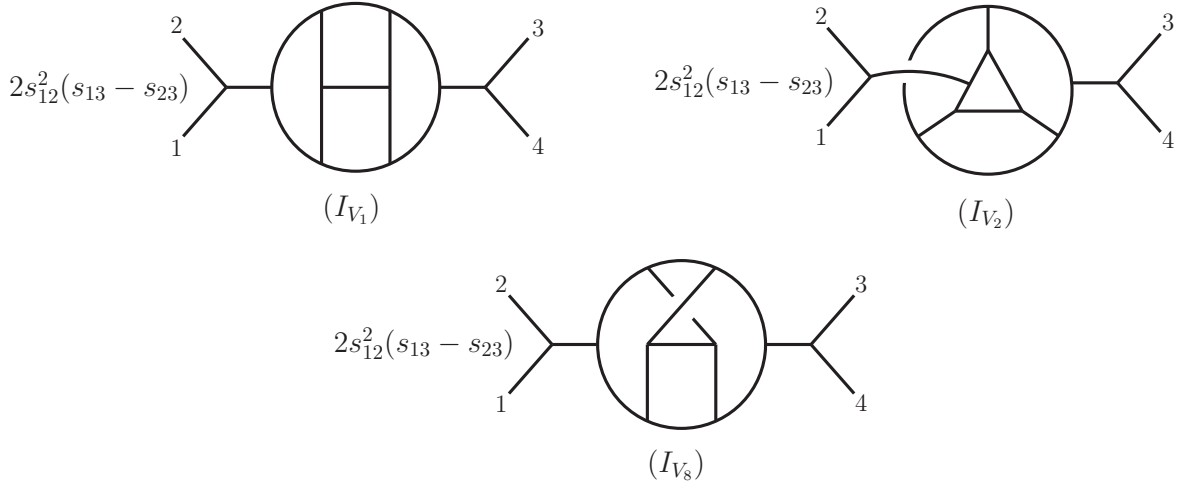}}
\caption[a]{\small At four loops the leading divergence can be
  rearranged so that it comes from graphs with two spurious
  external propagators. The graphs with $1/s_{12}$  propagators are
  displayed here.  This rearrangement
  makes manifest that the color factors of these graphs have no
  double-trace contributions. The complete contribution comes from
  summing over all 24 permutations of external legs and multiplication
  by a symmetry factor of $1/16$, $1/8$ and $1/16$, respectively, to
  remove double counts in both graphs.}
\label{PropagatorDiagrams4loopsFigure}
\end{figure}

For a general gauge group the leading UV divergence at four loops 
has a similarly simple structure, proportional to the tree-level
color tensor:
\begin{equation} 
{\cal A}_4^{\fourl}(1,2,3,4) \Bigr|^{G}_{\rm pole} =  
g^{10} \, {\cal K}\, {\cal V}^{(4)}\, 
\Bigl( s_{12} \, \colorf{a_2 a_3 b}\colorf{b a_4 a_1} + s_{23}
 \,\colorf{a_1 a_2 b}\colorf{b a_3 a_4} \Bigr) \,,
\label{FourLoopPoleG}
\end{equation}
where
\begin{equation} 
{\cal V}^{(4)} = 3 \, (C_{V_1} V_1+2C_{V_2}V_2+C_{V_8}V_8) \, .
\end{equation} 
The coefficients $C_{V_1}$, $C_{V_2}$ and $C_{V_8}$ are the group
invariants associated with the corresponding vacuum diagrams. Their
expressions in terms of structure constants are collected in
\app{ColorAppendix}.  
This structure is similar to that of the two- and three-loop UV poles in 
\eqns{ThreeLoopPoleG}{ThreeLoopCasimirs}.
As at three loops, the four-loop group invariants are not independent;
rather, they satisfy the following relations:
\begin{eqnarray} 
C_{V_1} - C_{V_2} &=& \frac{C_A^4}{8} \,,
\label{CV128A} \\
\frac{1}{3} C_{V_1} + \frac{2}{3} C_{V_2}
&=& \frac{d_A^{abcd}d_A^{abcd}}{N_A} \,,
\label{CV128B} \\
C_{V_8} &=& C_{V_2} \,.
\label{CV128C}
\end{eqnarray} 

As with the three-loop case discussed above, it is possible
to rearrange the UV-divergent contributions at four loops
into one-particle-reducible parent graphs.
This form manifestly exhibits the absence of double-trace
terms. In this case our representation exhibits two propagators in the
same momentum channel, which are canceled by numerator factors, as
depicted in \fig{PropagatorDiagrams4loopsFigure}. The divergent part
of the amplitude then has the simple form
\begin{eqnarray} 
{\cal A}_4^{\fourl}(1,2,3,4) 
&=&  -  g^{10} \, {\cal K} \sum_{S_4}
        \Bigl[ \Frac{1}{16} C^{V_1}_{1234} I_{V_1}
   + \Frac{1}{8} \, C^{V_2}_{1234} I_{V_2} 
   + \Frac{1}{16} C^{V_8}_{1234} I_{V_8} \Bigr]
\; +\; \hbox{subleading} . \nn\\ 
\end{eqnarray}
where the integrals correspond to the three graphs in
\fig{PropagatorDiagrams4loopsFigure}, and their color factors are
proportional to the tree-level color factors,
\begin{equation}
 C^{V_i}_{1234}\,=\,
C_{V_i} \, \colorf{a_1 a_2 b} \colorf{b a_3 a_4}  \,,
\qquad i = 1,2,8.
\label{generalGGcolor4loop}
\end{equation}

The next step is to evaluate the UV poles for the
four-loop vacuum integrals $V_1$, $V_2$ and $V_8$ in their
critical dimension $D=11/2 - 2\e$. To this end we use
the same infrared rearrangement~\cite{Vladimirov} 
(related to the $R^*$ operation~\cite{Rstar}) 
that was used~\cite{CompactThree} to evaluate 
the three-loop vacuum integrals $V^{\A}$ and $V^{\B}$:
We inject and remove momentum $k^\mu$, with $k^2 \neq0$,
at two of the vertices of the vacuum integral, thus
transforming it into a four-loop two-point integral,
which possesses the same UV poles,
but no infrared divergences.  (The infrared divergences
arise in the small-momentum limit from doubled internal propagators.)
We always take
the two vertices in question to be connected by a single propagator.
Then the four-loop two-point integral factorizes into the 
product of a finite three-loop two-point integral and a
UV-divergent one-loop two-point integral.

\begin{figure}[tb]
\centerline{\epsfxsize 3.8 truein \epsfbox{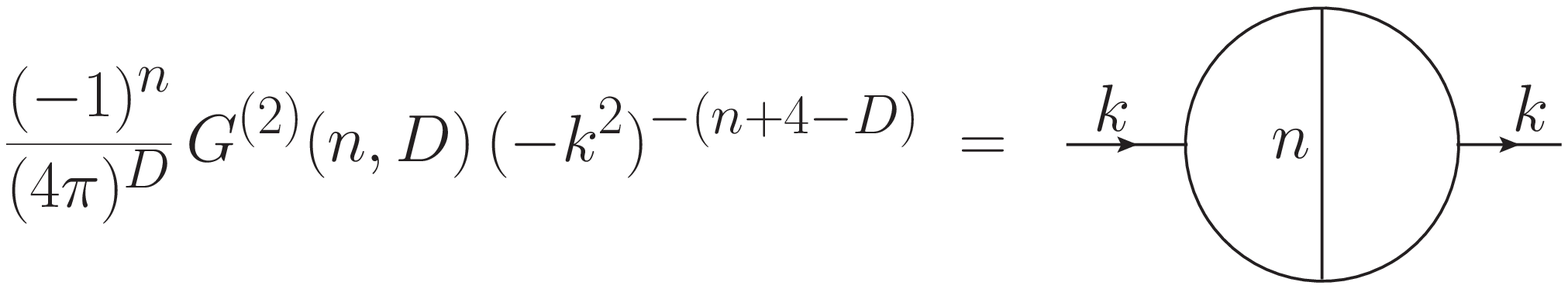}}
\caption[a]{\small A two-loop integral with central propagator 
raised to the power $n$.  If $n$ is not an integer, $G^{(2)}(n,D)$ 
cannot be reduced to one-loop integrals.}
\label{twoloopnFigure}
\end{figure}

Some of the three-loop two-point integrals can be
evaluated through a similar procedure, by factorizing them
into a product of two-loop and one-loop two-point integrals.
A few of the resulting two-loop integrals,
such as those shown in \fig{twoloopnFigure}, are not factorizable.
To evaluate them we employ the gluing relations~\cite{CTIBP}, which 
require consistency of the various ways of factorizing a higher-loop
UV-divergent integral into products of lower-loop integrals.

\begin{figure}[tbh]
\centerline{\epsfxsize 4.5 truein \epsfbox{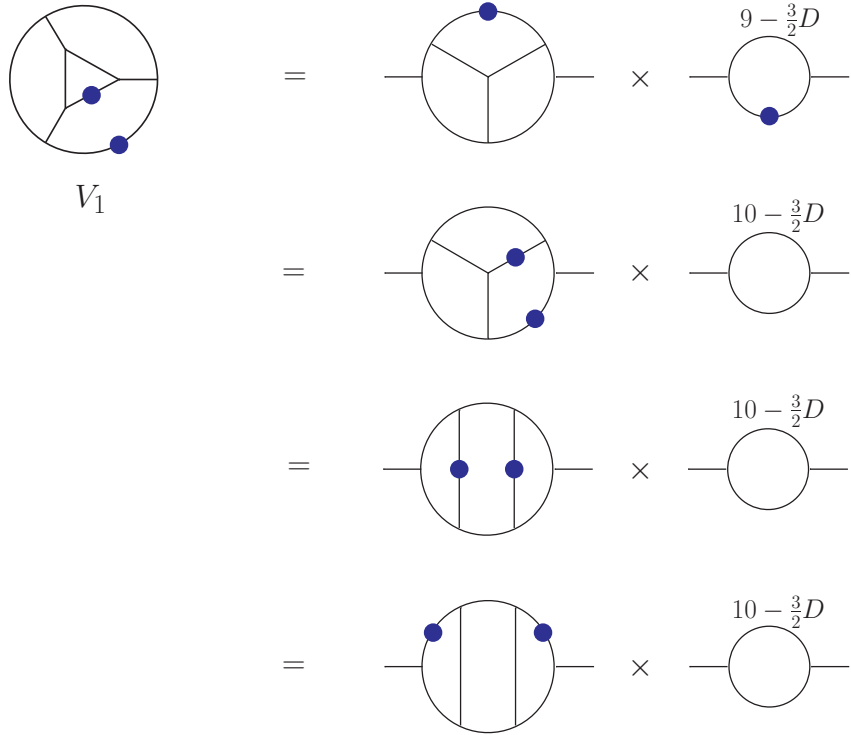}}
\caption[a]{\small The UV poles in the vacuum diagram $V_1$
can be determined from the product of a finite three-loop
two-point integral with a UV-divergent one-loop two-point integral.
There are four inequivalent ways of doing the reduction,
corresponding to different propagators connecting the two
points at which the momentum $k^\mu$ is injected.}
\label{V1reductionFigure}
\end{figure}

For example, the diagram $V_1$ has four
inequivalent propagators (not related by symmetry), 
leading to four inequivalent factorizations; they are
shown in \fig{V1reductionFigure}.  As usual,
a dot indicates that a given propagator appears squared
in the integral.  Similarly, the numbers $(9-3D/2)$ and $(10-3D/2)$ 
indicate the power to which that propagator is raised,
which is determined by dimensional analysis of the three-loop
integral. All four factorizations should give the same answer; 
this consistency condition is an example of a gluing relation~\cite{CTIBP}.

The one-loop bubble integral is simple to evaluate.
For an arbitrary dimension $D$ and powers $n_1$ and $n_2$ 
of the two propagators, it is given by~\cite{GrozinLectures}
\begin{equation}
I^{\rm bubble}(n_1,n_2)\ \equiv -i\int\frac{d^Dp}{(2\pi)^D}
\frac{1}{\bigl(\,(p+k)^2\,\bigr)^{n_1} (p^2){\vphantom{\big)}}^{n_2} }
\ =\ \frac{(-1)^{n_1+n_2}}{(4\pi)^{D/2}}
G(n_1,n_2) (-k^2)^{-(n_1+n_2-D/2)}\,,
\label{Bubble}
\end{equation}
where
\begin{equation}
G(n_1,n_2)\ =\ { \Gamma(-D/2+n_1+n_2) \Gamma(D/2-n_1)\Gamma(D/2-n_2)
                \over \Gamma(n_1) \Gamma(n_2) \Gamma(D-n_1-n_2) } \,.
\label{DefineG}
\end{equation}
In $D=11/2 -2\e$ dimensions, for the cases required in
\fig{V1reductionFigure} we have $(n_1,n_2) =
(\textstyle{\frac{3}{4}}+3\e,2)$ and $(n_1,n_2) =
(\textstyle{\frac{7}{4}}+3\e,1)$.  Inserting these values into
\eqn{Bubble} we find that both integrals have the same UV pole,
\begin{eqnarray}
G({\textstyle{\frac{3}{4}}}+3\e,2)
& =&  { 4 \over 21 } {1 \over \Gamma(\textstyle{\frac{3}{4}})}
                    {1 \over \e}\ +\ \Ord(1) \,,
\label{UVdivbubblea}\\
G({\textstyle{\frac{7}{4}}}+3\e,1)
&=& { 4 \over 21 } {1 \over \Gamma(\textstyle{\frac{3}{4}})}
                    {1 \over \e}\ +\ \Ord(1) \,.
\label{UVdivbubbleb}
\end{eqnarray}

The finite three-loop two-point integrals can be reduced to a set of
master integrals using the method of integration by parts
(IBP)~\cite{CTIBP}, in particular the algorithm MINCER, which is available
in {\sc FORM}~\cite{VermaserenFORM}.  For the two planar topologies occurring
in \fig{V1reductionFigure}, the so-called Benz and ladder topologies,
the IBP reduction procedure results in integrals that factorize into
the product of two-loop and one-loop two-point integrals.
\Eqn{Bubble} can be applied to the latter.  The two-loop integrals can
also be reduced to products of one-loop integrals, except for the
integrals $G^{(2)}(n,D)$ shown in \fig{twoloopnFigure}, in which the
power $n$ to which the central propagator is raised is not an integer.
However, a gluing relation can be used to solve for such integrals in
$D=11/2$.

For example, IBP reduction
of the top three lines in \fig{V1reductionFigure} results in
the following relations for the $1/\e$ pole terms,
\begin{eqnarray}
V_1 &=& \Biggl[ 
  \frac{6272}{25} \, \Gamma^5({\textstyle{\frac{3}{4}}})
- \frac{256}{5} \, \Gamma^4({\textstyle{\frac{3}{4}}})
 \Gamma({\textstyle{\frac{1}{2}}}) \Gamma({\textstyle{\frac{1}{4}}})
+ 8 \, { \Gamma^2({\textstyle{\frac{3}{4}}}) 
         \Gamma({\textstyle{\frac{1}{4}}})
       \over  \Gamma({\textstyle{\frac{1}{2}}}) }
  G^{(2)}({\textstyle{\frac{9}{4}}},{\textstyle{\frac{11}{2}}})
 \biggr] 
{ G({\textstyle{\frac{3}{4}}}+3\e,2) \over (4\pi)^{11} } \quad{~}
\label{V1a}\\
&=& \Biggl[
  \frac{12992}{25} \, \Gamma^5({\textstyle{\frac{3}{4}}})
- \frac{496}{5} \, \Gamma^4({\textstyle{\frac{3}{4}}})
 \Gamma({\textstyle{\frac{1}{2}}}) \Gamma({\textstyle{\frac{1}{4}}})
+ \frac{1}{2} \, { \Gamma^2({\textstyle{\frac{3}{4}}}) 
         \Gamma({\textstyle{\frac{1}{4}}})
       \over  \Gamma({\textstyle{\frac{1}{2}}}) }  
  G^{(2)}({\textstyle{\frac{9}{4}}},{\textstyle{\frac{11}{2}}})
 \biggr] 
{ G({\textstyle{\frac{7}{4}}}+3\e,1) \over (4\pi)^{11} }
\qquad{~}
\label{V1b}\\
&=& \Biggl[
  \frac{12352}{25} \, \Gamma^5({\textstyle{\frac{3}{4}}})
- \frac{288}{5} \, \Gamma^4({\textstyle{\frac{3}{4}}})
 \Gamma({\textstyle{\frac{1}{2}}}) \Gamma({\textstyle{\frac{1}{4}}})
- 5 \, { \Gamma^2({\textstyle{\frac{3}{4}}}) 
         \Gamma({\textstyle{\frac{1}{4}}})
       \over  \Gamma({\textstyle{\frac{1}{2}}}) }  
 \Bigl( G^{(2)}({\textstyle{\frac{9}{4}}},{\textstyle{\frac{11}{2}}})
 - 6 \, G^{(2)}({\textstyle{\frac{5}{4}}},{\textstyle{\frac{11}{2}}})
 \Bigr)
 \biggr] 
\nonumber\\ &&\hskip0.2cm \null \times
{ G({\textstyle{\frac{7}{4}}}+3\e,1) \over (4\pi)^{11} }
\,.\qquad{~}
\label{V1c}
\end{eqnarray}
Equating the three forms for $V_1$ at order $1/\e$ yields
\begin{eqnarray}
 G^{(2)}({\textstyle{\frac{5}{4}}},{\textstyle{\frac{11}{2}}})
&=& - \frac{64}{25} \, \Gamma^2({\textstyle{\frac{3}{4}}})
                 \Gamma^2({\textstyle{\frac{1}{2}}})
  + \frac{928}{125} \, { \Gamma^3({\textstyle{\frac{3}{4}}})
                 \Gamma({\textstyle{\frac{1}{2}}})
           \over \Gamma({\textstyle{\frac{1}{4}}}) }
\ +\ \Ord(\e) \,, 
\label{prop2l54}\\
 G^{(2)}({\textstyle{\frac{9}{4}}},{\textstyle{\frac{11}{2}}})
&=& - \frac{32}{5} \, \Gamma^2({\textstyle{\frac{3}{4}}})
                 \Gamma^2({\textstyle{\frac{1}{2}}})
  + \frac{896}{25} \, { \Gamma^3({\textstyle{\frac{3}{4}}})
                 \Gamma({\textstyle{\frac{1}{2}}})
           \over \Gamma({\textstyle{\frac{1}{4}}}) }
\ +\ \Ord(\e) \,, 
\label{prop2l94}
\end{eqnarray}
and
\begin{equation}
V_1 = {1\over (4\pi)^{11} \, \e}
 \Biggl[ \frac{512}{5} \, \Gamma^4({\textstyle{\frac{3}{4}}})
      - \frac{2048}{105} \, \Gamma^3({\textstyle{\frac{3}{4}}})
 \Gamma({\textstyle{\frac{1}{2}}}) \Gamma({\textstyle{\frac{1}{4}}})
 \Biggr]\ +\ \Ord(1) \,.
\label{V1full}
\end{equation}
The fourth line of \fig{V1reductionFigure} provides a redundant
equation.

\begin{figure}[tbh]
\centerline{\epsfxsize 4.2 truein \epsfbox{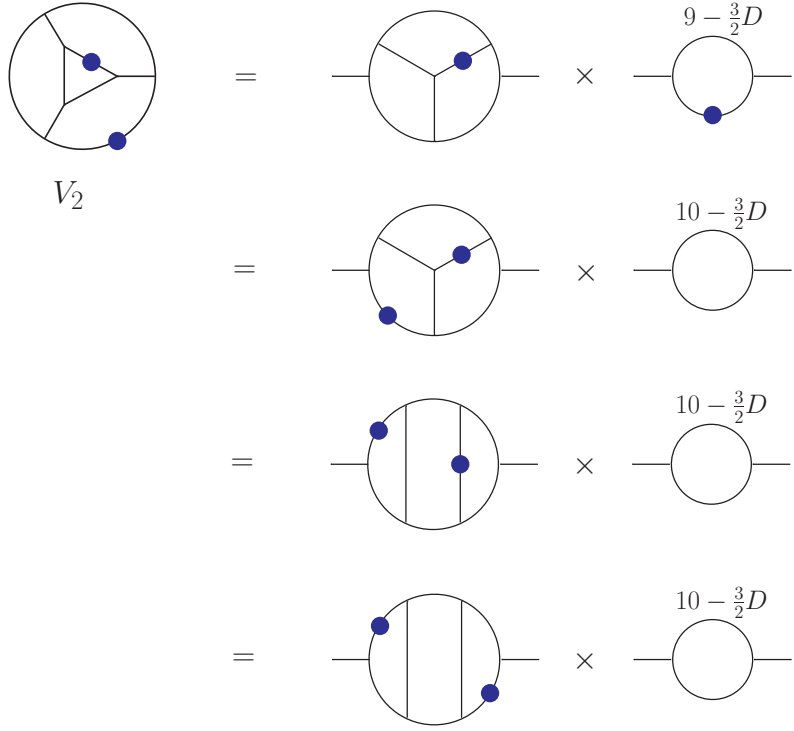}}
\caption[a]{\small Four inequivalent ways of reducing
vacuum diagram $V_2$ to a product of a finite three-loop
two-point integral with a UV-divergent one-loop two-point integral.}
\label{V2reductionFigure}
\end{figure}

A similar strategy leads to an analytic evaluation of $V_2$.
As depicted in \fig{V2reductionFigure}, there are four 
inequivalent ways of factorizing  this integral into a product of three-loop 
and one-loop two-point integrals.  The same Benz and ladder topologies
appear as for $V_1$, but with different configurations of double
propagators.  After using \eqns{prop2l54}{prop2l94}
for $G^{(2)}({\textstyle{\frac{5}{4}}},{\textstyle{\frac{11}{2}}})$ 
and $G^{(2)}({\textstyle{\frac{9}{4}}},{\textstyle{\frac{11}{2}}})$,
they all give the same result,
\begin{equation}
V_2 = {1\over (4\pi)^{11} \, \e}
 \Biggl[ - \frac{4352}{105} \, \Gamma^4({\textstyle{\frac{3}{4}}})
      + \frac{832}{105} \, \Gamma^3({\textstyle{\frac{3}{4}}})
 \Gamma({\textstyle{\frac{1}{2}}}) \Gamma({\textstyle{\frac{1}{4}}})
 \Biggr]\ +\ \Ord(1) \,.
\label{V2full}
\end{equation}
%

\begin{figure}[tbh]
\centerline{\epsfxsize 4.5 truein \epsfbox{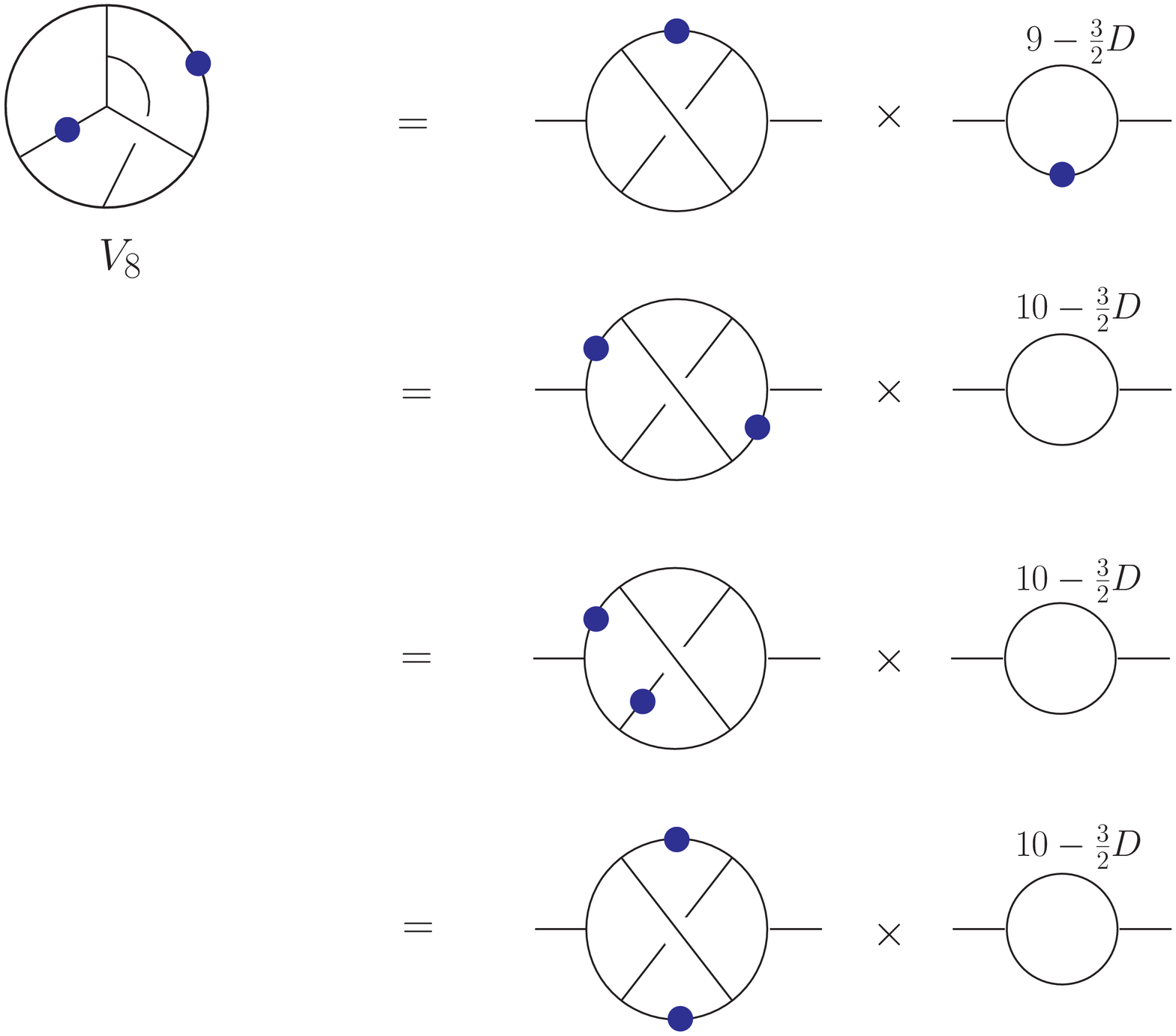}}
\caption[a]{\small Four inequivalent ways of reducing
vacuum diagram $V_8$ to a product of a finite non-planar three-loop
two-point integral with a UV-divergent one-loop two-point integral.}
\label{V8reductionFigure}
\end{figure}

There are four inequivalent ways of factorizing the non-planar
vacuum integral $V_8$ into a product of three-loop and 
one-loop two-point integrals, depicted in \fig{V8reductionFigure}. 
This time the non-planar three-loop two-point topology is obtained.
The IBP equations can be used to reduce any integral with this
topology down to the master integral ``NO$_m$'' of the same
topology, in which all propagators appear undoubled.
 In $D=4-2\e$, gluing relations
allow one to solve for the value of this master integral~\cite{CTIBP}
as $\e \to 0$ (which is proportional to $\zeta_5$).
However, in $D=11/2$, we find that
the gluing relations do not give new information.
That is, all four ways of factorizing $V_8$ lead to the same
expression,
\begin{equation}
V_8 = {1\over(4\pi)^{11}}
{4\over 21} {1\over\Gamma({\textstyle{\frac{3}{4}}})}
{V_8^{\rm fin} \over \e}\ +\ \Ord(1) \,,
\label{V8full}
\end{equation}
where
\begin{equation}
V_8^{\rm fin} = 
- \frac{5248}{125} \, \Gamma^5({\textstyle{\frac{3}{4}}})
+ \frac{224}{25} \, \Gamma^4({\textstyle{\frac{3}{4}}})
   \Gamma({\textstyle{\frac{1}{2}}}) 
   \Gamma({\textstyle{\frac{1}{4}}}) 
+ 2 \, {\rm NO}_m \,.
\label{V8fin}
\end{equation}
Although it is not needed for the four-loop 
\NeqFoursYM\ amplitude, a similar factorization
and reduction procedure for $V_9$ gives
\begin{equation}
V_9 = {1\over(4\pi)^{11}}
{4\over 21} {1\over\Gamma({\textstyle{\frac{3}{4}}})}
{V_9^{\rm fin} \over \e}\ +\ \Ord(1) \,,
\label{V9full}
\end{equation}
where
\begin{equation}
V_9^{\rm fin} = 
- \frac{15552}{125} \, \Gamma^5({\textstyle{\frac{3}{4}}})
+ \frac{576}{25} \, \Gamma^4({\textstyle{\frac{3}{4}}})
   \Gamma({\textstyle{\frac{1}{2}}}) 
   \Gamma({\textstyle{\frac{1}{4}}}) 
- 2 \, {\rm NO}_m \,.
\label{V9fin}
\end{equation}
%


\subsection{Gegenbauer sums for non-planar three-loop integral}   

Although we could not obtain an analytical value for $V_8$, or
equivalently for $V_9$ or NO$_m$, we could obtain the result to 13
digits using the Gegenbauer
polynomial $x$-space technique (GPXT)~\cite{CKTGPXT}.  This method is
based on the observation that, in position space, all propagators
depend only on the coordinates of the vertices they connect and thus,
after Wick rotation, they may be identified with the generation
function of the Gegenbauer (or ultraspherical) polynomials. Expanding
them and evaluating the integrals using properties of these
polynomials reduces Feynman integrals to finitely many (nested) sums.
A nice exposition of this technique is given in ref.~\cite{Bekavac};
in particular the integral NO$_m$ in any dimension $D$ is
reduced to a triple sum.  Using this approach for
$D={\textstyle\frac{11}{2}}$, we obtain
\bea
{\rm NO}_m &=& \frac{48}{50}
   \frac{\Gamma^3(\textstyle{\frac{3}{4}})}{\Gamma(\textstyle{\frac{1}{2}})}
\sum_{\ka=0}^\infty
   \frac{\Gamma(\ka+\textstyle{\frac{7}{4}})}{\ka!}
\sum_{n=0}^\infty
   \frac{\Gamma(\textstyle{n+\ka+\frac{7}{2}})}
        {\Gamma(\textstyle{n+\ka+\frac{11}{4}})}
\sum_{l=0}^n
   \frac{\Gamma(l+\textstyle{\frac{7}{4}})}{l!}
   \frac{\Gamma(n-l+\textstyle{\frac{7}{4}})}{(n-l)!}
\label{nomastersimp}\\ && \hskip0.0cm \null
\times
   \frac{1}{(l+\ka+\textstyle{\frac{7}{4}})(n-l+\ka+\textstyle{\frac{7}{4}})}
 \Biggl[ 
 - 2  \Biggl( \frac{1}{(n+\textstyle{\frac{1}{2}})
                       (n+\ka+\textstyle{\frac{5}{4}})
                       (n+\ka+\textstyle{\frac{11}{4}})
                       (n+\ka+2)}
\nonumber\\ && \hskip6.0cm \null
            - \frac{1}{(n+3)(n+\ka+\textstyle{\frac{5}{2}})
                       (n+\ka+\textstyle{\frac{13}{4}})(n+\ka+4)}  
      \Biggr)
\nonumber\\ && \hskip0.5cm \null
 + \frac{n+2\ka+\textstyle{\frac{11}{4}}}
        {(l+\ka+1)(n-l+\ka+1)}
   \Biggl(
      \frac{3n+4\ka+\textstyle{\frac{17}{2}}}
           {(n+3)(n+\ka+\textstyle{\frac{5}{2}}) 
            (n+\ka+\textstyle{\frac{13}{4}})(n+2\ka+\textstyle{\frac{7}{2}})}
\nonumber\\ && \hskip6.0cm \null
    - \frac{3n+4\ka+6}
           {(n+\textstyle{\frac{1}{2}})(n+\ka+\textstyle{\frac{5}{4}}) 
            (n+\ka+2)(n+2\ka+\textstyle{\frac{7}{2}})} \Biggr)
\nonumber\\ && \hskip0.5cm \null
 + \frac{n+2\ka+\textstyle{\frac{17}{4}}}
        {(l+\ka+\textstyle{\frac{5}{2}})(n-l+\ka+\textstyle{\frac{5}{2}})}
   \Biggl(
      \frac{3n+4\ka+\textstyle{\frac{23}{2}}}
           {(n+3)(n+\ka+\textstyle{\frac{13}{4}}) 
            (n+\ka+4)(n+2\ka+\textstyle{\frac{7}{2}})}
\nonumber\\ && \hskip5.5cm \null
    - \frac{3n+4\ka+9}
           {(n+\textstyle{\frac{1}{2}})(n+\ka+\textstyle{\frac{11}{4}}) 
            (n+\ka+2)(n+2\ka+\textstyle{\frac{7}{2}})} \Biggr)
 \Biggr] 
\,.
\nonumber
\eea
The sum over $l$ can be done in terms of hypergeometric functions:
\bea
{\rm NO}_m &=& \frac{2^5}{5^2}
   \frac{\Gamma^4(\textstyle{\frac{3}{4}})}{\Gamma(\textstyle{\frac{1}{2}})}
\sum_{n=0}^\infty \sum_{\ka=0}^\infty
  A(\ka,n) \biggl[ 
\nonumber\\ && \hskip0cm \null
  \Bigl( D(\ka,n) 
  - D(\ka-{\textstyle{\frac{5}{4}}},n+{\textstyle{\frac{5}{2}}})
  - D(\ka-{\textstyle{\frac{3}{4}}},n)
  + D(\ka-2,n+{\textstyle{\frac{5}{2}}}) \Bigr) H(\ka,n)
\nonumber\\ && \hskip-0.5cm \null
 + \Bigl( D(\ka,n) 
  - D(\ka-{\textstyle{\frac{5}{4}}},n+{\textstyle{\frac{5}{2}}})
  - D(\ka+{\textstyle{\frac{3}{4}}},n)
  + D(\ka-{\textstyle{\frac{1}{2}}},n+{\textstyle{\frac{5}{2}}}) 
  \Bigr) H(\ka+{\textstyle{\frac{3}{2}}},n) \biggr] \,,
\nonumber\\
{~} \label{NomNEW}
\eea
where
\be
  A(\ka,n) = \frac{\Gamma(\ka+{\textstyle{\frac{7}{4}}})
    \, \Gamma(n+{\textstyle{\frac{7}{4}}})
   \, \Gamma(\textstyle{n+\ka+\frac{7}{2}})}
    {\ka! \, n! \, \Gamma(\textstyle{n+{\ka+\frac{11}{4}}})} \,,
\label{Adef}
\ee
\be
 D(\ka,n) = {1\over (\ka+{\textstyle{\frac{3}{2}}})^2 }
\Biggl[ {1\over n+{\textstyle{\frac{1}{2}}}}
      - {2\over (n+{\textstyle{\frac{1}{2}}})
               +(\ka+{\textstyle{\frac{3}{2}}})}
      + {1\over (n+{\textstyle{\frac{1}{2}}})
               + 2(\ka+{\textstyle{\frac{3}{2}}})} \Biggr] \,,
\label{Ddef}
\ee
and
\be
H(\ka,n) = {1\over \ka+1}
 \ {}_{3}F_2({\textstyle{\frac{7}{4}}},\ka+1,-n;
           \ka+2,-n-{\textstyle{\frac{3}{4}}};1) \,.
\label{Hdef}
\ee
Using this representation and truncating the $\ka$ and $n$
sums at a value $N$ up to 6500, we found sequences of truncated values
NO$_m(N)$, such as 
\begin{eqnarray}
{\rm NO}_m(6000) &=& -6.197074209444889\, , \quad
{\rm NO}_m(6100)    =   -6.197095923684655\,, \nonumber\\
{\rm NO}_m(6200) &=& -6.197116937698505\,, \quad
{\rm NO}_m(6300)   =    -6.197137284819593\,,  \\
{\rm NO}_m(6400) &=& -6.197156996298421\,, \quad
{\rm NO}_m(6500)   =    -6.197176101462998\,, \nonumber
\label{NOmvalues}
\end{eqnarray}
which we then fit to a polynomial in $1/N$, obtaining
\begin{equation}
{\rm NO}_m = -6.198399226750(2),
\label{NOmbest}
\end{equation}
where the number in parentheses indicates the uncertainty in the last
digit.\footnote{Recently a much more accurate numerical value for
this integral has been obtained by Lee, Smirnov and
Smirnov~\cite{LSSPrivate}, using methods similar to those in
ref.~\cite{LSS}.}

We also applied GPXT to $V_8^{\rm fin}$
and $V_9^{\rm fin}$, obtaining similar sums, but with
somewhat more complicated summands.  The sequence for $V_8^{\rm fin}$
converges the fastest with $N$, but its numerical evaluation 
takes longer.  We obtain,
\begin{eqnarray}
V_8^{\rm fin} &=& 1.428452926283(3), 
\label{V8finbest}\\
V_9^{\rm fin} &=& 2.472370645275(3).
\label{V9finbest}
\end{eqnarray}
To the given accuracy, these values are compatible with 
\eqn{NOmbest} and the two analytic relations, 
\eqns{V8fin}{V9fin}.  

The combination appearing in the subleading-color part of
\eqn{FourLoopPoleVForm} is
\begin{equation}
V_1 + 2 V_2 + V_8 = {1\over(4\pi)^{11}}
{4\over 21} {1\over\Gamma({\textstyle{\frac{3}{4}}})}
{6.161859216543(3) \over \e} \,,
\label{V12V2V8best}
\end{equation}
Thus the ratio of the subleading-color term to
the leading-color term in \eqn{FourLoopPoleVForm} is
\begin{equation}
{ 12 \, (V_1 + 2 V_2 + V_8) \over N_c^2 V_1 }
= {44.40538395605(2) \over N_c^2} \,.
\label{sublnumerical}
\end{equation}
Amusingly, the large-$N_c$ approximation is strikingly bad, if
we take the gauge group to be $SU(3)$ as in QCD.  It is also strikingly
bad at three loops, where the ratio analogous to \eqn{sublnumerical}
can be extracted from \eqn{ThreeLoopPole} and is very similar in
magnitude, $36\, \zeta_3/N_c^2 = (43.274\ldots)/N_c^2$.

The fact that these numerical ratios are irrational (in the four-loop
case, apparently irrational) precludes the overall UV divergence
from canceling for any gauge group $G$ at three or four loops,
because ratios of group invariants are always rational.


\section{UV divergences of subleading-color structures}
\label{UVSubleadingColorSection}

As discussed in the previous section, the color double-trace terms are
better behaved in the ultraviolet than the single-trace terms.
Instead of the finiteness bound (\ref{SuperYangMillsPowerCount}), for
three and four loops the double-trace terms satisfy the
bound (\ref{SuperYangMillsDoubleTracePowerCount}).  The double-trace
terms of the three- and four-loop amplitudes are finite in the
dimensions where the corresponding single-trace amplitudes first
diverge, respectively in $D=6$ and $D=11/2$.  This result implies
that two-derivative double-trace operators of the form
$\Tr({\cal D}^2 F^2) \Tr(F^2)$ are not renormalized in these dimensions.

Suppose an $L$-loop cubic parent integral has $l$ powers of the loop momentum
in the numerator factor $N_i$.  Because it has $3L+1$ propagators
in the denominator, it behaves in the UV as,
\begin{equation}
\sim\ \int {d^{DL}\ell\ \ell^l \over (\ell^2)^{3L+1}}
\ \sim\ \ell^{DL+l-6L-2} \,.
\label{PowerCount1}
\end{equation}
By dimensional analysis, the number of powers $m$ of external momenta
sitting in front of this integral (including the four powers in the
prefactor ${\cal K}$ defined in \eqn{Prefactor}) is related to $l$
by $m = 2L + 2 -l$.  The critical dimension in which UV divergences
can first appear in \eqn{PowerCount1} is found by setting $DL+l-6L-2=0$.
Eliminating $l$ in factor of $m$, the critical dimension is
\begin{equation}
D(m,L)= 4+\frac{m}{L} \,.
\label{DivergenceDimension}
\end{equation}

By Lorentz invariance, only even powers of momenta give nonvanishing
results, implying that $m$ is effectively always even.
Furthermore, the three- and
four-loop integrals have at least six powers of external momenta that
are manifest in the prefactors, counting the four powers in ${\cal K}$
and two powers in the ``worst behaved'' integral numerators $N_i$;
therefore $m\ge6$.  We showed in the previous section that the
divergences in the single-trace terms indeed start at $m=6$, with no
further hidden cancellations.  We also found cancellations in the
double-traces terms in $D=4+6/L$.  If there are no further cancellations,
we expect the double-trace divergences to start with $m=8$,
corresponding to $D=20/3$ for the three-loop double-trace terms,
and implying $D=6$ for the four-loop double-trace terms.

An important question is whether the
bound (\ref{SuperYangMillsDoubleTracePowerCount})
is saturated for the double-trace terms, or whether further hidden 
cancellations remain.  To definitively answer this question we must
directly integrate our expressions, in order to extract the
coefficients of the potential double-trace UV divergences in
$D=20/3$ and $D=6$ at three and four loops, respectively.
Below we answer this question at three loops.


\subsection{Extracting UV divergences}
 
A systematic procedure for obtaining the potential divergences from
multi-loop integrals is based on differentiating with respect to
external momenta~\cite{Vladimirov,MarcusSagnotti}.  Here we
follow a related procedure, and expand the amplitudes for small external
momenta $k_i$. Formally this is achieved by introducing a single
infinitesimal parameter $\varepsilon$, giving formal amplitudes
$A(k_i)\rightarrow A(\varepsilon k_i)=\varepsilon^6 a_6+\varepsilon^7
a_7+\varepsilon^8 a_8+\ldots$, where the expansion starts at
$\varepsilon^6$ due to the manifest $m=6$ behavior of each integral.
At the integrand level this expansion corresponds to,
\begin{equation} 
I(k_i,\ell_j)\rightarrow I(\varepsilon k_i,\ell_j)=\varepsilon^2 v_6
     +\varepsilon^3 v_7+\varepsilon^4 v_8+\ldots \,.
\label{FormalExpansion}
\end{equation}
In \eqn{FormalExpansion} we have dropped the overall tree factor
${\cal K}$, and the $v_m$ correspond to sums of vacuum-like integrands,
depending only on the loop momenta $\ell_j$,
\begin{equation} 
v_m = \sum_p \rho_{mp}(k_i) \, \V_p(\ell_j)\, ,
\label{FactorizedForm}
\end{equation}
where the dependence on the external momenta factorizes into
polynomials $\rho_{mp}$ of
degree $(m-4)$.  The terms with odd powers of $\varepsilon$
in \eqn{FormalExpansion} may be dropped because they are zero
by Lorentz invariance.  To make sense of the
expansion under the integral sign, we formally take
$\varepsilon |k_i| \ll |\ell_j|$ (where $|\cdot|$ is the Euclidean norm).
This is not true in general, because the $\ell_j$ are integrated over
all values. Fortunately, this error affects only the
UV-finite part of the amplitude and is therefore not relevant to our
discussion. The UV-divergent contributions indeed come from the region
$|k_i| \ll |\ell_j|$ (if the amplitude contains no UV-divergent
subdiagrams).  We thus interpret the integrand expansion
(\ref{FormalExpansion}) as a formal series where the integrals
over $v_m$ encode the UV divergence of the integral in
dimension $D(m,L)$.  If
subdivergences appear, which happens for the potential color
double-trace divergence at four loops, they must be accounted for;
a procedure for doing so may be found in
refs.~\cite{Vladimirov,MarcusSagnotti}.
Here we only evaluate the three-loop case, which has no
subdivergences.

When evaluating terms in the expansion~(\ref{FormalExpansion}) one
encounters tensor vacuum integrands that are not quite in the
factorized form of \eqn{FactorizedForm}, but contain numerator
factors in which external and loop momenta are contracted,
of the form $k_i \cdot \ell_j$.  These integrands are easily converted 
to the factorized form~(\ref{FactorizedForm}) using the identity,
\begin{equation}
\ell_i^{\mu}\ell_j^{\nu}
\rightarrow \eta^{\mu \nu} \, \frac{\ell_i \cdot \ell_j}{D}\,.
\end{equation}
This identity is valid for tensor vacuum integrals with two free space-time
indices, which must be proportional to the metric tensor by Lorentz
invariance. (Similar identities for tensors with more free indices are
easily constructed, but we will not need them here.)

The various vacuum integrands $\V_j$ are distinguished by their
propagator structure and possibly numerator factors $\ell_i\cdot \ell_j$,
and can be represented as Feynman-like diagrams of various
topologies. There are many hidden relations between the
integrals, some of them generated by IBP identities~\cite{CTIBP}. 
These identities complicate the analysis slightly, because the 
representation of $v_m$ in terms of the $\V_j$, as
given in \eqn{FactorizedForm}, is not unique. To expose the identities
we can make use of the invariance of the integrals under
reparametrizations,
\begin{equation}
\int \prod_{j=1}^L d^{D}\ell_j \, I\Bigl(\varepsilon k_i,\ell_j \Bigr)
=  \int \prod_{j=1}^L d^{D}\ell_j \, 
I\Bigl(\varepsilon k_i,\ell_j +\sum_p c_{jp} \varepsilon k_p \Bigr) \, ,
\label{Reparam1}
\end{equation}
where $c_{jp}$ are arbitrary numbers. After such a reparametrization
the $\varepsilon$-expansion of the integrands look different,
but the total UV pole after integration must be the same.
We equate the different representations of $v_m$ integrals,
for a sufficient number of reparametrized expansions of the
form~(\ref{Reparam1}),
\begin{equation} 
\int v_m = \sum_p \rho_{mp}\, V_p = \sum_p \rho'_{mp} \, V_p = \ldots,
\label{ConsistencyEquation}
\end{equation} 
where $V_p$ stands for the integral of $\V_p$.  The resulting system
of linear equations for the poles of the $V_j$ can be solved easily.
In contrast to IBP identities, the relations we find are between the
pole parts of vacuum integrals that are relevant to our calculation,
and no others.


\subsection{The logarithmic three-loop divergence in $D=20/3-2\e$}
\label{3loopSubDivSection}

\begin{figure}[tb]
\centerline{\epsfxsize 1.2 truein \epsfbox{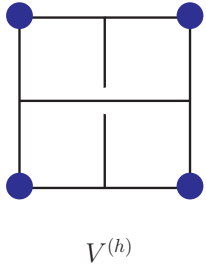}}
\caption[a]{\small The three-loop vacuum diagram appearing in the
UV divergence of the three-loop four-point \NeqFoursYM\ amplitude in
$D=20/3-2\e$.  Dots indicate the appearance of squared 
propagator factors in the integral. 
}
\label{Vacuum3loops2Figure}
\end{figure}

Consider now the amplitude in $D=20/3$, which is the lowest dimension
for which a potential divergence can appear in the three-loop color
double-trace terms, and corresponds to the $v_8$ terms in
\eqn{FormalExpansion}.
In the expansion of the nine three-loop integrals we obtain 11
different scalar vacuum integrals. We will not present them
explicitly here.  However, we note that nine of them have a propagator
topology very similar to the nine parent graphs of the three-loop
amplitude, but where the external momenta is set to zero, 
$k_i \rightarrow 0$, effectively replacing a leg insertion by a
two-point vertex, and thereby doubling a propagator.
(An example is provided by $V^{(h)}$ in
\fig{Vacuum3loops2Figure}.)  The two remaining scalar vacuum diagrams
both have a contact interaction. Remarkably, the consistency of the expansion
(\ref{ConsistencyEquation}) demands that the eleven vacuum integrals
are all proportional to each other in $D=20/3$.  The proportionality
constants are simple rational numbers. We can therefore
express the results in terms of the single vacuum-like diagram
displayed in \fig{Vacuum3loops2Figure}.  We obtain the following
contributions to the three-loop divergence in $D=20/3-2\e$,
\begin{eqnarray} 
{\cal A}_4^{\threl}(1,2,3,4) \Bigr|_{\Tra{1234}-\rm pole} ^{(D=20/3)}
&=& -g^{8} \,{\cal K} \,V^{(h)}  \Tra{1234} \,
\Bigl(\frac{2}{5}N_c^3\,(7s^2 + 7t^2 + 6u^2)
\cr
&&\qquad\qquad\qquad\quad\;\null
 - \frac{5}{2}N_c\,{}
     (23s^2 + 23t^2 - 126u^2)\Bigr)
\,,
\label{3loopSingleTrace}
\end{eqnarray}
\begin{eqnarray} 
{\cal A}_4^{\threl}(1,2,3,4) \Bigr|_{\Tra{12}\Tra{34}-\rm pole} ^{(D=20/3)}
&=& -g^{8} \,{\cal K} \,V^{(h)} \Tra{12}\Tra{34} 
\Bigl(7N_c^2\,(58s^2 + 3t^2 + 3u^2)
\cr 
&&\qquad\qquad\qquad\qquad\;\null
+ 400\,(s^2 + t^2 + u^2)\Bigr)
\,,
\label{3loopDoubleTrace}
\end{eqnarray}
where we have isolated the pieces of the divergence proportional to
two characteristic four-point color structures of $SU(N_c)$: the
single trace $\Tra{1234}$ and double trace $\Tra{12}\Tra{34}$,
respectively. The divergences of all other single- and double-trace
structures are simply related to \eqn{3loopSingleTrace} and
\eqn{3loopDoubleTrace} by crossing symmetry. This proves that no
further cancellations remain in the double-trace terms and at least
for $L=3$, the finiteness formula
(\ref{SuperYangMillsDoubleTracePowerCount}) is saturated.
The scalar vacuum integral appearing in \eqn{3loopSingleTrace} and
\eqn{3loopDoubleTrace} is shown in \fig{Vacuum3loops2Figure}, and the
UV divergence is given by,
\begin{equation} 
V^{(h)}= - {1\over(4\pi)^{10}}
 {\Gamma^3(-\textstyle{\frac{2}{3}}) \over 2310 \, \eps} 
\ +\ \Ord(1)\, .
\end{equation}

We can similarly determine the $D=20/3$ logarithmic divergence for a
general gauge group $G$. For clarity, we split up the divergence into
two contributions: one containing the tree color tensors and one
containing the irreducible loop color tensors (the latter was shown
to be finite in $D=6$). The result is,
\begin{eqnarray} 
{\cal A}_4^{\threl}(1,2,3,4) \Bigr|_{\rm tree\hbox{-}pole}^{(D=20/3)}
&=& 
- \frac{1}{320} g^{8} \,{\cal K} \,V^{(h)}\Bigl[ b_1^{(0)} 
\Bigl(200 C_{V^{\B}} (7 s^2 - 94 t^2 + 7 u^2) \nn \\
 && \null \hskip 4 cm 
- 16 C_A^3 (7 s^2 + 6 t^2 + 7 u^2)\Bigr) \nn\\
&& \null \hskip 2.3 cm 
- b_2^{(0)} \Bigl(200 C_{V^{\B}}(14 s^2 - 87 t^2 - 87 u^2) \nn \\
&& \null \hskip 4 cm 
+ 16 C_A^3 (6 s^2 + 7 t^2 + 7 u^2)\Bigr) \Bigr]
\,,
\label{3loopGtree}
\end{eqnarray}
\begin{eqnarray} 
{\cal A}_4^{\threl}(1,2,3,4) \Bigr|_{\rm loop\hbox{-}pole}^{(D=20/3)}
&=& 
- \frac{1}{24} g^{8} \,{\cal K} \,V^{(h)}
\Bigl[ 21 b_1^{(1)} C_A^2 (3 s^2 + 3 t^2 + 58 u^2) 
+770 b_2^{(2)} C_A (t^2 - u^2)  \nn\\
&& \null
 - 70 b_1^{(2)} C_A (9 s^2 + 20 t^2 + 31 u^2) 
+ 1200 b_1^{(3)} (s^2 + t^2 + u^2)\Bigr]
\,, \nn \\
\label{3loopGloop}
\end{eqnarray}
where the basis color tensors $b_i^{(L)}$ can be found in
\app{ColorAppendix}. For the three-loop four-point amplitude,
any color tensor can be written in terms of the six
basis tensors that feature in the above formul\ae. Note that none of
the coefficients of  the basis elements vanish. The full divergence is
simply the sum of the tree (\ref{3loopGtree}) and loop
(\ref{3loopGloop}) color-tensor contributions.
Note also that this division into tree and loop contributions is
somewhat arbitrary, and certainly not
unique (unless the loop contribution vanishes, as in the case of the
three-loop divergence in $D=6$); it depends on the choice of basis
for the color tensors.  (Specifically, choosing
different basis tensors $b_i^{(L)}$ at $L$ loops affects the 
coefficients of the $b_i^{(L)}$ tensors, as well as of
lower-loop basis tensors, but not the coefficients of higher-loop
basis tensors.)  The split in \eqns{3loopGtree}{3loopGloop} 
does not even respect the Bose symmetry of the divergence.

By inspecting the above expressions, it is clear that at
three loops the coefficients of all independent color tensors,
for $SU(N_c)$ as well as for a general group $G$, develop UV divergences in
$D=20/3$.  In other words, the double-trace terms for $SU(N_c)$, and
the loop color-tensor terms for general $G$, which are finite in $D=6$,
exhibit no further cancellations; their divergences start at $m=8$, or
equivalently at $D=20/3$.

Here we will not evaluate the four-loop divergence in $D=6$.  Its
evaluation is more involved than that of the three-loop case,
not only because of the greater complexity of the four-loop integrands,
but also because of a three-loop subdivergence.  A calculation of the
divergences in $D=6$ would, of course, answer the question of whether
there may be further hidden UV cancellations in the four-loop
double-trace terms.


\section{Conclusions}
\label{ConclusionSection}

In this paper we obtained the four-loop four-point amplitude of
$\NeqFour$ super-Yang-Mills theory, {\it i.e.} both the planar and the
non-planar contributions, in terms of a set of 50 loop integrals
(one of which has vanishing color coefficient and also integrates to
zero).  The planar part of the amplitude was already known~\cite{BCDKS}.
Here we presented the more complicated non-planar terms.
At the core of our computation lies the
unitarity method~\cite{UnitarityMethod}, and in particular the method
of maximal cuts~\cite{BCDKS,FiveLoop}. For the planar amplitude, hidden
symmetries~\cite{MagicIdentities} lead to major simplifications.  No
comparable considerations are presently available for the non-planar
terms.

An important problem is therefore to develop new tools for
constructing multi-loop integrands, which are valid also for 
non-planar amplitudes.
In this paper we extended and developed graphical rules applicable to
such amplitudes, to speed up their construction.
The simplest of these rules is a generalization of the rung
rule~\cite{BRY,BDDPR}, which allows us to write down all contributions
having two-particle cuts directly from lower-loop results.  However,
not all terms have two-particle cuts. A more powerful tool, which
partly bypasses this limitation, is the box cut, which yields terms
having four-point subgraphs. This rule has a simple graphical
formulation.  Another rule, based on the color-kinematic
duality~\cite{BCJ}, allows us to generate many non-planar
contributions from much simpler planar ones.  In fact, it appears
that these relations can be applied directly at loop level with no cut
conditions imposed, though this has been confirmed only through three
loops~\cite{BCJLoop}.  Although the graphical rules presented in 
this paper do not determine all
contributions, they determine most terms, allowing us to focus on the
remaining ones, for which we employed generalized unitarity,
in particular maximal and near-maximal cuts.  Nevertheless, 
identifying further tools and structures would be extremely helpful
for future studies.

The results in this paper, as well as those of
refs.~\cite{GravityThree,FiveLoop,CompactThree,GravityFour}, provide a
wealth of information on the explicit form of planar and non-planar
amplitudes in \NeqFoursYM.  However, the structures we are seeking, which
could allow the development of new tools, are obscured by ambiguities in
assigning contact terms to the parent graphs.  Each form of the amplitude
may expose one property or structure while hiding others. For example, the
representations of the complete three- and four-loop four-point amplitudes
presented in this paper do not highlight the fact that double-trace terms
are better behaved in the UV than single-trace terms. As we discussed, a
nontrivial rearrangement is necessary to manifestly expose this
feature. As another example, nontrivial rearrangements would be needed to
expose a recently proposed loop-level color-kinematics duality in the
four-loop four-point \NeqFoursYM{} amplitude. It is quite likely that
other important features will be revealed through appropriate
reorganizations of the amplitudes.

Generally it is simplest to construct an ansatz for the amplitude in
four dimensions, where powerful helicity and on-shell superspace
methods can be used.  Because we are interested in the
higher-dimensional UV structure of the theory, it is important that
our construction of the four-loop amplitude be valid in higher
dimensions.  In this paper, we evaluated a complete set of cuts in
four dimensions, but only an incomplete set in $D$ dimensions.  Strong
checks that our expressions are valid in $D$ dimensions come from
two-particle cuts, box cuts, and the color-kinematic
duality~\cite{BCJ}.  The planar contributions to the four-loop
four-point amplitude have been evaluated in $D$ dimensions, subject
only to the mild condition that no terms violate the expected power
count~\cite{BCDKS}.  It is nevertheless still important to compare our
results against a complete evaluation of the $D$-dimensional cuts. At
least in $D=6$, efficient tools for doing so now exist~\cite{OConnell,
SixDimSusy, FutureD6}.

A key reason for computing the four-loop \NeqFoursYM\
amplitude is to study its UV behavior as a function of dimension.
Our expressions are manifestly finite for $D<11/2$ in
accordance with the expectation~\cite{BDDPR,HoweStelleRevisited} that
$L$-loop \NeqFoursYM\ amplitudes are finite for
$D<4+6/L$.  Direct evaluation of the integrals reveals that the
theory is indeed UV divergent in $D=11/2$ for a general non-abelian
gauge group $G$; we have therefore demonstrated that no hidden UV
cancellations remain for generic $G$, through at least four loops.

In the course of our analysis of the color structure of UV
divergences, we have found, however, that certain terms are more
convergent than the general expectation: in particular, for
$G=SU(N_c)$ the terms with double-trace color factors have no
divergences in the critical dimension for which single-trace
divergences first appear, at three and four
loops~\cite{DurhamAndCopenhagenTalks}.  This curious improved behavior
has been discussed from the vantage point of the pure-spinor formalism
in string theory~\cite{DoubleTraceBerkovits} and field
theory~\cite{BG2010}, as well as from the standpoint of
field-theoretic algebraic non-renormalization
theorems~\cite{DoubleTraceNonrenormalization}.  In the present paper,
we give a field-theoretic clarification, motivated by the structure of
the leading contributions in the string theory analysis of
ref.~\cite{DoubleTraceBerkovits}:  It is possible to rearrange all
leading-divergence contributions of the amplitude into a
one-particle-reducible form, which has propagators depending only on
external momenta.  The color dressing of such terms forbids the
presence of double-trace color factors.  In contrast to the
string-based approach, this also clarifies the absence of
subleading-color double-trace terms.  This rearrangement can first be
performed at three loops because this is the first loop order where
inverse propagators can appear in numerators of integrals.  The
first potential divergence of double-trace terms is then at $D=4+8/L$,
in contrast to the $D=4+6/L$ divergences for single-trace terms.
We showed that the $D=4+8/L$ bound is saturated at three loops;
it remains an interesting open question whether it is saturated at
four loops.

An equally thorough understanding of the UV properties of $\NeqEight$
supergravity is absent.  However, the complete four-loop four-point
amplitude constructed in this paper helps shed light on this
issue. This is another important motivation for our work.  The
\NeqFoursYM\ amplitude is the basic input into the construction of the
corresponding $\NeqEight$ supergravity amplitude~\cite{GravityFour}, 
following the
strategy of refs.~\cite{BDDPR,GravityThree,CompactThree}:
as a consequence of generalized
unitarity~\cite{GeneralizedUnitarity,TwoLoopSplit} and of the
KLT~\cite{KLT} and graphical numerator double-copy
relations~\cite{BCJ}, cuts of $\NeqEight$ supergravity amplitudes can
be expressed directly in terms of the cuts of \NeqFoursYM\ amplitudes.
Although $\NeqEight$ supergravity is known to have surprisingly good
UV properties~\cite{BDDPR,Finite,
DualityArguments,HoweStelleRevisited,Berkovits,GravityThree,
CompactThree,BHS,GravityFour} and may even be UV
finite~\cite{Finite}, its UV behavior beyond four loops is still
unclear.  Recently, a consensus has formed~\cite{UnExpectedCancel,%
DoubleTraceBerkovits,DoubleTraceNonrenormalization,GRV2010,
FreedmanCounterTerms,KalloshRamond,Vanhove2010}
that supersymmetry alone cannot protect the theory beyond seven
loops in $D=4$ and that an additional mechanism must be at play if the
theory is finite.  Clearly, further high loop studies would be helpful for
resolving this issue.  Apart from its direct relevance for answering the
question of ultraviolet finiteness of $\NeqEight$ supergravity,
\NeqFoursYM\ offers a much simpler arena for sharpening our
understanding of such divergences in supersymmetric theories, as well
as for explicitly testing general arguments.

The results described in this paper may also shed light on the
infrared singularities of gauge-theory amplitudes, by allowing
tests of proposed structures for the soft anomalous-dimension
matrix~\cite{CompactThree,BN,GM} and the cusp anomalous dimension.  The
evaluation of the non-planar integrals appearing in the expression of
the amplitude in $D=4-2\e$ is a necessary step for the extraction of
its infrared divergences.  Unfortunately, such four-point integrals
are notoriously difficult to evaluate; they have not been computed
even at three loops. Thus, the use of our results for this purpose
must await the development of new techniques for evaluating
higher-loop non-planar integrals.

In summary, the results and tools presented here should provide new
handles on a number of important unanswered questions, including the
structure of multi-loop infrared divergences in gauge theories and the
UV properties of gravity theories.  They may also help expose hitherto
unexpected structures in these theories.


\section*{Acknowledgments}
\vskip -.3 cm

We thank Nima Arkani-Hamed, Nathan Berkovits, Kostja Chetyrkin,
Andrzej Czarnecki, Tristan Dennen, Michael Green, 
Paul Howe, Yu-tin Huang, Harald Ita, David Kosower, 
Pierre Vanhove, Volodya Smirnov and Kelly Stelle for many
stimulating discussions.  We are particularly grateful to Henriette Elvang,
Dan Freedman and Michael Kiermaier for providing expressions for 
various super-sums appearing in generalized cuts involving NMHV tree
amplitudes.  We thank Academic Technology Services at UCLA for computer
support.  This research was supported by the US Department of Energy
under contracts DE--AC02--76SF00515, DE--FG03--91ER40662 and
DE-FG02-90ER40577, and by the US National Science Foundation under
grants PHY-0455649 and PHY-0608114. R.~R. acknowledges support by the
A.~P. Sloan Foundation.  J.~J.~M.~C. gratefully acknowledges
the financial support of Guy Weyl Physics and Astronomy Alumni Grants.
H.J.'s research is supported by the European Research Council under 
Advanced Investigator Grant ERC-AdG-228301.
The figures were generated using Jaxodraw~\cite{Jaxo1and2}, based on
Axodraw~\cite{Axo}.

\appendix

\section{Sample evaluation of a non-planar cut}
\label{NonTrivialCutAppendix}

In this appendix we describe the details of the evaluation of a
nontrivial cut which appears in the construction of
\eqn{FourLoopYMAmplitude}.  The cut presented here can be used to
confirm large parts of our construction, and is sensitive to the most
complicated non-planar terms in the amplitude.  Seven-particle cuts,
which break the four-loop amplitude into four- and five-point MHV and
\MHVbar{} tree amplitudes are especially helpful in our construction,
because they are sensitive to all terms actually present in the
amplitudes, and yet are relatively simple to work with.  As mentioned in
the text, we checked that our expression for the amplitudes matches
all such cuts.

\begin{figure}[ht!]
\centering \centerline{\epsfxsize 6.4 truein
  \epsfbox{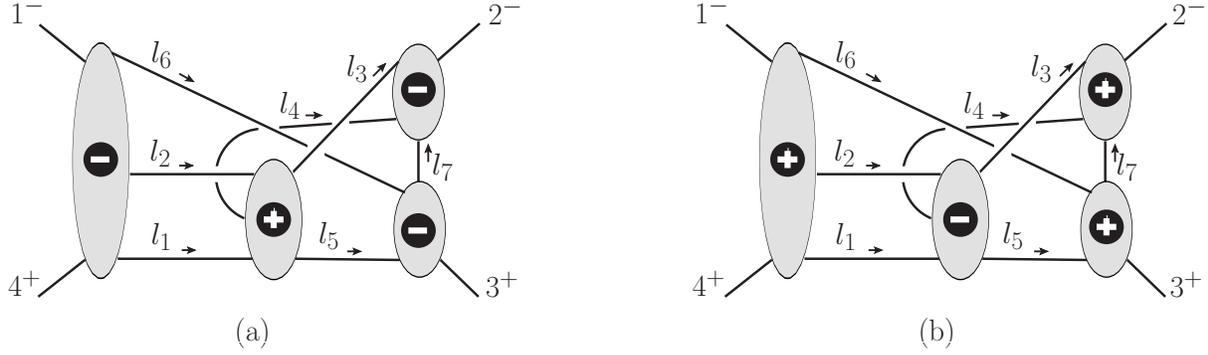}}
\caption{A nontrivial non-planar cut at four loops. 
The cuts (a) and (b) represent the two distinct internal helicity
configurations contributing to this seven-particle cut. The
blobs representing MHV trees are labeled by a ``$+$'' and the
blobs representing \MHVbar{} blobs are labeled by a ``$-$''. For a four-point 
amplitude either assignment can be made; the assignment in the figure 
makes the supersum calculation most transparent.}
\label{AppFourLoopNonPlanarExampleFigure}
\end{figure}

Here we present the steps necessary to test
eq.~(\ref{FourLoopYMAmplitude}) and the numerator factors assigned to
the integrals, using the seven-particle cut displayed in
\fig{AppFourLoopNonPlanarExampleFigure}.  This cut has already been
discussed in considerable detail in sect.~6C of ref.~\cite{SuperSum},
so here we will simply quote the result, and focus instead on the cut
of our result in \eqn{FourLoopYMAmplitude}.  As noted in
\fig{AppFourLoopNonPlanarExampleFigure}, the contributions to the cut
comes in two distinct sectors, depending on the configuration of MHV 
and \MHVbar{} assignment of the tree amplitudes composing the 
cut.  
The fact that all factors of tree amplitudes are either MHV or
\MHVbar{} leads to compact expressions for these two components. 
From ref.~\cite{SuperSum}, we have the simplified result:
\begin{eqnarray}
C^{\ref{AppFourLoopNonPlanarExampleFigure}\rm (a)} &=& \sum_{\rm states}
A_5^{\overline{\rm MHV}}(1^-,l_6,l_2,l_1,4^+) \,
A_5^{{\rm MHV}}(-l_1,l_4,-l_2,l_3,l_5) \, \nn \\
&&\null \times 
A_4(-l_4,-l_3,2^-,-l_7) \, A_4(-l_5,-l_6,l_7,3^+) \nn \\
&=&
\big(\spa1.2 \spb{l_7}.2 \spb{l_6}.3 \spb1.4 \big)^4
\frac{1}{[41][1l_6][l_6l_2][l_2l_1][l_1 4]} \nn\\
&&\null \times 
\frac{1}{\spa{l_1}.{l_4} \spa{l_4}.{l_2}\spa{l_2}.{l_3}
\spa{l_3}.{l_5}\spa{l_5}.{l_1}}
\frac{1}{[3l_5][l_5l_6][l_6l_7][l_73]}
\frac{1}{[2l_7][l_7l_4][l_4l_3][l_32]} \,, \nn \\
C^{\ref{AppFourLoopNonPlanarExampleFigure}\rm (b)} &=& \sum_{\rm states}
A_5^{{\rm MHV}}((1^-,l_6,l_2,l_1,4^+) \,
A_5^{\overline{\rm MHV}}(-l_1,l_4,-l_2,l_3,l_5)
 \, \nn \\
&&\null \times 
A_4(-l_4,-l_3,2^-,-l_7) \, A_4(-l_5,-l_6,l_7,3^+)\cr
&=&
\big( \spa1.2 \spb2.3 \spb{l_3}.{l_4}\spb{l_5}.{l_7} \big)^4
\frac{1}{\spa4.1 \spa1.{l_6}\spa{l_6}.{l_2}\spa{l_2}.{l_1}\spa{l_1}.4}
\cr
&&\null \times 
\frac{1}{[l_1l_4][l_4l_2][l_2l_3][l_3l_5][l_5 l_1]}
\frac{1}{[3l_5][l_5l_6][l_6l_7][l_73]}
\frac{1}{[2l_7][l_7l_4][l_4l_3][l_32]} \,.
\label{NonPlanarCutExample}
\end{eqnarray}
This simple result is determined by the purely gluonic configurations
crossing the cuts; in the first case there are seven such
configurations and in the second eight.

The above two expressions are related by complex conjugation: 
\begin{equation}
\frac{C^{\ref{AppFourLoopNonPlanarExampleFigure}\rm (b)}}{A^\tree_4}
= \Biggl(
\frac{C^{\ref{AppFourLoopNonPlanarExampleFigure}\rm (a)}}{A^\tree_4}
\Biggr)^\dagger \,.
\label{DaggerRelation4}
\end{equation}
This relation guarantees from the outset that after dividing by 
the tree amplitude, the sum of the two expressions is a function
of only Lorentz dot products, with all spinors or Levi-Civita tensors
dropping out.  (This property is special to four-point amplitudes
and does not hold for higher-point amplitudes.)

The cut in \eqn{NonPlanarCutExample} needs to be compared to the cut of
our expression for the four-loop amplitude in \eqn{FourLoopYMAmplitude}.
To identify the possible contributions to this cut we write out all
possible tree graphs with only three-point vertices, for the tree
amplitudes composing the cut.  These are depicted in
\fig{AppFourLoopNonPlanarExampleFigure}.  (We need only track graphs with
three-vertices, following our organization of contributions according to
such graphs.)  The graphs labeled by (a) and (b) correspond to the two
five-point amplitudes in the cut, and those labeled by (c) and (d) to the
four-point amplitudes.  Two graphs were dropped from (a), in which
external legs 1 and 4 fuse together, because they would generate
three-point subgraphs, which we know do not occur.

\begin{figure}[ht!]
\centering
\centerline{\epsfxsize 6.6 truein \epsfbox{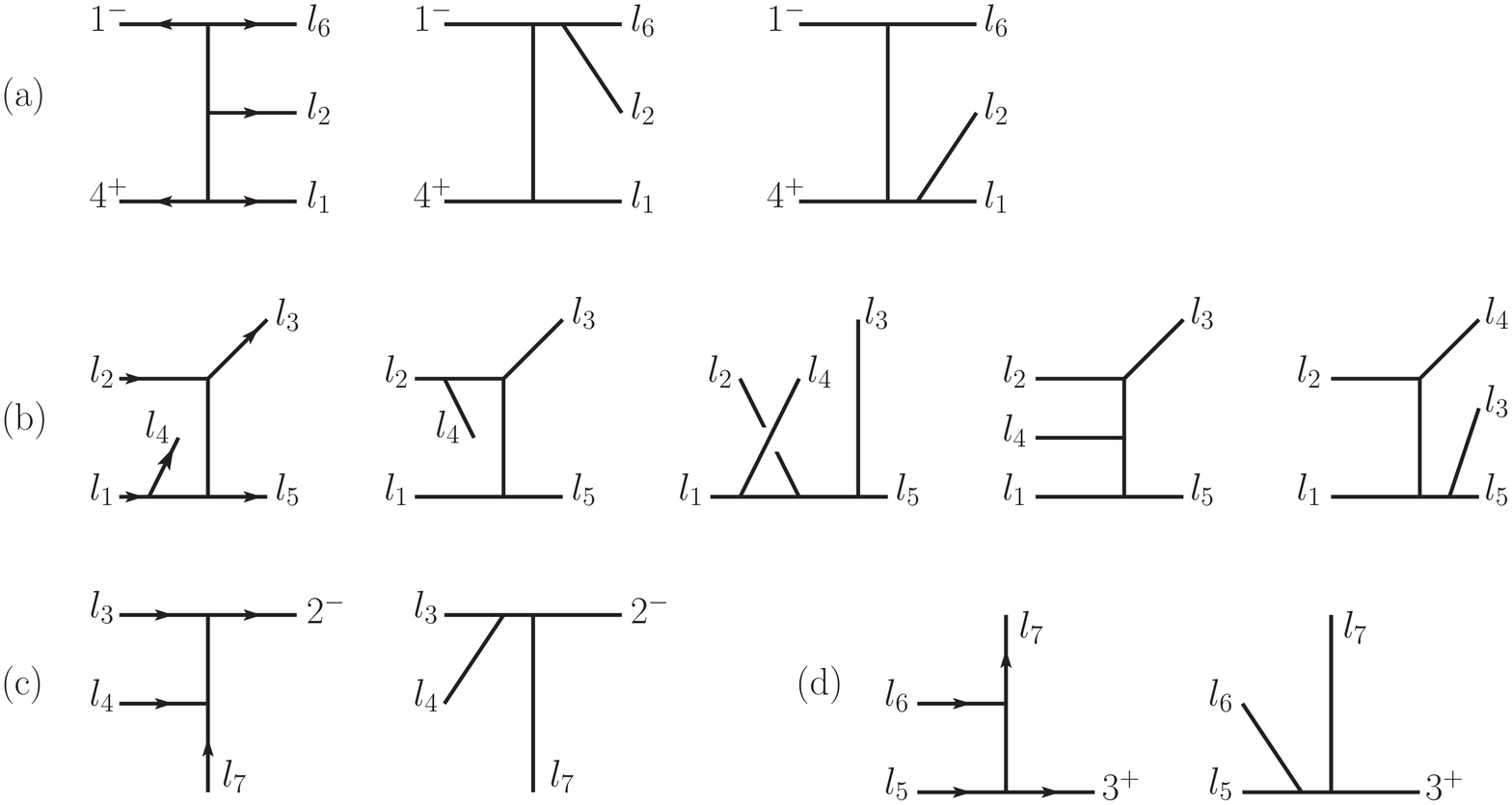}}
\caption{Tree structures building up the cut in
\fig{AppFourLoopNonPlanarExampleFigure}.}
\label{treesFigure}
\end{figure}

After sewing together the four trees, we can identify the integrals and
labelings of internal and external lines that contribute to this cut.
Of the 60 possible such products,
22 may be ignored as they do not appear in the list of
integrals composing the answer.  The unnecessary ones would have triangle
or bubble subgraphs, and would violate the known no-triangle
property of \NeqFoursYM\ at one, two or three loops.
Organized in this way, the contributions to the cut in
\fig{AppFourLoopNonPlanarExampleFigure} are listed in \tab{MonsterTable},
which is structured as follows. The first column represents the
product of trees --- $(ijkl)$ standing for the product of the $i$th
tree from \fig{treesFigure}(a), the $j$th tree from figure
\ref{treesFigure}(b), the $k$th tree from \fig{treesFigure}(c)
and the $l$th tree from \fig{treesFigure}(d). The second column
is the number $n$ of the integral ${\cal I}_n$ for which $(ijkl)$
represents a cut contribution.  A given integral may appear multiple
times in the same cut, Several such products may represent cuts of the
same integral. The third column contains the mapping of (internal and
external) momenta between figs.~\ref{BC1Figure}--\ref{E2Figure} and the
momenta of the tree amplitudes assigned in
figs.~\ref{AppFourLoopNonPlanarExampleFigure} and \ref{treesFigure}.
The last column represents the numerator factors from
figs.~\ref{BC1Figure}--\ref{E2Figure}, evaluated for the cut conditions
specified in \fig{AppFourLoopNonPlanarExampleFigure}.  We
specify the mapping only for the external momenta or when the
numerator contains a given internal momentum. For convenience
we have introduced the notation,
\begin{eqnarray}
k_{i,j} = k_i + k_j\,, \hskip 2cm l_{i,j} = l_i + l_j \,, \hskip 2 cm 
 l_{i,\bar j} = l_i - l_j \,, \cr
\hskip 2cm  l_{i,j,m} = l_i + l_j +l_m \,, \hskip 2 cm 
 l_{i,\bar j, \bar m} = l_i - l_j - l_m \,.
\end{eqnarray}
In general, subscripts separated by commas denote sums of momenta, while a bar
over the subscripts denotes that the sign of the corresponding momentum
is minus.

In \tab{MonsterTable} there are a total of 38 contributions
to the cut.  The first such contribution is 
\begin{equation}
C^{(1321)} = {\cal K} \,
{-(k_2+l_5 - l_7)^2 \Bral s_{13} (k_2 + l_5 - l_7)^2 
          -s_{23} (k_1 + k_3 + l_7)^2 \Brar \over 
 (k_1 + l_6)^2 (l_1 + k_4)^2 (l_4 - l_1)^2  (l_3+ l_5)^2 (l_3+l_4)^2 
(l_5 - k_3)^2} \,,
\end{equation}
where the numerator is given in \tab{MonsterTable} and the
denominators can be read off from the propagators of
the diagrams in \fig{treesFigure}.  We have put back the 
overall prefactor $\cal K$ defined in
\eqn{Prefactor}.  The reader may check that the sum over all 38
contributions adds up to give the sum of the two contributions in
\eqn{NonPlanarCutExample}.  It is convenient to make this comparison
numerically, but it is also straightforward to carry out analytically
after imposing momentum conservation.

\vfill

\def\hsa {$\null$ \hskip -.4 cm}
\def\hsb {\hskip  2.3 cm}
\def\hsc {\hskip  .4 cm}

\newpage

\begin{table}[t]
\caption{The numerator factors and tree structures appearing in the 
seven particle cut of \fig{AppFourLoopNonPlanarExampleFigure}.
The first column gives the tree structure according to 
\fig{AppFourLoopNonPlanarExampleFigure} and the notation 
described in the text.  The second column gives the integral number
$n$ whose cut gives the tree structure.  The third column gives
the relabelings needed to go from the integral labels to 
the cut labels. The last column gives the numerator factors 
in the labeling of the cut.
The associated propagator factors can be read off from the 
tree structures.
\label{MonsterTable}
}

\vskip .6 cm

\def\bar{\overline}
\begin{tabular}{lc|p{1.70in}|p{3.5in}}

Trees & ${\cal I}_n$& Momentum relabeling & Numerator factor \cr
\hline
$(1321)$ &
34\,
&
\begin{tabular}{ll}
\vphantom{${}^{\big|}$}
\hlp
$k_1\rightarrow k_3,$    & $k_2\rightarrow k_1,$\cr
$k_3\rightarrow k_4,$    & $k_4\rightarrow k_2,$\cr
$l_5\rightarrow k_2 - l_7,$& $l_6\rightarrow l_5$
\hrp
\end{tabular}
&

$-(k_2+l_{5,\bar 7})^2 \Bral s_{13} (k_2 + l_{5,\bar 7})^2 
          -s_{23} (k_{1,3}+ l_7)^2 \Brar $

\cr
\hline

$(1322)$&
18\,& 
\begin{tabular}{ll}
\vphantom{${}^{\big|}$}
\hlp
$k_1\rightarrow k_2,$ & $k_2\rightarrow k_3,$ \cr
$k_3\rightarrow k_1,$ & $k_4\rightarrow k_4,$ \cr
$l_5 \rightarrow k_3 + l_7,$ \cr
$l_6 \rightarrow k_1 + l_6,$\cr
$l_7 \rightarrow l_4$
\hrp
\end{tabular}

& 
~$s_{23}(k_{1,3}+l_7)^2(k_{1,4}+l_6)^2$

\cr
\hline

$(3412)$&
18\, &  
\begin{tabular}{ll}
\vphantom{${}^{\big|}$}
\hlp
$k_1\rightarrow k_4,$ & $k_2\rightarrow k_1,$ \cr
$k_3\rightarrow k_3,$ & $k_4\rightarrow k_2,$ \cr
$l_5\rightarrow -l_{6},$ & $l_6\rightarrow -l_7,$\cr
$l_7 \rightarrow -l_2$
\hrp
\end{tabular}
& 
$~s_{14}(k_3-l_6)^2(k_2-l_7)^2$

\cr
\hline

$(1522)$ & 
20\, &
\begin{tabular}{ll}
\vphantom{${}^{\big|}$}
\hlp
$k_1\rightarrow k_2,$&  $k_2\rightarrow k_3,$\cr
$k_3\rightarrow k_4,$&  $k_4\rightarrow k_1,$\cr
$l_5\rightarrow l_{3,5},$&  $l_6\rightarrow l_4,$ \cr
$l_7\rightarrow k_3+l_7$& 
\hrp
\end{tabular}
& 
$~\,s_{23}\, l_{3,4,5}^2(k_{1,3}+l_7)^2$
\cr
\hline
$(3212)$ & 
20\, &
\begin{tabular}{ll}
\vphantom{${}^{\big|}$}
\hlp
$k_1\rightarrow k_4,$&  $k_2\rightarrow k_1,$\cr
$k_3\rightarrow k_2,$&  $k_4\rightarrow k_3,$\cr
$l_5\rightarrow -l_{1,{\bar 5}},$&  $l_6\rightarrow -l_2,$ \cr
$l_7\rightarrow -l_6$
\hrp
\end{tabular}
& 
$~\,s_{14} l_{\bar 1,\bar 2,5}^2 (k_3-l_6)^2$
\cr
\hline

\end{tabular}
\end{table}


\newpage

\begin{tabular}{lc|p{1.75in}|p{3.5in}}

Trees & ${\cal I}_n$& Momentum relabeling & Numerator factor \cr
\hline

$(2212)$ &
25\, & 
\begin{tabular}{ll}
\vphantom{${}^{\big|}$}
\hlp
$k_1\rightarrow k_4,$ &  $k_2\rightarrow k_1,$\cr
$k_3\rightarrow k_2,$     & $k_4\rightarrow k_3,$\cr
$l_6\rightarrow -l_2,$ & $l_7\rightarrow -l_1,$  \cr
$l_8\rightarrow -l_{1,{\bar5}},$\cr
$l_9\rightarrow -l_7$
\hrp
\end{tabular}
& 
$~\,s_{14}[ (k_2-l_1)^2 l_{\bar 1,\bar 2,5}^2-s_{14}(l_{\bar 1,5}+k_2)^2 ]$
\cr
\hline

$(2412)$ & 
25\, & 
\begin{tabular}{ll}
\vphantom{${}^{\big|}$}
\hlp
$k_1\rightarrow k_1,$ & $k_2\rightarrow k_4,$\cr
$k_3\rightarrow k_2,$ & $k_4\rightarrow k_3,$\cr
$l_6\rightarrow -l_{1,{\bar5}},$ & $l_7\rightarrow -l_{2,6},$\cr
$l_8\rightarrow -l_2,$ & $l_9\rightarrow -l_7$
\hrp
\end{tabular}
& 
%
~$\, -s_{14} \Bral {} ( k_2-l_{2,6})^2 (l_5-l_{1,2})^2-s_{14}(k_2-l_2)^2 \Brar$

~$\null + s_{14} l_{5, \bar 1}^2 (k_2 - l_{2,6})^2$

\cr
\hline

$(1511)$ & 
33\,  & 
\begin{tabular}{ll}
\vphantom{${}^{\big|}$}
\hlp
$k_1\rightarrow k_4,$ & $k_2\rightarrow k_1,$\cr
$k_3\rightarrow k_3,$ & $k_4\rightarrow k_2,$\cr
$l_5\rightarrow l_{3,5},$ & $l_6\rightarrow l_4,$ \cr
$l_7\rightarrow l_5,$ & $l_8\rightarrow -l_7,$ \cr
$l_9\rightarrow k_3-l_5,$ \cr
$l_{10} \rightarrow l_{4,7},$ \cr
$l_{11} \rightarrow k_1 + l_6,$ \cr 
$l_{12} \rightarrow -l_6$
\hrp
\end{tabular}

& 
~$ s_{14} \Bral l_{3,5}^2 (k_{1,3} - l_5)^2 
      + l_{5,\bar 7}^2 (k_4 + l_{3,5})^2 \Brar $

~$\null -l_{3,4,5}^2 \Bral s_{23}  (k_{1,3}-l_5)^2
       -s_{13} (k_3+l_{4,\bar 5,7})^2
       +s_{13} l_{4,7}^2 \Brar
$

\cr
\hline
$(1312)$ & 
33\, & 
\begin{tabular}{ll}
\vphantom{${}^{\big|}$}
\hlp
$k_1\rightarrow k_4,$ & $k_2\rightarrow k_1,$\cr
$k_3\rightarrow k_3,$ & $k_4\rightarrow k_2,$\cr
$l_5\rightarrow l_4,$ & $l_6\rightarrow l_{3,5},$ \cr
$l_7\rightarrow -l_7,$ & $l_8\rightarrow l_5,$\cr
$l_9\rightarrow k_3+l_7,$ & $l_{10}\rightarrow l_3,$ \cr
$l_{11} \rightarrow k_1 + l_6,$\cr
$ l_{12} \rightarrow -l_6$
\hrp
\end{tabular}
& 

~$\, l_{3,4,5}^2 \Bral s_{14} (k_{1,3}+l_7)^2 - s_{13} (k_3+l_{3,7})^2 \Brar$

$\null -s_{14}(k_4+l_4)^2 l_{5,\bar 7}^2$

\cr
\hline

\end{tabular}

\vfill
\newpage

\begin{tabular}{lc|p{1.75in}|p{3.5in}}
Trees & ${\cal I}_n$& Momentum relabeling & Numerator factor \cr
\hline

$(1412)$ & 
33\,  & 
\begin{tabular}{ll}
\vphantom{${}^{\big|}$}
\hlp
$k_1\rightarrow k_2,$ & $k_2\rightarrow k_3,$\cr
$k_3\rightarrow k_1,$ & $k_4\rightarrow k_4,$\cr
$l_5\rightarrow -l_2,$ & $l_6\rightarrow -l_{1,{\bar 5}},$ \cr
$l_7\rightarrow k_1+l_6,$ & $l_8\rightarrow l_5,$\cr
$l_9\rightarrow -l_6,$ & $l_{10}\rightarrow -l_1,$ \cr
$l_{11}\rightarrow -l_7,$  \cr
$l_{12}\rightarrow k_3 + l_7$
\hrp
\end{tabular}
& 

\vspace{-8truemm}
~$-(l_5-l_1-l_2)^2 \Bral s_{13} (l_6+l_1)^2-s_{23}(k_3-l_6)^2\Brar$

~$\null -s_{23}(k_2-l_2)^2(k_1+l_5+l_6)^2$

\cr
\hline

$(1311)$ & 
36\,  &
\begin{tabular}{ll}
\vphantom{${}^{\big|}$}
\hlp
$k_1\rightarrow k_2,$ & $k_2\rightarrow k_3,$\cr
$k_3\rightarrow k_1,$ & $k_4\rightarrow k_4,$\cr
$l_5\rightarrow l_4,$ & $l_6\rightarrow l_5,$ \cr
$l_7\rightarrow k_1+l_6,$ \cr 
$l_8\rightarrow k_3-l_5,$ & 
$l_9\rightarrow l_{4,7},$ \cr 
$l_{10}\rightarrow -l_7,$& $l_{11}\rightarrow l_{3,5},$\cr
$l_{12} \rightarrow -l_{6}$
\hrp
\end{tabular}

& 

\vspace{-15truemm}
~$s_{14} (k_1+l_6)^2  (k_2-l_4)^2 
  - l_{5,\bar 7}^2 s_{14} (k_4+l_4)^2$

~$\null + l_{3,5}^2 s_{14}  \Bral s_{13} - (k_1+l_{5,6})^2 \Brar $

~$\,\null +(k_{1,4} + l_6)^2 \Bral s_{14} (k_1+l_{5,6})^2 $

~$ \null\hsb  +s_{13} ((k_3+l_{4,\bar 5,7})^2 - s_{14}) \Brar $

\cr

\hline

$(1212)$ &  
36\, &
\begin{tabular}{ll}
\vphantom{${}^{\big|}$}
\hlp
$k_1\rightarrow k_4,$ & \hsa $k_2\rightarrow k_1,$\cr
$k_3\rightarrow k_3,$ & \hsa $k_4\rightarrow k_2,$\cr
$l_5\rightarrow -l_{1,{\bar 5}},$ & \hsa $l_6\rightarrow k_1+l_6,$ \cr
$l_7\rightarrow -l_7,$ & \hsa $l_8\rightarrow -l_6,$\cr
$l_9\rightarrow -l_1,$ & \hsa $l_{10}\rightarrow l_5,$\cr
$l_{11} \rightarrow -l_2,$ \cr
$l_{12} \rightarrow k_3 + l_7$ \cr
\hrp
\end{tabular}
& 

\vspace{-11truemm}

~$-(k_2-l_7)^2 \Bral s_{14} (k_1+l_{6,\bar 7})^2 
 + s_{13} l_{1,6}^2 - s_{14} s_{13} \Brar$

~$\null +s_{14}(k_2+l_{\bar 1,5})^2 (k_1+l_{5,6})^2$

~$\null -l_{\bar 1,5}^2 s_{14} \Bral {} 
     (k_1+l_{5,6})^2 - \frac{1}{2} s_{13}\Brar$

\cr

\hline

$(1512)$ &  
36\,  &
\begin{tabular}{ll}
\vphantom{${}^{\big|}$}
\hlp
$k_1\rightarrow k_2,$ & $k_2\rightarrow k_3,$\cr
$k_3\rightarrow k_1,$ & $k_4\rightarrow k_4,$\cr
$l_5\rightarrow l_{3,5},$ & $l_6\rightarrow -l_7,$  \cr
$l_7\rightarrow k_1+l_6,$ \cr
$l_8\rightarrow k_3+l_7,$&\; $l_9\rightarrow l_3,$\cr
$l_{10}\rightarrow l_5,$&\, $l_{11}\rightarrow l_4,$\cr
$l_{12} \rightarrow -l_6$
\hrp

\end{tabular}
& 

\vspace{-11truemm}
~$ \frac{1}{2} l_{3,5}^2 s_{14} \Bral s_{13}+ 2 (k_1+l_6)^2
                  -2 l_{5,\bar 7}^2 \Brar $

~$\null + s_{23} \Bral l_{5,\bar 7}^2 (k_4+l_{3,5})^2
       -(k_1+l_6)^2 (k_2- l_{3,5})^2 \Brar$

~$\null +(k_{1,4}+l_6)^2 \Bral s_{23} (s_{13} - (k_1+l_{6,\bar 7})^2)$

~$\null\hsb -s_{13} (k_3 + l_{3,7})^2 \Brar$
\cr
\hline

\end{tabular}

\vfill


\newpage

\begin{tabular}{lc|p{1.75in}|p{3.5in}}
Trees & ${\cal I}_n$& Momentum relabeling & Numerator factor \cr
\hline

$(1521)$ &  
32\,  &
\begin{tabular}{ll}
\vphantom{${}^{\big|}$}
\hlp
$k_1\rightarrow k_2,$ &\, $k_2\rightarrow k_1,$\cr
$k_3\rightarrow k_4,$ &\, $k_4\rightarrow k_3,$\cr
$l_5\rightarrow l_5,$ &\, $l_6\rightarrow k_2 - l_7$\cr
\hrp
\end{tabular}
& 

~$(k_2 + l_{5,\bar 7})^2 \Bral s_{12} (k_2 + l_{5,\bar 7} )^2
                                - s_{12} (k_2 - l_7)^2$

$\null \hsb     - s_{14}(k_4+l_5)^2  \Brar $

\cr
\hline
$(2411)$ &  
21\,  &
\begin{tabular}{ll}
\vphantom{${}^{\big|}$}
\hlp
$k_1\rightarrow k_4,$ & $k_2\rightarrow k_1,$\cr
$k_3\rightarrow k_2,$ & $k_4\rightarrow k_3,$\cr
$l_5\rightarrow -l_2,$ & $l_6\rightarrow -l_7,$ \cr
$l_7\rightarrow -l_1$
\hrp
\end{tabular}
& 

~$s_{14} l_{2,7}^2(k_3-l_1)^2$

\cr
\hline
$(3411)$ &  
22\,  
&
\begin{tabular}{ll}
\vphantom{${}^{\big|}$}
\hlp
$k_1\rightarrow k_1,$ & $k_2\rightarrow k_4,$\cr
$k_3\rightarrow k_2,$ & $k_4\rightarrow k_3,$\cr
$l_5\rightarrow -l_{1{\bar5}},$ & $l_6\rightarrow -l_2,$ \cr
$l_7\rightarrow -l_6$
\hrp
\end{tabular}
& 

$~~\,s_{14} l_{5, \bar 1, \bar 2}^2 (k_3-l_6)^2$ 

\cr
\hline

$(2211)$ &  
23\,  &
\begin{tabular}{ll}
\vphantom{${}^{\big|}$}
\hlp
$k_1\rightarrow k_4,$ & $k_2\rightarrow k_1,$\cr
$k_3\rightarrow k_2,$ & $k_4\rightarrow k_3,$\cr
$l_5\rightarrow -l_7,$ & $l_6\rightarrow -l_2,$ \cr
$l_7\rightarrow -l_1$
\hrp
\end{tabular}
& 

~$s_{14}l_{2,7}^2(k_3-l_1)^2$
\cr
\hline

$(3211)$ &  
23\,  &
\begin{tabular}{ll}
\vphantom{${}^{\big|}$}
\hlp
$k_1\rightarrow k_1,$ & $k_2\rightarrow k_4,$\cr
$k_3\rightarrow k_2,$ & $k_4\rightarrow k_3,$\cr
$l_5\rightarrow -l_{1,{\bar5}},$ & $l_6\rightarrow -l_2,$ \cr
$l_7\rightarrow -l_6$
\hrp
\end{tabular}
& 

$~~\,s_{14} l_{5, \bar 1, \bar 2}^2 (k_3-l_6)^2$ 

\cr
\hline

$(1211)$ &  
37\,  &
\begin{tabular}{ll}
\vphantom{${}^{\big|}$}
\hlp
$k_1\rightarrow k_3,$ & $k_2\rightarrow k_4,$\cr
$k_3\rightarrow k_1,$ & $k_4\rightarrow k_2,$\cr
$l_5\rightarrow -l_7,$ & $l_6\rightarrow -l_{1,{\bar 5}},$ \cr
$l_7\rightarrow k_4+l_1,$  \cr
$l_8\rightarrow k_1+l_6,$ \cr
$l_9 \rightarrow -l_2$
\hrp
\end{tabular}
& 

\vspace{-11truemm}

~$\,(k_2+l_{\bar 1,5})^2[s_{34}(k_{1,4}+l_6)^2 - s_{14}(k_4+l_5)^2]$

~$\null +(k_2-l_7)^2[s_{13}(k_{1,4}+l_1)^2-s_{14}(k_1+l_{6,\bar 7})^2]$

~$\null +s_{14}(k_2+l_{\bar 1,5})^2 (k_2-l_7)^2$

\cr

\hline

\end{tabular}

\vfill


\newpage

\begin{tabular}{lc|p{1.75in}|p{3.5in}}

Trees & ${\cal I}_n$& Momentum relabeling & Numerator factor \cr
\hline

$(1411)$ &  
37\,  &
\begin{tabular}{ll}
\vphantom{${}^{\big|}$}
\hlp
$k_1\rightarrow k_1,$ & $k_2\rightarrow k_3,$\cr
$k_3\rightarrow k_4,$ & $k_4\rightarrow k_2,$\cr
$l_5\rightarrow -l_2,$ & $l_6\rightarrow -l_7,$\cr
$l_7\rightarrow l_5,$ & $l_8\rightarrow -l_1,$\cr
$l_9 \rightarrow l_{\bar 1,5}$
\hrp
\end{tabular}
& 

\vspace{-8truemm}

~$(k_2-l_7)^2[s_{13}(k_3-l_1)^2 - s_{34} l_{5,\bar 7}^2]$

~$\null +(k_2-l_2)^2[s_{14}(k_4+l_5)^2-s_{34} l_{1,2}^2]$

~$\null +s_{34}(k_2-l_7)^2 (k_2-l_2)^2 +  s_{13} s_{14} l_{\bar 1,5}^2 $

\cr
\hline
$(2522)$ &  
13\,  &
\begin{tabular}{ll}
\vphantom{${}^{\big|}$}
\hlp
$k_1\rightarrow k_1,$ & $k_2\rightarrow k_4,$\cr
$k_3\rightarrow k_2,$ & $k_4\rightarrow k_3,$\cr
$l_5\rightarrow -l_{2,6},$ \cr
$\,l_6\rightarrow k_3+l_7$
\hrp
\end{tabular}

& 

~$s_{14}^2(k_3 + l_{7, \bar 2, \bar 6})^2$

\cr
\hline

$(2311)$ &  
27\,  &
\begin{tabular}{ll}
\vphantom{${}^{\big|}$}
\hlp
$k_1\rightarrow k_1,$ & $k_2\rightarrow k_4,$\cr
$k_3\rightarrow k_2,$ & $k_4\rightarrow k_3,$\cr
$l_5\rightarrow -l_{2,6},$\cr
$l_7\rightarrow l_5,$ & $l_8\rightarrow -l_2,$\cr
$l_9\rightarrow -l_6,$ & $l_{10}\rightarrow -l_4,$\cr
$l_{11}\rightarrow l_7,$ & $l_{12}\rightarrow -l_{3,5}$
\hrp
\end{tabular}
& 


~$  s_{14}[s_{14} l_{\bar 2, 5}^2 - s_{14}(k_3-l_{2,6})^2  ]
 +s_{14}(k_3-l_{2,6})^2 [l_{4,6}^2 - l_{3,5}^2 ]$

\cr
\hline

$(2512)$ &  
27\, &
\begin{tabular}{ll}
\vphantom{${}^{\big|}$}
\hlp
$k_1\rightarrow k_1,$ & $k_2\rightarrow k_4,$\cr
$k_3\rightarrow k_2,$ & $k_4\rightarrow k_3,$\cr
$l_5\rightarrow -l_{2,6},$ & $\,l_7\rightarrow -l_7,$ \cr
$l_8\rightarrow -l_{2},$& $l_9\rightarrow -l_6,$ \cr 
$l_{10}\rightarrow -l_{3,5},$& $l_{11}\rightarrow -l_5,$ \cr 
$l_{12}\rightarrow -l_4$
\hrp
\end{tabular}
& 


~$ s_{23}^2 (k_3-l_{2,6})^2
   -s_{23}\Bral s_{23} l_{2,7}^2 + (k_3-l_{2,6})^2
    l_{3,5,6}^2 \Brar $ 

\cr
\hline

$(2511)$ &  
26\,  &
\begin{tabular}{ll}
\vphantom{${}^{\big|}$}
\hlp
$k_1\rightarrow k_4,$ & $k_2\rightarrow k_1,$\cr
$k_3\rightarrow k_3,$ & $k_4\rightarrow k_2,$\cr
$l_5\rightarrow -l_{2,6},$ & $l_6\rightarrow -l_6,$\cr
$l_7\rightarrow -l_4,$ & $l_8\rightarrow l_{2,\bar 4},$\cr
$l_9\rightarrow l_{4,7},$ & $l_{10}\rightarrow -l_{3,5},$\cr
$l_{11} \rightarrow l_5$
\hrp
\end{tabular}
& 

\vspace{-8truemm}

~$ s_{14} l_{4,7}^2 \Bral {}(k_3+l_2)^2+(k_3+l_6)^2 -l_2^2-l_6^2 \Brar $

~$\null -s_{14}^2 l_{2,7}^2 
  -s_{14}l_{2,6}^2 (k_2-l_{3,5})^2 +  s_{14} l_{4,6}^2 (k_3-l_{2,6})^2 $


\cr
\hline

$\vphantom{\Bigg{|}}$\cr

\end{tabular}

\vfill


\newpage

\begin{tabular}{lc|p{1.75in}|p{3.5in}}

Trees & ${\cal I}_n$& Momentum relabeling & Numerator factor \cr
\hline

$(2321)$ &  
26\,  &
\begin{tabular}{ll}
\vphantom{${}^{\big|}$}
\hlp
$k_1\rightarrow k_1,$ & $k_2\rightarrow k_4,$\cr
$k_3\rightarrow k_2,$ & $k_4\rightarrow k_3,$\cr
$l_5\rightarrow -l_1,$ & $l_6\rightarrow -l_4,$\cr
$l_7\rightarrow -l_{3,5},$ & $l_8\rightarrow -l_2,$\cr
$l_9\rightarrow l_5,$ & $l_{10}\rightarrow -l_6,$ \cr
$l_{11}\rightarrow l_7$ 
\hrp
\end{tabular}
& 

~$s_{14}^2 l_{5,\bar 2}^2-s_{14} (k_2-l_1)^2\, l_{3,4,5}^2$ 

\cr
\hline
$(2521)$ &  
26\, &
\begin{tabular}{ll}
\vphantom{${}^{\big|}$}
\hlp
$k_1\rightarrow k_1,$ & $k_2\rightarrow k_4,$\cr
$k_3\rightarrow k_3,$ & $k_4\rightarrow k_2,$\cr
$l_5\rightarrow -l_1,$ & $l_6\rightarrow -l_{3,5},$\cr
$l_7\rightarrow -l_4,$ & $l_8\rightarrow -l_2,$\cr
$l_9\rightarrow k_2-l_7,$ & $l_{10}\rightarrow -l_6,$\cr
$l_{11}\rightarrow k_3 - l_5$
\hrp
\end{tabular}

& 

\vspace{-6truemm}

~$s_{14} (k_3-l_1)^2 l_{3,4,5}^2 - s_{14}^2 (k_2-l_{2,7})^2 $

~$\null + s_{14} (k_2-l_7)^2 \Bral {} (k_3+l_1)^2 - l_1^2 \Brar $

\cr
\hline

$(2312)$ &  
26\,  &
\begin{tabular}{ll}
\vphantom{${}^{\big|}$}
\hlp
$k_1\rightarrow k_4,$& \,$k_2\rightarrow k_1,$\cr
$k_3\rightarrow k_3,$& \,$k_4\rightarrow k_2,$\cr
$l_5\rightarrow -l_{2,6},$& \, $l_6\rightarrow -l_6,$\cr
$l_7\rightarrow -l_{3,5},$& \, $l_8\rightarrow -l_{1,{\bar 4}},$\cr
$l_9\rightarrow l_3,$ & \, $l_{10}\rightarrow -l_4,$\cr
$l_{11} \rightarrow -l_7$
\hrp
\end{tabular}
& 

\vspace{-8truemm}

~$s_{14}^2 l_{3,4,\bar 1}^2 - s_{14} (k_3-l_{2,6})^2\, l_{3,5,6}^2$ 

~$\null +s_{14} l_{2,6}^2 (k_2-l_4)^2$

\cr
\hline

$(2322)$ &  
12\,  &
\begin{tabular}{ll}
\vphantom{${}^{\big|}$}
\hlp
$k_1\rightarrow k_1,$& \,$k_2\rightarrow k_4,$\cr
$k_3\rightarrow k_2,$& \,$k_4\rightarrow k_3,$\cr
$l_5\rightarrow -l_{2,6},\, $& $l_6\rightarrow k_3+l_7$
\hrp
\end{tabular}
& 

~$s_{14}^2 (k_3+l_{\bar 2, \bar 6, 7})^2$

\cr
\hline
$(2122)$ &  
40\,  &
\begin{tabular}{ll}
\vphantom{${}^{\big|}$}
\hlp
$k_1\rightarrow k_1,$& \hsc $k_2\rightarrow k_4,$\cr
$k_3\rightarrow k_3,$& \hsc $k_4\rightarrow k_2,$\cr
$l_5\rightarrow l_3,$& \hsc $l_6\rightarrow l_{\bar 1, 4}$
\hrp
\end{tabular}
& 

~$s_{23}^2 l_{\bar 1,3,4}^2$

\cr
\hline
$(1122)$ &  
42\,  &
\begin{tabular}{ll}
\vphantom{${}^{\big|}$}
\hlp
$k_1\rightarrow k_3,$& $k_2\rightarrow k_2,$\cr
$k_3\rightarrow k_4,$& $k_4\rightarrow k_1,$\cr
$l_6\rightarrow k_3+l_7,$ \cr
$l_7\rightarrow k_1+l_6$
\hrp
\end{tabular}

& 


~$s_{14}(k_{1,3}+l_7)^2 (k_{1,4}+l_6)^2$

\cr
\hline

\end{tabular}

\vfill


\newpage

\begin{tabular}{lc|p{1.75in}|p{3.5in}}

Trees & ${\cal I}_n$& Momentum relabeling & Numerator factor \cr
\hline

$(3112)$ &  
42\,  &
\begin{tabular}{ll}
\vphantom{${}^{\big|}$}
\hlp
$k_1\rightarrow k_1,$&\hsc $k_2\rightarrow k_4,$\cr
$k_3\rightarrow k_2,$& \hsc $k_4\rightarrow k_3,$\cr
$l_6\rightarrow -l_6,$ & \hsc $l_7\rightarrow -l_7$
\hrp
\end{tabular}
& 

~$s_{14} (k_3-l_6)^2 (k_2-l_7)^2$

\cr
\hline

$(2121)$ &  
44\,  &
\begin{tabular}{ll}
\vphantom{${}^{\big|}$}
\hlp
$k_1\rightarrow k_1,$& $k_2\rightarrow k_4,$\cr
$k_3\rightarrow k_2,$& $k_4\rightarrow k_3,$\cr
$l_5\rightarrow -l_1,$&  \cr
$\,l_7\rightarrow k_2-l_7,$ \cr
$l_8\rightarrow l_5$
\hrp
\end{tabular}
& 


~$s_{14}[ (k_2-l_1)^2(l_5-l_1)^2 - (k_3-l_1)^2 (k_2 - l_{1,7})^2]$

\cr
\hline
$(2112)$ &  
44\,  &
\begin{tabular}{ll}
\vphantom{${}^{\big|}$}
\hlp
$k_1\rightarrow k_4,$& $k_2\rightarrow k_1,$\cr
$k_3\rightarrow k_2,$& $k_4\rightarrow k_3,$\cr
$l_5\rightarrow -l_{2,6},$ & $l_7\rightarrow l_3,$ \cr
$l_8\rightarrow k_3+l_7$\cr
\hrp
\end{tabular}
& 


~$ - s_{14}(k_2-l_{2,6})^2 (k_3 - l_{2,6,\bar 7})^2
  +s_{14}(k_3-l_{2,6})^2 l_{\bar 2,3,\bar 6}^2$

\cr
\hline
$(3111)$ &  
43\,  &
\begin{tabular}{ll}
\vphantom{${}^{\big|}$}
\hlp
$k_1\rightarrow k_4,$& \hsa $k_2\rightarrow k_1,$\cr
$k_3\rightarrow k_3,$& \hsa $k_4\rightarrow k_2,$\cr
$l_5\rightarrow -l_6,$& \hsa $l_6\rightarrow -l_7,$\cr
$l_7 \rightarrow -k_1 - l_6,$ \cr
$l_8 \rightarrow l_{1,2}$\cr
\hrp
\end{tabular}
& 

~$s_{14} (k_3-l_6)^2(k_2-l_7)^2$

\cr
\hline

$(2111)$ &  
45\,  &
\begin{tabular}{ll}
\vphantom{${}^{\big|}$}
\hlp
$k_1\rightarrow k_1,$& $k_2\rightarrow k_4,$\cr
$k_3\rightarrow k_2,$& $k_4\rightarrow k_3,$\cr
$l_5\rightarrow -l_1,$& $l_6\rightarrow -l_{2,6},$\cr
$l_7\rightarrow -l_2,$& $l_8\rightarrow l_5,$\cr
$l_9\rightarrow -l_{4,7},$& $l_{10}\rightarrow l_6,$\cr
$l_{11}\rightarrow l_{2,\bar 3},$& $l_{12}\rightarrow -l_{7},$ \cr
$l_{13} \rightarrow k_4 + l_1$
\hrp
\end{tabular}
& 

~$-s_{14}[(k_3-l_1)^2 l_{\bar 2,5}^2 + (k_2-l_1)^2 l_{\bar 4,6,\bar 7}^2]$

$ + s_{14}(k_2-l_1)^2 (k_3-l_1)^2 +s_{14}(k_2-l_1)^2 l_{\bar 2,3}^2$

\cr
\hline
$(1121)$ & 
46\,  &
\begin{tabular}{ll}
\vphantom{${}^{\big|}$}
\hlp
$k_1\rightarrow k_3,$& $k_2\rightarrow k_2,$\cr
$k_3\rightarrow k_1,$& $k_4\rightarrow k_4,$\cr
$l_5\rightarrow k_1+l_6,$& $l_6\rightarrow l_5,$\cr
$l_7\rightarrow k_2-l_7,$& $l_8\rightarrow -l_2,$\cr
$l_9 \rightarrow -l_6$ 
\hrp
\end{tabular}
& 

\vspace{-6truemm}

~$(l_5-l_2)^2 \Bral s_{23}(k_4+l_5)^2 - s_{12} (k_{1,4}+l_6)^2 \Brar$

~$ \null +(k_2-l_{2,7})^2\Bral  s_{13}(k_{1,4}+l_6)^2
                             - s_{23} (k_{1,3}+l_7)^2 \Brar $

$\null - s_{13}(k_1+l_6)^2(k_2-l_7)^2$

\cr
\hline

\end{tabular}


\newpage

\begin{tabular}{lc|p{1.75in}|p{3.5in}}

Trees & ${\cal I}_n$& Momentum relabeling & Numerator factor \cr
\hline

$(1112)$ & 
48\, &
\begin{tabular}{ll}
\vphantom{${}^{\big|}$}
\hlp
$k_1\rightarrow k_4,$& $k_2\rightarrow k_1,$\cr
$k_3\rightarrow k_3,$& $k_4\rightarrow k_2,$\cr
$l_5\rightarrow -l_6,$\cr
$\,l_6\rightarrow k_3+l_7,$ \cr
$l_7\rightarrow -l_2,$& $l_8\rightarrow l_4,$\cr
$l_9\rightarrow -l_1,$& $l_{10}\rightarrow l_3,$\cr
$l_{11}\rightarrow -l_{1, \bar 4},$& $l_{12}\rightarrow l_{\bar 2,3},$ \cr
$\,l_{13}\rightarrow k_1+l_6,$ \cr
$l_{14}\rightarrow -l_7,$ \cr
$\,l_{15}\rightarrow k_4+l_1,$ \cr
$\,l_{16}\rightarrow l_{4,7}$
\hrp
\end{tabular}

& 

\vspace{-25truemm}
~$s_{14}(k_1+l_3)^2(k_3-l_1)^2$ $-s_{14}(k_2-l_2)^2 (k_4+l_4)^2$

~$+s_{14}(k_1+l_3)^2(l_1+l_6)^2$ $+s_{14}(k_3-l_1)^2(k_3+l_{3,7})^2$

~$+s_{14}s_{13}(k_3 + l_{\bar 1,4,7})^2$ $-s_{13}l_{2,6}^2(k_3+l_{4,7})^2$ 

~$-s_{12}l_{1,6}^2(k_3+l_{3,7})^2$ $+s_{14}(k_3+l_7)^2  (k_3-l_6)^2$

~$+s_{14}(k_3+l_7)^2 (k_2+l_{2,\bar 3})^2$ $-s_{13}(k_3+l_7)^2 l_{1,6}^2$

~$-s_{14}(k_1+l_6)^2 l_{\bar 1,4,7}^2$ $-s_{14}(k_3+l_7)^2 l_{4,7}^2$

~$-s_{14}(k_3+l_7)^2 l_{\bar 2,3}^2$

\cr
\hline

$(1111)$ & 
49\, &
\begin{tabular}{ll}
\vphantom{${}^{\big|}$}
\hlp
$k_1\rightarrow k_4,$& \hsa $k_2\rightarrow k_1,$\cr
$k_3\rightarrow k_3,$& \hsa $k_4\rightarrow k_2,$\cr
$l_5\rightarrow k_1+l_6,$& \hsa $l_6\rightarrow -l_6,$\cr
$l_7\rightarrow -l_2,$& \hsa $l_8\rightarrow -l_7,$\cr
$l_9\rightarrow l_3,$& \hsa $l_{10}\rightarrow l_{4,7},$\cr
$l_{11}\rightarrow -l_1,$&\hsa  $l_{12}\rightarrow l_5,$\cr
$l_{13}\rightarrow -l_4,$&\hsa $l_{14}\rightarrow l_{2,{\bar 3}},$\cr
$\,l_{15}\rightarrow l_{{\bar 1},4} ,$ \cr
$\,l_{16}\rightarrow -l_{1}-k_4,$ \hsa  $\null$  \cr
$\,l_{17}\rightarrow l_5 -k_3$ 
\hrp
\end{tabular}

& 

\vspace{-30truemm}
~$ s_{14} (k_2-l_7)^2 (k_3-l_6)^2
    +s_{14} (k_4+l_3)^2 (k_3-l_6)^2 $

~$\null -l_{\bar 2,3}^2 s_{13} (k_1+l_6)^2
   +l_{\bar 2,3}^2 s_{13} (k_4+l_1)^2$

~$\null +l_{4,7}^2 s_{14} (k_1+l_6)^2
   -l_{4,7}^2 s_{12} (k_4+l_1)^2$

~$\null -l_{1,6}^2 s_{13}(k_2-l_7)^2
   -l_{\bar 2,3}^2 \Bral s_{13}(k_4-l_2)^2$

~$\null +s_{14} (s_{13}-(k_1+l_{5,6})^2) \Brar
   -s_{14} (k_2-l_2)^2 (k_1+l_{5,6})^2$

~$\null +s_{13} (k_2-l_2)^2 (k_{1,4}+l_6)^2
   +s_{13} (k_3+l_{4,7})^2 (k_{1,4}+l_6)^2 $

~$\null +l_{4,7}^2 \Bral s_{13} (k_4+l_2)^2
   -s_{14} ( (k_1+l_{3,6})^2 +(k_4+l_{1,\bar 4})^2 ) \Brar$

~$\null -(k_1+l_6)^2 \Bral -s_{13}(k_2-l_{1,\bar 4})^2+l_{\bar 2,5}^2 s_{13}
          -l_{\bar 1,4,7}^2 s_{12} \Brar$

~$\null +l_{\bar 1,5}^2 \, s_{13} s_{14}
   +l_{\bar 2,3}^2\, l_{4,7}^2\,  s_{14}$

\cr
\hline

\end{tabular}


\newpage

\section{Color factors: tensors and Casimirs}
\label{ColorAppendix}

In this paper we encountered several types of scalar color factors,
or Casimir invariants, as well as group invariants
associated with various vacuum graphs.  These factors are defined by,
\begin{eqnarray}
C_A &=&
\frac{1}{N_A} \colorf{a_1a_2a_3} \colorf{a_3a_2a_1}=
\frac{1}{N_A} \colorc{1,2,3} \colorc{3,2,1}
\,, \nn \\
C_{V^{\A}}&=&
\frac{1}{N_A C_A} \colorc{1, 2, 3} \colorc{3, 4, 5}\colorc{5, 6, 7}\colorc{7, 8, 1}
\colorc{9, 2, 10}\colorc{10, 4, 11}\colorc{11, 6, 12}\colorc{12, 8, 9}
\,, \nn \\
C_{V^{\B}}&=&
\frac{1}{N_A C_A} \colorc{1, 4, 3}\colorc{3, 2, 5}\colorc{5, 6, 7}\colorc{7, 8, 1}
\colorc{9, 2, 10}\colorc{10, 4, 11}\colorc{11, 6, 12}\colorc{12, 8, 9}
\,, \nn \\
C_{V_1}&=&
C_A C_{V^{\A}}
\,, \nn \\
C_{V_2}&=&
C_A C_{V^{\B}}
\,, \nn \\
C_{V_8}&=&
C_A C_{V^{\B}}
\,, \nn \\
d_A^{abcd}d_A^{abcd}&=&
\frac{1}{3}N_A C_A {} (C_{V^{\A}}+2 C_{V^{\B}}) 
\,,
\label{Casimirs}
\end{eqnarray}
where $N_A=\Tr_A(1)$ is the number of gluon states, or the dimension
of the adjoint representation. The structure constants are written as
\begin{equation}
\colorc{i,j,k} = \colorf{a_i a_j a_k} = \Tr([T^{a_i}, \, T^{a_j}] \, T^{a_k}) \,,
\label{cijkdef}
\end{equation}
and $i,j,k$ label internal or external lines.

For $G=SU(N_c)$, the quantities defined in \eqn{Casimirs} evaluate to,
\begin{eqnarray}
C_A &=&
2N_c 
\,, \nn \\
C_{V^{\A}}&=&
N_c {} (N_c^2 + 12)
 \,, \nn \\
C_{V^{\B}}&=&
12 N_c 
\,,  \nn \\
C_{V_1}&=&
2N_c^2 {} (N_c^2 + 12)
 \,, \nn \\
C_{V_2}&=&
24 N_c^2 
\,,  \nn \\
C_{V_8}&=&
24 N_c^2
 \,,  \nn \\
d_A^{abcd}d_A^{abcd}&=&
\frac{2}{3} {} (N_c^2-1) N_c^2 {} (N_c^2 + 36) 
\,.
\label{CasimirsForSUNc}
\end{eqnarray}
Note that with our normalization~(\ref{Trdef}) of the generators $T^a$
for $SU(N_c)$, namely $\Tr(T^aT^b)=\delta^{ab}$, the value of $C_A$ is
twice the more conventional value.

For four-point amplitudes in any purely adjoint gauge theory, with
gauge group $G$, we take the basis of color tensors through four
loops ($L \le 4$) to be $\{b_i^{(L)}\}$, where $i$ is a label
distinguishing color structures of same loop order. An explicit choice
of basis is,
\begin{align}
b_1^{(0)} &= 
\colorc{1, 2, 5} \colorc{5, 3, 4} 
\,,\nn \\ 
b_2^{(0)} &= 
\colorc{2, 3, 5} \colorc{5, 4, 1} 
\,,\nn \\ 
b_1^{(1)} &= 
\colorc{1, 5, 8} \colorc{2, 6, 5} \colorc{3, 7, 6} \colorc{4, 8, 7} 
\,,\nn\\
b_1^{(2)} &= 
\colorc{1, 7, 5} \colorc{2, 6, 7} \colorc{3, 11, 9} \colorc{4, 8, 11} \colorc{6, 9, 10} \colorc{5, 10, 8} 
\,,\nn\\
b_2^{(2)} &= 
\colorc{2, 7, 5} \colorc{3, 6, 7} \colorc{4, 11, 9} \colorc{1, 8, 11} \colorc{6, 9, 10} \colorc{5, 10, 8} 
\,,\nn\\
b_1^{(3)} &= 
\colorc{1, 5, 6} \colorc{2, 9, 5} \colorc{3, 13, 14} \colorc{4, 11, 13} \colorc{6, 7, 8}
\colorc{8, 10, 11} \colorc{7, 9, 12} \colorc{10, 12, 14} 
\,,\nn\\
b_1^{(4)} &= 
\colorc{1, 5, 6} \colorc{2, 7, 5} \colorc{3, 17, 14} \colorc{4, 16, 17} \colorc{6, 9, 10}
\colorc{7, 8, 9} \colorc{8, 11, 12} \colorc{10, 12, 13} \colorc{11, 14, 15} \colorc{13, 15, 16}
\,,\nn\\
b_2^{(4)} &= 
\colorc{2, 5, 6} \colorc{3, 7, 5} \colorc{4, 17, 14} \colorc{1, 16, 17} \colorc{6, 9, 10}
\colorc{7, 8, 9} \colorc{8, 11, 12} \colorc{10, 12, 13} \colorc{11, 14, 15} \colorc{13, 15, 16}
\,.
\label{ColorBasis}
\end{align}
 $b_1^{(0)}$ and $b_2^{(0)}$ are the $s$- and $t$-channel four-point
 trees, respectively; $b_1^{(1)}$ is the one-loop box; $b_1^{(L)}$ are
 the $s$-channel ladder graphs; $b_2^{(L)}$ are the $t$-channel
 ladder graphs.

We have demonstrated that this basis is complete for four-point
amplitudes with only adjoint states and interactions proportional
to $\colorf{a b c}$.  We have done so by explicitly solving the Jacobi
identities for all color tensors
appearing at $0 \le L \le 4$ for a generic amplitude, and converting
the reducible tensor structures to lower-loop color factors
multiplied by group invariants.
(Typical reducible tensor structures are color
graphs with two- and three-point subgraphs.) Thus, any
$L$-loop color tensor can be expressed in terms of a linear combination
of the basis tensors $b_i^{(0)},\ldots,b_i^{(L)}$.  Through four loops
the pattern is that the number of new independent (irreducible) $L$-loop
tensor structures are: two for even $L$, and one for odd $L$. The full
basis then consists of the independent $L$-loop basis
tensors, plus all the independent basis tensors for lower
loop orders, which gives a total of $({3\over2} L+ 2)$ basis tensors for
even loop order, and ${3\over2}(L+1)$ for odd $L$. We believe that this
pattern continues for $L>4$ for the case of four-point tensors, but we
have not proven it.

In the standard case of $G=SU(N_c)$ the basis color tensors are,
\begin{eqnarray}
b_1^{(0)} &=& 
\Tra{1234} + \Tra{1432} - \Tra{1243} - \Tra{1342} 
\,,\nn \\ 
b_2^{(0)} &=& 
\Tra{1234} + \Tra{1432}  - \Tra{1324} - \Tra{1423}  
\,,\nn \\ 
b_1^{(1)} &=&  
N_c {} (\Tra{1234} + \Tra{1432} )
+ 2 ( \Tra{12} \Tra{34} + \Tra{14} \Tra{23} + \Tra{13} \Tra{24} ) 
\,,\nn
\end{eqnarray}
\begin{eqnarray}
b_1^{(2)} &=& 
(N_c^2+2) (\Tra{1234} + \Tra{1432}) 
 + 2 \Tra{1243} + 2 \Tra{1342} - 4 \Tra{1423} - 4 \Tra{1324}
 + 6 N_c \Tra{12} \Tra{34}  
\,,\nn\\
b_2^{(2)} &=& 
(N_c^2+2) (\Tra{1234} + \Tra{1432})
 + 2 \Tra{1423} + 2 \Tra{1324} - 4 \Tra{1342} - 4 \Tra{1243}
 + 6 N_c \Tra{14} \Tra{23}  
\,,\nn\\
b_1^{(3)} &=& 
(N_c^3+2 N_c) (\Tra{1234} + \Tra{1432}) + 2 N_c {} (\Tra{1243} +  \Tra{1342})
 + (14N_c^2+8) \Tra{12} \Tra{34} \nn\\
& & \null +8 \Tra{14} \Tra{23} + 8 \Tra{13} \Tra{24}   
\,,\nn\\
b_1^{(4)} &=& 
(N_c^4 + 2 N_c^2+8)(\Tra{1234} + \Tra{1432})
 + (2 N_c^2+8) (\Tra{1243} + \Tra{1342}) \nn\\
&& \null - 16 (\Tra{1324} + \Tra{1423})+(30 N_c^3+24 N_c) \Tra{12} \Tra{34} 
 \,,\nn\\
b_2^{(4)} &=&
 (N_c^4 + 2 N_c^2+8)(\Tra{1234} + \Tra{1432})
 + (2 N_c^2+8) (\Tra{1423} + \Tra{1324}) \nn\\
&& \null - 16 (\Tra{1243} + \Tra{1342})+(30 N_c^3+24 N_c)  \Tra{14} \Tra{23}  
\,.
\end{eqnarray}

Using the Jacobi identity, the four-loop four-point
color tensors $C_i$ appearing in \eqn{FourLoopYMAmplitude}
can be reduced to the eight-dimensional basis~(\ref{ColorBasis}).
For completeness, we first give the corresponding reductions of the
one-loop four-point tensor,
\begin{equation}
C^{\rm box} = b^{(1)}_1,
\label{OneLoopColorReduction}
\end{equation}
the two-loop ones,
\begin{eqnarray}
C^{\P} &=& b^{(2)}_1, \nn\\
C^{\NP} &=& b^{(2)}_1-\Frac{1}{2} C_A b^{(1)}_1, 
\label{TwoLoopColorReduction}
\end{eqnarray}
and the three-loop ones,
\begin{eqnarray}
C^{\rm (a)} &=& b^{(3)}_1, \nn\\
C^{\rm (b)} &=& b^{(3)}_1
  -  \Frac{1}{2} C_A b^{(2)}_1, \nn\\
C^{\rm (c)} &=& b^{(3)}_1
  -  \Frac{1}{2} C_A b^{(2)}_1, \nn\\
C^{\rm (d)} &=& b^{(3)}_1
  -  C_A b^{(2)}_1 + \Frac{1}{4} C_A^2 b^{(1)}_1,\nn\\
C^{\rm (e)} &=& -\Frac{1}{2}b^{(3)}_1
  + \Frac{3}{4} C_A  b^{(2)}_1 + \Frac{1}{2} C_{V^{\B}} b^{(0)}_1,
\nn\\
C^{\rm (f)} &=& -\Frac{1}{2}b^{(3)}_1 
  + \Frac{3}{4} C_A b^{(2)}_1 - \Frac{1}{4} C_A^2 b^{(1)}_1
  + \Frac{1}{2} C_{V^{\B}} b^{(0)}_1, \nn\\
C^{\rm (g)} &=& -\Frac{1}{2}b^{(3)}_1
  + \Frac{1}{4} C_A b^{(2)}_1
  + \Frac{1}{2} C_{V^{\B}} b^{(0)}_1 , \nn\\
C^{\rm (h)} &=& \Frac{1}{2} b^{(3)}_1
  - \Frac{5}{12} C_A b^{(2)}_1 + \Frac{1}{6} C_A b^{(2)}_2
  - \Frac{1}{2} C_{V^{\B}} b^{(0)}_1 ,\nn\\
C^{\rm (i)} &=& \Frac{1}{6} C_A b^{(2)}_1
  - \Frac{1}{6} C_A b^{(2)}_2 .
\label{ThreeLoopColorReduction}
\end{eqnarray}
The four-loop reduction gives,
\begin{align} 
C_{1}&=
b^{(4)}_1
,\nn\\
C_{2}&=
b^{(4)}_1-\Frac{1}{2} C_A {} b^{(3)}_1
,\nn\\
C_{3}&=
b^{(4)}_1-C_A b^{(3)}_1+\Frac{1}{4}C_A^2 b^{(2)}_1
,\nn\\
C_{4}&=
b^{(4)}_1-\Frac{1}{2}C_A b^{(3)}_1
,\nn\\
C_{5}&=
b^{(4)}_1-C_A b^{(3)}_1+\Frac{1}{4}C_A^2 b^{(2)}_1
,\nn\\
C_{6}&=
-\Frac{1}{2}b^{(4)}_1+\Frac{3}{4}C_A b^{(3)}_1+\Frac{1}{4}C_{V_2}b^{(0)}_1
,\nn\\
C_{7}&=
-\Frac{1}{2}b^{(4)}_1+\Frac{3}{4}C_A b^{(3)}_1-\Frac{1}{4}C_A^2 b^{(2)}_1
+\Frac{1}{4}C_{V_2}b^{(0)}_1
,\nn\\
C_{8}&=
-\Frac{1}{2}b^{(4)}_1+\Frac{1}{4}C_A b^{(3)}_1+\Frac{1}{4}C_{V_2} b^{(0)}_1
,\nn\\
C_{9}&=
-\Frac{1}{2}b^{(4)}_1+C_A b^{(3)}_1-\Frac{3}{8}C_A^2 b^{(2)}_1
,\nn\\
C_{10}&=
-\Frac{1}{2}b^{(4)}_1+C_A b^{(3)}_1-\Frac{5}{8}C_A^2b^{(2)}_1
+\Frac{1}{8}C_A^3b^{(1)}_1
,\nn\\
C_{11}&=
-\Frac{1}{2}b^{(4)}_1+\Frac{1}{2}C_A b^{(3)}_1-\Frac{1}{8}C_A^2 b^{(2)}_1
,\nn\\
C_{12}&=
-\Frac{1}{2}b^{(4)}_1+\Frac{3}{4}C_A b^{(3)}_1+\Frac{1}{4}C_{V_2}b^{(0)}_1
,\nn\\
C_{13}&=
\Frac{1}{2}b^{(4)}_1-\Frac{1}{4}C_A b^{(3)}_1-\Frac{1}{4}C_{V_2}b^{(0)}_1
,\nn\\
C_{14}&=
b^{(4)}_1-\Frac{7}{4}C_A b^{(3)}_1
+\Frac{7}{8}C_A^2b^{(2)}_1+C_{V^{\B}}b^{(1)}_1
+\Frac{1}{12}C_{V_2}b^{(0)}_1
,\nn\\
C_{15}&=
-b^{(4)}_1+\Frac{5}{4}C_A b^{(3)}_1-\Frac{3}{8}C_A^2b^{(2)}_1
-C_{V^{\B}}b^{(1)}_1+\Frac{5}{12}C_{V_2}b^{(0)}_1
,\nn\\
C_{16}&=
-b^{(4)}_1+\Frac{3}{2}C_A b^{(3)}_1-\Frac{1}{2}C_A^2b^{(2)}_1
-C_{V^{\B}}b^{(1)}_1+\Frac{1}{6}C_{V_2}b^{(0)}_1
,\nn\\
C_{17}&=
-b^{(4)}_1+\Frac{7}{4}C_A b^{(3)}_1-\Frac{7}{8}C_A^2b^{(2)}_1
+(\Frac{1}{8}C_A^3-C_{V^{\B}})b^{(1)}_1+\Frac{7}{12}C_{V_2}b^{(0)}_1
,\nn\\
C_{18}&=
-\Frac{1}{2}b^{(4)}_1+\Frac{1}{2}C_A b^{(3)}_1+\Frac{1}{8}C_A^2b^{(2)}_1
+\Frac{1}{3}C_{V_2}b^{(0)}_1
,\nn\\
C_{19}&=
\Frac{1}{2}b^{(4)}_1-\Frac{3}{4}C_A b^{(3)}_1+\Frac{1}{4}C_A^2b^{(2)}_1
-\Frac{1}{12}C_{V_2}b^{(0)}_1
,\nn\\
C_{20}&=
-\Frac{1}{2}b^{(4)}_1+\Frac{1}{2}C_A b^{(3)}_1-\Frac{1}{8}C_A^2b^{(2)}_1
-\Frac{1}{3}C_{V_2}b^{(0)}_1
,\nn\\
C_{21}&=
-\Frac{1}{2}b^{(4)}_1+\Frac{3}{4}C_A b^{(3)}_1
-\Frac{1}{4}C_A^2b^{(2)}_1-C_{V^{\B}}b^{(1)}_1+\Frac{1}{12}C_{V_2}b^{(0)}_1
,\nn\\
C_{22}&=
-\Frac{1}{2}b^{(4)}_1+\Frac{1}{2}C_A b^{(3)}_1+\Frac{1}{8}C_A^2b^{(2)}_1
+(-\Frac{1}{8}C_A^3-C_{V^{\B}})b^{(1)}_1+\Frac{1}{3}C_{V_2}b^{(0)}_1
,\nn\\
C_{23}&=
-\Frac{1}{2}b^{(4)}_1+\Frac{1}{2}C_A b^{(3)}_1-\Frac{1}{8}C_A^2b^{(2)}_1
-C_{V^{\B}}b^{(1)}_1+\Frac{1}{3}C_{V_2}b^{(0)}_1
,\nn\\
C_{24}&=
\Frac{1}{6}C_{V_2}b^{(0)}_1
,\nn\\
C_{25}&=
-\Frac{1}{4}C_A b^{(3)}_1+\Frac{1}{8}C_A^2b^{(2)}_1
-\Frac{1}{12}C_{V_2}b^{(0)}_1
,\nn\\\nn
\end{align}
\begin{align}
C_{26}&=
\Frac{1}{4}C_A b^{(3)}_1-\Frac{1}{8}C_A^2b^{(2)}_1-\Frac{1}{12}C_{V_2}b^{(0)}_1
,\nn\\
C_{27}&=
\Frac{1}{2}b^{(4)}_1-\Frac{1}{2}C_A b^{(3)}_1+\Frac{1}{8}C_A^2b^{(2)}_1
-\Frac{1}{6}C_{V_2}b^{(0)}_1
,\nn\\
C_{28}&=
-\Frac{1}{2}b^{(4)}_1-\Frac{1}{2}b^{(4)}_2+\Frac{9}{8}C_A b^{(3)}_1
-\Frac{11}{16}C_A^2b^{(2)}_1+\Frac{5}{8}C_A^2b^{(2)}_2
-\Frac{1}{2}C_{V^{\B}}b^{(1)}_1-\Frac{7}{24}C_{V_2}b^{(0)}_1
+\Frac{5}{6}C_{V_2}b^{(0)}_2
,\nn\\
C_{29}&=
-\Frac{1}{2}b^{(4)}_1-\Frac{1}{2}b^{(4)}_2+\Frac{11}{8}C_A b^{(3)}_1
-\Frac{17}{16}C_A^2b^{(2)}_1+\Frac{5}{8}C_A^2b^{(2)}_2
-\Frac{1}{2}C_{V^{\B}}b^{(1)}_1-\Frac{13}{24}C_{V_2}b^{(0)}_1
+\Frac{5}{6}C_{V_2}b^{(0)}_2
,\nn\\
C_{30}&=
-\Frac{1}{2}b^{(4)}_1+\Frac{5}{8}C_A b^{(3)}_1-\Frac{3}{16}C_A^2b^{(2)}_1
-\Frac{1}{2}C_{V^{\B}}b^{(1)}_1+\Frac{5}{24}C_{V_2}b^{(0)}_1
-\Frac{1}{6}C_{V_2}b^{(0)}_2
,\nn\\
C_{31}&=
\Frac{1}{2}b^{(4)}_2-\Frac{5}{8}C_A b^{(3)}_1+\Frac{13}{16}C_A^2b^{(2)}_1
-\Frac{5}{8}C_A^2b^{(2)}_2+\Frac{1}{2}C_{V^{\B}}b^{(1)}_1
+\Frac{19}{24}C_{V_2}b^{(0)}_1-\Frac{5}{6}C_{V_2}b^{(0)}_2
,\nn\\
C_{32}&=
b^{(4)}_1+ \Frac{1}{2}b^{(4)}_2-\Frac{9}{4}C_A b^{(3)}_1
+\Frac{35}{24}C_A^2b^{(2)}_1-\Frac{17}{24}C_A^2b^{(2)}_2+C_{V^{\B}}b^{(1)}_1
+\Frac{7}{12}C_{V_2}b^{(0)}_1-\Frac{5}{6}C_{V_2}b^{(0)}_2
,\nn\\
C_{33}&=
-\Frac{1}{2}b^{(4)}_2+\Frac{5}{8}C_A b^{(3)}_1-\Frac{13}{16}C_A^2b^{(2)}_1
+\Frac{5}{8}C_A^2b^{(2)}_2-\Frac{1}{2}C_{V^{\B}}b^{(1)}_1
-\Frac{5}{8}C_{V_2}b^{(0)}_1+\Frac{5}{6}C_{V_2}b^{(0)}_2
,\nn\\
C_{34}&=
\Frac{1}{2}b^{(4)}_1+b^{(4)}_2-\Frac{9}{4}C_A b^{(3)}_1+2C_A^2b^{(2)}_1
-\Frac{5}{4}C_A^2b^{(2)}_2+C_{V^{\B}}b^{(1)}_1+\Frac{17}{12}C_{V_2}b^{(0)}_1
-\Frac{5}{3}C_{V_2}b^{(0)}_2
,\nn\\
C_{35}&=
-\Frac{1}{2}b^{(4)}_2+\Frac{5}{8}C_A b^{(3)}_1-\Frac{35}{48}C_A^2b^{(2)}_1
+\Frac{13}{24}C_A^2b^{(2)}_2-\Frac{1}{2}C_{V^{\B}}b^{(1)}_1
-\Frac{5}{8}C_{V_2}b^{(0)}_1+\Frac{5}{6}C_{V_2}b^{(0)}_2
,\nn\\
C_{36}&=
\Frac{1}{2}b^{(4)}_2-\Frac{7}{8}C_A b^{(3)}_1+\Frac{15}{16}C_A^2b^{(2)}_1
-\Frac{5}{8}C_A^2b^{(2)}_2+\Frac{1}{2}C_{V^{\B}}b^{(1)}_1
+\Frac{7}{8}C_{V_2}b^{(0)}_1-\Frac{5}{6}C_{V_2}b^{(0)}_2
,\nn\\
C_{37}&=
-\Frac{1}{8}C_A b^{(3)}_1+\Frac{7}{48}C_A^2b^{(2)}_1
-\Frac{1}{12}C_A^2b^{(2)}_2-\Frac{1}{2}C_{V^{\B}}b^{(1)}_1
+\Frac{1}{8}C_{V_2}b^{(0)}_1
,\nn\\
C_{38}&=
\Frac{1}{2}b^{(4)}_1+b^{(4)}_2-2C_A b^{(3)}_1+\Frac{43}{24}C_A^2b^{(2)}_1
-\Frac{7}{6}C_A^2b^{(2)}_2+C_{V^{\B}}b^{(1)}_1+\Frac{4}{3}C_{V_2}b^{(0)}_1
-\Frac{5}{3}C_{V_2}b^{(0)}_2
,\nn\\
C_{39}&=
\Frac{1}{2}b^{(4)}_1+\Frac{1}{2}b^{(4)}_2-\Frac{11}{8}C_A b^{(3)}_1
+\Frac{17}{16}C_A^2b^{(2)}_1-\Frac{5}{8}C_A^2b^{(2)}_2
+\Frac{1}{2}C_{V^{\B}}b^{(1)}_1+\Frac{17}{24}C_{V_2}b^{(0)}_1
-\Frac{5}{6}C_{V_2}b^{(0)}_2
,\nn\\
C_{40}&=
\Frac{1}{2}b^{(4)}_1-\Frac{3}{4}C_A b^{(3)}_1+\Frac{1}{4}C_A^2b^{(2)}_1
+\Frac{1}{4}C_{V_2}b^{(0)}_1
,\nn\\
C_{41}&=
b^{(4)}_1-\Frac{3}{2}C_A b^{(3)}_1+\Frac{1}{2}C_A^2b^{(2)}_1
+C_{V^{\B}}b^{(1)}_1-\Frac{1}{3}C_{V_2}b^{(0)}_1
,\nn\\
C_{42}&=
-\Frac{1}{2}b^{(4)}_1+\Frac{3}{4}C_A b^{(3)}_1-\Frac{1}{4}C_A^2b^{(2)}_1
+\Frac{1}{12}C_{V_2}b^{(0)}_1
,\nn\\
C_{43}&=
\Frac{1}{2}b^{(4)}_1-\Frac{3}{4}C_A b^{(3)}_1+\Frac{1}{4}C_A^2b^{(2)}_1
+C_{V^{\B}}b^{(1)}_1-\Frac{1}{12}C_{V_2}b^{(0)}_1
,\nn\\
C_{44}&=
-\Frac{1}{6}C_{V_2}b^{(0)}_1
,\nn\\
C_{45}&=
\Frac{1}{2}b^{(4)}_1-\Frac{3}{4}C_A b^{(3)}_1+\Frac{1}{4}C_A^2b^{(2)}_1
+C_{V^{\B}}b^{(1)}_1-\Frac{1}{4}C_{V_2}b^{(0)}_1
,\nn\\
C_{46}&=
-\Frac{1}{2}b^{(4)}_1-b^{(4)}_2+\Frac{9}{4}C_A b^{(3)}_1-2C_A^2b^{(2)}_1
+\Frac{5}{4}C_A^2b^{(2)}_2-C_{V^{\B}}b^{(1)}_1-\Frac{17}{12}C_{V_2}b^{(0)}_1
+\Frac{5}{3}C_{V_2}b^{(0)}_2
,\nn\\
C_{47}&=
\Frac{1}{2}b^{(4)}_1+b^{(4)}_2-\Frac{9}{4}C_A b^{(3)}_1+2C_A^2b^{(2)}_1
-\Frac{5}{4}C_A^2b^{(2)}_2+C_{V^{\B}}b^{(1)}_1+\Frac{19}{12}C_{V_2}b^{(0)}_1
-\Frac{5}{3}C_{V_2}b^{(0)}_2
,\nn\\
C_{48}&=
-\Frac{1}{2}b^{(4)}_1+\Frac{3}{4}C_A b^{(3)}_1-\Frac{1}{4}C_A^2b^{(2)}_1
+\Frac{1}{12}C_{V_2}b^{(0)}_1
,\nn\\
C_{49}&=
\Frac{1}{2}b^{(4)}_1+\Frac{1}{2}b^{(4)}_2-\Frac{3}{2}C_A b^{(3)}_1
+\Frac{9}{8}C_A^2b^{(2)}_1-\Frac{5}{8}C_A^2b^{(2)}_2+C_{V^{\B}}b^{(1)}_1
+\Frac{2}{3}C_{V_2}b^{(0)}_1-\Frac{5}{6}C_{V_2}b^{(0)}_2
,\nn\\
C_{50}&=
0
.
\label{FourLoopColorReduction}
\end{align} 
In these equations, $C_{V_2}$ is the Casimir for the four-loop vacuum
graph $V_2$ and $C_{V^{\B}}$ is the one for three-loop vacuum
graph $V^{\B}$, as defined in \eqn{Casimirs}.

Finally, because the color tensors $C_i$ may appear with arbitrary
permutations of the external legs, we also need to give the reductions
of the basis elements for arbitrary permutations.  In the following,
we quote particular permutations, as well as symmetries
of the tensors that allow us to reduce all other permutations:
\begin{align} 
b^{(0)}_1(1,2,3,4) &= b^{(0)}_1 , \nn\\
b^{(0)}_1(2,3,4,1) &= b^{(0)}_2 , \nn\\
b^{(0)}_1(1,3,4,2) &= b^{(0)}_2 - b^{(0)}_1,\nn\\
b^{(0)}_1(a,b,c,d) &= - b^{(0)}_1(b,a,c,d) = b^{(0)}_1(c,d,a,b),
\label{TreeBasisPerms}
\end{align}
\begin{align} 
b^{(1)}_1(1,2,3,4) &= b^{(1)}_1 , \nn\\
b^{(1)}_1(1,3,4,2) &= b^{(1)}_1 - \Frac{1}{2} C_A b^{(0)}_1 , \nn\\
b^{(1)}_1(a,b,c,d) &= b^{(1)}_1(b,c,d,a) = b^{(1)}_1(a,d,c,b),
\label{OneLoopBasisPerms}
\end{align} 
\begin{align} 
b^{(2)}_1(1,2,3,4) &= b^{(2)}_1 , \nn\\
b^{(2)}_1(2,3,4,1) &= b^{(2)}_2 , \nn\\
b^{(2)}_1(1,2,4,3) &= b^{(2)}_1
   - \Frac{1}{4} C_A^2 b^{(0)}_1 , \nn\\
b^{(2)}_1(2,3,1,4) &= b^{(2)}_2
   - \Frac{1}{4} C_A^2 b^{(0)}_2 , \nn\\
b^{(2)}_1(1,3,4,2) &= - b^{(2)}_1 - b^{(2)}_2
   + \Frac{3}{2} C_A b^{(1)}_1 - \Frac{1}{4} C_A^2 b^{(0)}_1 , \nn\\
b^{(2)}_1(1,3,2,4) &= -b^{(2)}_1 - b^{(2)}_2
   + \Frac{3}{2} C_A b^{(1)}_1 - \Frac{1}{4} C_A^2 b^{(0)}_2 , \nn\\
b^{(2)}_1(a,b,c,d) &= b^{(2)}_1(b,a,d,c) = b^{(2)}_1(c,d,a,b) ,
\label{TwoLoopBasisPerms}
\end{align} 
\begin{align} 
b^{(3)}_1(1,2,3,4) &= b^{(3)}_1 , \nn\\
b^{(3)}_1(1,2,4,3) &= b^{(3)}_1 - \Frac{1}{8} C_A^3 b^{(0)}_1 , \nn\\
b^{(3)}_1(2,3,4,1) &= b^{(3)}_1
  - \Frac{7}{6} C_A (b^{(2)}_1-b^{(2)}_2)
  - C_{V^{\B}} (b^{(0)}_1-b^{(0)}_2) , \nn\\
b^{(3)}_1(2,3,1,4) &= b^{(3)}_1
  - \Frac{7}{6} C_A (b^{(2)}_1-b^{(2)}_2)
  - C_{V^{\B}} (b^{(0)}_1-b^{(0)}_2)
  - \Frac{1}{8} C_A^3 b^{(0)}_2 , \nn\\
b^{(3)}_1(1,3,2,4) &= b^{(3)}_1
  - \Frac{7}{6} C_A (2 b^{(2)}_1+b^{(2)}_2)
  + \Frac{7}{4} C_A^2 b^{(1)}_1
  + (C_{V^{\B}} - \Frac{1}{8} C_A^3) b^{(0)}_2 , \nn\\
b^{(3)}_1(1,3,4,2) &= b^{(3)}_1
  - \Frac{7}{6} C_A (2 b^{(2)}_1+b^{(2)}_2)
  + \Frac{7}{4} C_A^2 b^{(1)}_1 + C_{V^{\B}} b^{(0)}_2
  - \Frac{1}{8} C_A^3 b^{(0)}_1 , \nn\\
b^{(3)}_1(a,b,c,d) &= b^{(3)}_1(b,a,d,c) = b^{(3)}_1(c,d,a,b) ,
\label{ThreeLoopBasisPerms}
\end{align} 
\begin{align} 
b^{(4)}_1(1,2,3,4) &= b^{(4)}_1,\nn\\
b^{(4)}_1(2,3,4,1) &= b^{(4)}_2,\nn\\
b^{(4)}_1(1,2,4,3) &= b^{(4)}_1 - \Frac{1}{16} C_A^4 b^{(0)}_1 , \nn\\
b^{(4)}_1(2,3,1,4) &= b^{(4)}_2 - \Frac{1}{16} C_A^4 b^{(0)}_2 , \nn\\
b^{(4)}_1(1,3,4,2) &= - b^{(4)}_1 - b^{(4)}_2
  + \Frac{9}{2} C_A b^{(3)}_1 - \Frac{21}{4}C_A^2 b^{(2)}_1
  + \Frac{1}{8} (15 C_A^3 - 16 C_{V^{\B}}) b^{(1)}_1 \nn\\
&\null\hskip0.5cm
  - \Frac{1}{48} (3 C_A^4 + 56 C_{V_2}) b^{(0)}_1
  + \Frac{10}{3} C_{V_2} b^{(0)}_2 , \nn\\
b^{(4)}_1(1,3,2,4) &= -b^{(4)}_1 - b^{(4)}_2
  + \Frac{9}{2} C_A b^{(3)}_1 - \Frac{21}{4}C_A^2 b^{(2)}_1
  + \Frac{1}{8} (15 C_A^3 - 16 C_{V^{\B}}) b^{(1)}_1 \nn\\
&\null\hskip0.5cm
  - \Frac{1}{48} (3 C_A^4 - 160 C_{V_2})b^{(0)}_2
  - \Frac{7}{6} C_{V_2}b^{(0)}_1 , \nn\\
b^{(4)}_1(a,b,c,d) &= b^{(4)}_1(b,a,d,c) = b^{(4)}_1(c,d,a,b) .
\label{FourLoopBasisPerms}
\end{align} 

\newpage

\section{Numerator and Color Factors of the parent integrals}
\label{NumeratorAppendix}

In this appendix we collect all 50 numerator factors of the parent
integrals, displayed in figs.~\ref{BC1Figure}-\ref{E2Figure}. 
These polynomials are,
\def\bar{\overline}
\begin{align}
& N_1,\, N_2, \, N_3,\, N_4,\, N_{5} = {}
s_{12}^3
\,,
\nn\\
&N_6, \, N_7, \, N_8,\, N_9,\, N_{10},\, N_{11}={}
s_{12}^2 s_{45}
\,,
\nn \\
& N_{12}={}
s_{12}^2 s_{56}
\,,
\nn\\
& N_{13}={}
-s_{12}^2 s_{56}
\,,
\nn\\
&N_{14}={}
s_{12}s_{35}^2
\,,
\nn\\
&N_{15},\, N_{16},\, N_{17}={}
-s_{12}s_{56}^2
\,,
\nn\\
&N_{18}={}
s_{12}s_{35}s_{46}
\,,
\nn\\
& N_{19}={}
-s_{12}s_{56}s_{47}
\,,
\nn\\
& N_{20},\, N_{21}, \, N_{22},\, N_{23}={}
s_{12}s_{56}s_{47}
\,,
\nn \\
& N_{24}={}
s_{12}(s_{35}s_{38}-s_{37}s_{36})
\,,
\nn\\
& N_{25}={}
s_{12}(s_{45}s_{38}-s_{12}s_{37})-l_9^2 s_{12}s_{45}-l_6^2 l_8^2 s_{12}
\,,
\nn\\
& N_{26}={}
s_{12}^2 s_{89} - s_{12}s_{35}s_{67}+ l_9^2 s_{12}\Tau{35}
 +l_5^2s_{12}(s_{4,10}-l_{11}^2 ) 
\,,
\nn\\
& N_{27}={} 
 s_{12} (s_{12} s_{78} + s_{45} s_{9, 10})-s_{12}^2 s_{45} - s_{12} (s_{12}l_{11}^2  +  s_{45}l_{12}^2)
\,,
\nn\\
& N_{28}={}
s_{12}s_{17}s_{36}+s_{23}s_{15}s_{38}-\frac{1}{2}l_9^2 s_{12}s_{23} 
  - l_5^2 l_8^2 s_{12} - l_6^2 l_7^2 s_{23} 
\,,
\nn\\
& N_{29}={}
s_{12}s_{17}s_{68}+s_{23}s_{38}s_{15}-l_6^2 s_{12}s_{23} +l_6^2 l_8^2s_{12}
   -l_7^2 l_9^2s_{23} 
\,,
\nn\\
& N_{30}={}
s_{23}s_{15}\Tau{29}-s_{13}s_{26}\Tau{19}-s_{12}s_{15}s_{26}- l_9^2 
   (l_5^2s_{13}+l_6^2s_{23}) 
\,,
\nn\\
& N_{31}={}
-s_{12}s_{36}s_{5\bar{6}}-s_{12}s_{2\bar{6}}s_{68}
-s_{12}s_{36}s_{2\bar{6}}-s_{23}s_{38}s_{25}
+l_6^2 s_{12} (s_{57} +s_{58}+s_{89})-l_6^2 s_{12} (l_5^2+l_8^2)
\,,
\nn \\
& N_{32}={}
s_{56}(s_{12}s_{56}-s_{23}s_{35}-l_6^2 s_{12})
\,,
\nn \\
& N_{33}={}
s_{12}(s_{15}s_{78}-s_{29}s_{56})+
s_{23}s_{56}(s_{9,10}-l_{10}^2)
+l_{5}^2(s_{12}s_{29}-l_{12}^2 s_{23})
\,,
\nn \\
& N_{34}={}
s_{15}(s_{13}s_{56}-s_{12}s_{78}-l_8^2s_{13})
-s_{16}(s_{23}s_{56}-s_{12}s_{79}-l_9^2s_{23})
+l_7^2s_{12}\tau_{1,10}
\,,
\nn 
\end{align}

\begin{align}
& N_{35}={}
s_{23}(s_{10,11}(s_{17}-l_6^2)-l_5^2l_{11}^2)
+s_{12}s_{15}(s_{23}-s_{78}+l_9^2)
\nn \\ &
+s_{16}(s_{12}s_{79}+l_{10}^2 s_{23})
-l_7^2(s_{12}s_{1,10}-l_{12}^2s_{12}+l_{10}^2s_{23})
\,,
\nn \\
& N_{36}={}
s_{12}s_{47}(s_{23}-s_{67}+l_{10}^2)
+s_{12}s_{45}s_{6,10}-s_{23}s_{47}s_{89}
+l_{11}^2s_{12}(s_{67}-s_{23})
\nn \\ &
+s_{12}(l_5^2l_7^2-l_5^2s_{6,10}-l_{7}^2s_{1\bar{5}})
+l_{11}^2l_{12}^2s_{23}+\frac{1}{2}l_{5}^2s_{12}s_{23}
\,,
\nn \\
& N_{37}={}
s_{46}(s_{12}s_{28}-s_{23}s_{67})+s_{45}(s_{13}s_{37}-s_{23}s_{58})
+s_{23}s_{45}s_{46}+l_9^2s_{12}s_{13}-s_{23} l_{5}^2 l_{6}^2
\,,
\nn \\
& N_{38}={} 
s_{56}(s_{12}s_{78}+s_{23}s_{56}-s_{12}s_{23}-s_{12}s_{10,11}-s_{23}l_{9}^2)
+s_{12}^2s_{10,11}+s_{13}(l_{12}^2l_{14}^2+l_{13}^2l_{15}^2)
\,,
\nn \\
& N_{39}={}
s_{12}(s_{12} s_{6, \overline{14}} - s_{29}s_{38} - s_{35} s_{16,17} + s_{1,15} s_{4, \overline{17}}- s_{2,12} s_{15,16})
 + s_{13} s_{15,16} s_{16,17} 
\nn \\ &
+ s_{12}(l_{14}^2 s_{5, \overline{10}} + l_{13}^2s_{7, \overline{12}}- l_{9}^2s_{8, \overline{14}} - l_{8}^2s_{9, \overline{13}} )
- s_{13}(l_{5}^2 s_{12,14} + l_{12}^2 s_{5,13})\nn \\ &
 + s_{23} (l_{15}^2 l_{17}^2-l_{5}^2 l_{12}^2)
+s_{12}(l_{8}^2l_{9}^2+l_{7}^2l_{10}^2-l_{7}^2l_{13}^2+l_{8}^2l_{13}^2 +l_{9}^2l_{14}^2-l_{10}^2l_{14}^2+l_{13}^2l_{14}^2) 
\,,
\nn \\
& N_{40}={}
-s_{12}^2 s_{56}
\,,
\nn \\  
& N_{41}={}
s_{12}s_{45}^2
\,,
\nn \\
& N_{42}={}
s_{12}s_{46}s_{37}
\,,
\nn \\ 
& N_{43}={}
-s_{12}s_{35}s_{46}
\,,
\nn \\ 
& N_{44}={}
s_{12}(s_{45}s_{57}-s_{35}s_{58})
\,,
\nn \\  
& N_{45}={}
s_{12}(s_{45}s_{78}+s_{35}s_{9,10}-s_{45}s_{35}-l_{11}^2s_{35}-l_{12}^2 s_{45})
\,,
\nn \\ 
& N_{46}={}
s_{68}(s_{23}s_{45}-s_{12}s_{46})+s_{78}(s_{12}s_{47}-s_{13}s_{45})+l_{8}^2
(l_{6}^2 s_{13}-l_{7}^2s_{23})+l_{5}^2(l_{7}^2 s_{13}- l_{6}^2 s_{23})
\,,
\nn \\ 
& N_{47}={}
s_{13}s_{6,10}(s_{68}-s_{12}-l_{11}^2)
-s_{23}s_{8,10}(s_{68}-s_{12}-l_{11}^2)\nn \\ &
+s_{12}(s_{79}s_{6,10}-s_{59}s_{8,10})
+2s_{12}(l_{6}^2-l_{8}^2)(l_{12}^2-l_{10}^2)
\,,
\nn \\
& N_{48}={}
s_{12}(s_{2,10}s_{39}-s_{47}s_{18}+s_{2,10}s_{59}+s_{39}s_{6,10}+s_{23}s_{6,11})-s_{23}s_{57}s_{68}-s_{13}s_{59}s_{6,10}
\nn \\ &
+l_{6}^2(s_{12}s_{35}+s_{12}s_{4,\overline{12}}-s_{23}s_{59})+l_{5}^2(s_{12}s_{26}+s_{12}s_{1,\overline{11}}-s_{23}s_{6,10})\nn \\ &
+l_{9}^2(s_{12}s_{12,\overline{13}}-s_{13}s_{10,11})+l_{10}^2(s_{12}s_{11,\overline{14}}-s_{13}s_{9,12})\nn \\ &
-l_{13}^2s_{12}s_{11,\overline{14}}-l_{14}^2s_{12}s_{12,\overline{13}} +(s_{13}-2 s_{12})l_{9}^2 l_{10}^2 \nn \\ &
+s_{23}(l_{5}^2 l_{6}^2 -l_{7}^2 l_{8}^2 + l_{6}^2 l_{7}^2 +l_{5}^2 l_{8}^2)+s_{12}l_{13}^2 l_{14}^2 -s_{12}l_{5}^2 l_{6}^2 
\nn \\ &
+s_{12}(-l_{5}^2l_{8}^2+l_{5}^2l_{9}^2-l_{5}^2l_{11}^2 - l_{5}^2 l_{15}^2 - l_{9}^2 l_{15}^2)\nn \\ &
+s_{12}(-l_{6}^2l_{7}^2+l_{6}^2 l_{10}^2-l_{6}^2l_{12}^2-l_{6}^2 l_{16}^2 -l_{10}^2 l_{16}^2)\nn \\ &
+s_{23}(l_{9}^2 l_{12}^2 + l_{10}^2 l_{11}^2 -l_{7}^2 l_{9}^2-l_{8}^2 l_{10}^2 ) + s_{13}(l_{9}^2 l_{11}^2 +l_{10}^2 l_{12}^2)
\,,
\nn
\end{align}
\begin{align}
& N_{49}={}
s_{12}(s_{47}s_{5,12}-s_{19}s_{36}-s_{48}s_{36})+s_{23}(s_{48}s_{6,11}-s_{15}s_{3,10}-s_{15}s_{47})-s_{12}s_{23}s_{11,12}\nn \\ &
+l_{5}^2 (s_{23} s_{7, 12}-s_{23} s_{4, 15}  - s_{13} s_{10, 11})+l_{6}^2 (s_{12} s_{8, 11}-s_{12} s_{4, \overline{15}}- s_{13} s_{9, 12})\nn \\ &
+l_{9}^2 (s_{23} s_{3, 15}-s_{12} s_{3\overline{8}}+ s_{23} s_{6, 10})+l_{10}^2 (s_{12}s_{1,  \overline{15}} -s_{23} s_{1\overline{7}} + s_{12} s_{59})\nn \\ &
+l_{13}^2 ( s_{12}s_{23} + s_{12}s_{38}
 - s_{23} s_{6, 11})
+l_{14}^2 (s_{23} s_{12} + s_{23} s_{17}- s_{12} s_{5, 12})\nn \\ &
+l_{11}^2 s_{23} (s_{4, 12} - s_{6, 10})
 + l_{12}^2 s_{12} (s_{4, 11} - s_{59})\nn \\ &
 +s_{13} (l_{7}^2 l_{8}^2  + l_{5}^2 l_{8}^2+ l_{6}^2 l_{7}^2 + l_{11}^2 l_{12}^2 +  l_{10}^2 l_{16}^2
 + l_{9}^2 l_{17}^2 - l_{9}^2 l_{12}^2 - l_{10}^2 l_{11}^2)\nn \\ &
+s_{12} (- l_{5}^2 l_{10}^2 + l_{6}^2( l_{14}^2+ l_{13}^2- l_{10}^2)
  + l_{12}^2( l_{9}^2 + l_{5}^2  - l_{7}^2 +  l_{14}^2)+l_{8}^2 (l_{9}^2+l_{16}^2) )\nn \\ &
+s_{23} (-l_{6}^2 l_{9}^2  + l_{5}^2 (l_{13}^2+  l_{14}^2- l_{9}^2)
+ l_{11}^2(l_{10}^2  + l_{6}^2 - l_{8}^2+  l_{13}^2) 
+l_{7}^2( l_{10}^2+ l_{17}^2) )\nn \\ &
+s_{12} (l_{12}^2 l_{13}^2 -l_{8}^2 l_{13}^2 - l_{10}^2 l_{13}^2 
- l_{10}^2 l_{14}^2 -l_{13}^2 l_{17}^2)+s_{23} (l_{11}^2 l_{14}^2 
-l_{7}^2 l_{14}^2 - l_{9}^2 l_{14}^2 - l_{9}^2 l_{13}^2 -l_{14}^2 l_{16}^2)
\,,
\nn \\
& N_{50}={}
s_{12}s_{28}s_{4,12}-s_{12}s_{37}s_{1,11} 
-s_{23}s_{16}s_{3,10}+s_{23}s_{25}s_{49}
+\frac{1}{2}s_{12}s_{23}(s_{13,15}-s_{13,14})\nn \\ &
+s_{12}(l_{6}^2 l_{10}^2 - l_{5}^2l_{9}^2)
 + s_{23}(l_{7}^2 l_{11}^2 - l_{8}^2 l_{12}^2)
\,.
\end{align}
We note that graph 50 has a vanishing color factor and therefore
does not contribute; the integral also vanishes under integration.

We now give the tensor color factors for the 50 four-loop integrals
describing the amplitude.  Although it is simple to read off 
these color factors from the graphs, we include them here
because they contain signs.  The color-factor sign (and hence the
sign of the numerator factor multiplying it) depends on how the
graph is drawn.  The signs here are properly correlated with the
signs of the numerator factors given above, as well as with
the graphs in figs.~\ref{BC1Figure}-\ref{E2Figure}. We have,
\begin{align}
 C_{1} &= 
\colorc{1, 5, 6} \colorc{2, 7, 5} \colorc{3, 17, 14} \colorc{4, 16, 17} \colorc{6, 9, 10}
\colorc{7, 8, 9} \colorc{8, 11, 12} \colorc{10, 12, 13} \colorc{11, 14, 15} \colorc{13, 15, 16}
\,,\nn \\ 
 C_{2} &= 
\colorc{1, 5, 6} \colorc{2, 7, 5} \colorc{3, 15, 16} \colorc{4, 17, 14} \colorc{6, 9, 10}
\colorc{7, 8, 9} \colorc{8, 11, 12} \colorc{10, 12, 13} \colorc{11, 14, 15} \colorc{13, 16, 17}
\,,\nn \\ 
 C_{3} &= 
\colorc{1, 5, 6} \colorc{2, 9, 8} \colorc{3, 14, 15} \colorc{4, 17, 16} \colorc{5, 7, 8}
\colorc{6, 9, 10} \colorc{7, 11, 12} \colorc{10, 12, 13} \colorc{11, 16, 14} \colorc{13, 15, 17}
\,,\nn \\ 
 C_{4} &= 
\colorc{1, 5, 6} \colorc{2, 7, 5} \colorc{3, 13, 16} \colorc{4, 16, 17} \colorc{6, 9, 10}
\colorc{7, 8, 9} \colorc{8, 11, 12} \colorc{10, 14, 15} \colorc{11, 15, 17} \colorc{12, 13, 14}
\,,\nn \\ 
 C_{5} &= 
\colorc{1, 5, 6} \colorc{2, 9, 8} \colorc{3, 13, 16} \colorc{4, 16, 17} \colorc{5, 7, 8}
\colorc{6, 9, 10} \colorc{7, 11, 12} \colorc{10, 14, 15} \colorc{11, 15, 17} \colorc{12, 13, 14}
\,,\nn \\ 
 C_{6} &= 
\colorc{1, 6, 7} \colorc{2, 8, 6} \colorc{3, 16, 11} \colorc{4, 15, 17} \colorc{5, 14, 15}
\colorc{7, 10, 5} \colorc{8, 9, 10} \colorc{9, 11, 12} \colorc{12, 13, 14} \colorc{13, 16, 17}
\,,\nn \\ 
 C_{7} &= 
\colorc{1, 6, 7} \colorc{2, 8, 6} \colorc{3, 15, 13} \colorc{4, 12, 17} \colorc{5, 11, 12}
\colorc{7, 10, 5} \colorc{8, 9, 10} \colorc{9, 16, 15} \colorc{11, 13, 14} \colorc{14, 16, 17}
\,,\nn \\ 
 C_{8} &= 
\colorc{1, 6, 7} \colorc{2, 8, 6} \colorc{3, 13, 16} \colorc{4, 15, 17} \colorc{5, 14, 15}
\colorc{7, 10, 5} \colorc{8, 9, 10} \colorc{9, 11, 12} \colorc{11, 17, 16} \colorc{12, 13, 14}
\,,\nn \\ 
 C_{9} &= 
\colorc{1, 6, 7} \colorc{2, 10, 9} \colorc{3, 17, 11} \colorc{4, 15, 16} \colorc{5, 14, 15}
\colorc{6, 8, 9} \colorc{7, 10, 5} \colorc{8, 11, 12} \colorc{12, 13, 14} \colorc{13, 17, 16}
\,,\nn 
\end{align}
\begin{align}
 C_{10} &= 
\colorc{1, 6, 7} \colorc{2, 8, 10} \colorc{3, 12, 13} \colorc{4, 16, 17} \colorc{5, 15, 16}
\colorc{6, 9, 10} \colorc{7, 8, 5} \colorc{9, 11, 12} \colorc{11, 17, 14} \colorc{13, 14, 15}
\,,\nn \\ 
 C_{11} &= 
\colorc{1, 6, 7} \colorc{2, 10, 9} \colorc{3, 13, 16} \colorc{4, 15, 17} \colorc{5, 14, 15}
\colorc{6, 8, 9} \colorc{7, 10, 5} \colorc{8, 11, 12} \colorc{11, 17, 16} \colorc{12, 13, 14}
\,,\nn \\ 
 C_{12} &= 
\colorc{1, 7, 5} \colorc{2, 9, 7} \colorc{3, 15, 16} \colorc{4, 6, 15} \colorc{5, 8, 13}
\colorc{9, 11, 10} \colorc{10, 12, 8} \colorc{11, 16, 17} \colorc{12, 17, 14} \colorc{13, 14, 6}
\,,\nn \\ 
 C_{13} &= 
\colorc{1, 7, 5} \colorc{2, 8, 7} \colorc{3, 17, 16} \colorc{4, 6, 17} \colorc{5, 13, 11}
\colorc{8, 9, 10} \colorc{9, 11, 12} \colorc{10, 15, 14} \colorc{12, 16, 15} \colorc{13, 14, 6}
\,,\nn \\ 
 C_{14} &= 
\colorc{1, 8, 9} \colorc{2, 5, 8} \colorc{3, 14, 16} \colorc{4, 10, 13} \colorc{7, 17, 15}
\colorc{6, 15, 14} \colorc{5, 16, 17} \colorc{9, 11, 10} \colorc{11, 7, 12} \colorc{12, 6, 13}
\,,\nn \\ 
 C_{15} &= 
\colorc{1, 7, 8} \colorc{2, 9, 7} \colorc{3, 15, 17} \colorc{4, 14, 16} \colorc{8, 12, 13}
\colorc{9, 10, 11} \colorc{10, 6, 15} \colorc{11, 17, 5} \colorc{12, 5, 16} \colorc{13, 14, 6}
\,,\nn \\ 
 C_{16} &= 
\colorc{1, 7, 8} \colorc{2, 9, 7} \colorc{3, 17, 16} \colorc{4, 12, 13} \colorc{8, 13, 14}
\colorc{9, 10, 11} \colorc{10, 6, 17} \colorc{11, 16, 5} \colorc{12, 5, 15} \colorc{14, 15, 6}
\,,\nn \\ 
 C_{17} &= 
\colorc{1, 7, 8} \colorc{2, 8, 9} \colorc{3, 14, 16} \colorc{4, 11, 12} \colorc{5, 15, 17}
\colorc{6, 14, 15} \colorc{7, 17, 16} \colorc{9, 10, 11} \colorc{10, 6, 13} \colorc{12, 13, 5}
\,,\nn \\ 
 C_{18} &= 
\colorc{1, 8, 9} \colorc{2, 5, 8} \colorc{3, 6, 12} \colorc{4, 16, 14} \colorc{7, 17, 16}
\colorc{13, 15, 17} \colorc{5, 12, 11} \colorc{9, 10, 7} \colorc{10, 11, 13} \colorc{6, 14, 15}
\,,\nn \\ 
 C_{19} &= 
\colorc{1, 8, 7} \colorc{2, 9, 8} \colorc{3, 15, 11} \colorc{4, 16, 14} \colorc{5, 11, 12}
\colorc{7, 16, 17} \colorc{9, 10, 5} \colorc{10, 17, 6} \colorc{12, 13, 14} \colorc{13, 15, 6}
\,,\nn \\ 
 C_{20} &= 
\colorc{1, 8, 9} \colorc{2, 7, 8} \colorc{3, 15, 13} \colorc{4, 11, 12} \colorc{7, 10, 11}
\colorc{9, 6, 17} \colorc{10, 17, 5} \colorc{12, 13, 14} \colorc{14, 16, 6} \colorc{15, 5, 16}
\,,\nn \\ 
 C_{21} &= 
\colorc{1, 8, 7} \colorc{2, 9, 8} \colorc{3, 13, 14} \colorc{4, 17, 11} \colorc{5, 14, 15}
\colorc{6, 12, 13} \colorc{7, 16, 17} \colorc{9, 10, 5} \colorc{10, 11, 6} \colorc{12, 16, 15}
\,,\nn \\ 
 C_{22} &= 
\colorc{1, 8, 7} \colorc{2, 9, 8} \colorc{3, 16, 14} \colorc{4, 12, 11} \colorc{5, 15, 17}
\colorc{6, 17, 16} \colorc{7, 13, 12} \colorc{9, 10, 6} \colorc{10, 11, 5} \colorc{13, 14, 15}
\,,\nn \\ 
 C_{23} &= 
\colorc{1, 8, 7} \colorc{2, 9, 8} \colorc{3, 15, 17} \colorc{4, 12, 11} \colorc{5, 17, 16}
\colorc{6, 16, 14} \colorc{7, 13, 12} \colorc{9, 10, 6} \colorc{10, 11, 5} \colorc{13, 14, 15}
\,,\nn \\ 
 C_{24} &= 
\colorc{1, 10, 7} \colorc{2, 5, 10} \colorc{3, 16, 14} \colorc{4, 13, 9} \colorc{5, 11, 6}
\colorc{6, 17, 16} \colorc{7, 8, 12} \colorc{8, 14, 15} \colorc{11, 12, 13} \colorc{15, 17, 9}
\,,\nn \\ 
 C_{25} &= 
\colorc{1, 10, 6} \colorc{2, 5, 10} \colorc{3, 14, 16} \colorc{4, 17, 8} \colorc{5, 11, 9}
\colorc{9, 12, 13} \colorc{6, 7, 15} \colorc{11, 15, 17} \colorc{12, 8, 16} \colorc{13, 14, 7}
\,,\nn \\ 
 C_{26} &= 
\colorc{1, 12, 13} \colorc{2, 5, 12} \colorc{3, 16, 11} \colorc{4, 15, 9} \colorc{5, 14, 6}
\colorc{6, 17, 16} \colorc{7, 9, 17} \colorc{10, 11, 15} \colorc{13, 10, 8} \colorc{14, 8, 7}
\,,\nn \\ 
 C_{27} &= 
\colorc{1, 13, 5} \colorc{2, 6, 13} \colorc{3, 16, 15} \colorc{4, 17, 7} \colorc{5, 9, 8}
\colorc{6, 14, 10} \colorc{10, 15, 11} \colorc{11, 17, 9} \colorc{14, 8, 12} \colorc{16, 12, 7}
\,,\nn \\ 
 C_{28} &= 
\colorc{1, 10, 11} \colorc{2, 6, 5} \colorc{3, 15, 13} \colorc{4, 7, 8} \colorc{6, 13, 14}
\colorc{10, 5, 12} \colorc{11, 17, 7} \colorc{12, 9, 17} \colorc{14, 16, 9} \colorc{15, 8, 16}
\,,\nn \\ 
 C_{29} &= 
\colorc{1, 10, 11} \colorc{2, 9, 5} \colorc{3, 17, 15} \colorc{4, 7, 8} \colorc{9, 13, 14}
\colorc{10, 5, 12} \colorc{11, 16, 7} \colorc{12, 6, 16} \colorc{13, 8, 15} \colorc{14, 17, 6}
\,,\nn \\ 
 C_{30} &= 
\colorc{1, 10, 11} \colorc{2, 15, 13} \colorc{3, 6, 5} \colorc{4, 8, 7} \colorc{6, 13, 14}
\colorc{10, 5, 12} \colorc{11, 17, 9} \colorc{12, 7, 17} \colorc{14, 16, 8} \colorc{15, 9, 16}
\,,\nn \\ 
 C_{31} &= 
\colorc{1, 9, 5} \colorc{2, 16, 10} \colorc{3, 17, 14} \colorc{4, 7, 8} \colorc{5, 10, 11}
\colorc{9, 12, 13} \colorc{11, 15, 7} \colorc{12, 8, 14} \colorc{13, 17, 6} \colorc{15, 16, 6}
\,,\nn \\ 
 C_{32} &= 
\colorc{1, 7, 6} \colorc{2, 9, 15} \colorc{3, 17, 16} \colorc{4, 5, 8} \colorc{5, 10, 11}
\colorc{6, 12, 10} \colorc{7, 8, 9} \colorc{11, 14, 17} \colorc{12, 13, 14} \colorc{13, 15, 16}
\,,\nn \\ 
 C_{33} &= 
\colorc{1, 15, 17} \colorc{2, 12, 11} \colorc{3, 7, 9} \colorc{4, 10, 13} \colorc{6, 16, 14}
\colorc{8, 9, 12} \colorc{10, 8, 6} \colorc{13, 5, 7} \colorc{14, 15, 5} \colorc{16, 11, 17}
\,,\nn \\ 
 C_{34} &= 
\colorc{1, 16, 14} \colorc{2, 13, 7} \colorc{3, 11, 9} \colorc{4, 8, 12} \colorc{5, 14, 15}
\colorc{6, 17, 16} \colorc{10, 5, 8} \colorc{10, 9, 6} \colorc{11, 12, 13} \colorc{15, 17, 7}
\,,\nn \\ 
 C_{35} &= 
\colorc{1, 16, 14} \colorc{2, 13, 7} \colorc{3, 8, 11} \colorc{4, 10, 12} \colorc{5, 14, 15}
\colorc{6, 17, 16} \colorc{9, 11, 13} \colorc{10, 9, 6} \colorc{12, 5, 8} \colorc{15, 17, 7}
\,,\nn \\ 
 C_{36} &= 
\colorc{1, 13, 9} \colorc{2, 8, 6} \colorc{3, 12, 7} \colorc{4, 15, 17} \colorc{5, 14, 15}
\colorc{8, 10, 12} \colorc{9, 5, 10} \colorc{11, 16, 14} \colorc{13, 6, 11} \colorc{16, 7, 17}
\,,\nn 
\end{align}
\begin{align}
 C_{37} &= 
\colorc{1, 10, 11} \colorc{2, 7, 12} \colorc{3, 13, 8} \colorc{4, 16, 14} \colorc{6, 17, 16}
\colorc{7, 8, 9} \colorc{9, 15, 17} \colorc{10, 12, 6} \colorc{11, 5, 13} \colorc{14, 15, 5}
\,,\nn \\ 
 C_{38} &= 
\colorc{1, 16, 12} \colorc{2, 13, 8} \colorc{3, 14, 7} \colorc{4, 17, 15} \colorc{5, 15, 10}
\colorc{6, 10, 14} \colorc{9, 7, 17} \colorc{12, 5, 11} \colorc{13, 11, 6} \colorc{16, 8, 9}
\,,\nn \\ 
 C_{39} &= 
\colorc{1, 5, 9} \colorc{2, 13, 10} \colorc{3, 14, 7} \colorc{4, 12, 8} \colorc{5, 15, 6}
\colorc{9, 10, 17} \colorc{13, 6, 16} \colorc{15, 8, 7} \colorc{16, 14, 11} \colorc{17, 11, 12}
\,,\nn \\ 
 C_{40} &= 
\colorc{1, 7, 8} \colorc{2, 9, 7} \colorc{3, 14, 15} \colorc{4, 16, 14} \colorc{5, 17, 16}
\colorc{8, 12, 13} \colorc{9, 6, 17} \colorc{10, 5, 12} \colorc{10, 6, 11} \colorc{11, 13, 15}
\,,\nn \\ 
 C_{41} &= 
\colorc{1, 6, 7} \colorc{2, 5, 6} \colorc{3, 16, 15} \colorc{4, 10, 8} \colorc{5, 8, 9}
\colorc{7, 15, 13} \colorc{9, 14, 17} \colorc{10, 11, 12} \colorc{11, 13, 14} \colorc{12, 16, 17}
\,,\nn \\ 
 C_{42} &= 
\colorc{1, 11, 6} \colorc{2, 5, 11} \colorc{3, 16, 12} \colorc{4, 9, 7} \colorc{5, 13, 14}
\colorc{7, 12, 8} \colorc{8, 15, 13} \colorc{9, 6, 10} \colorc{10, 15, 17} \colorc{14, 16, 17}
\,,\nn \\ 
 C_{43} &= 
\colorc{1, 7, 8} \colorc{2, 5, 7} \colorc{3, 15, 9} \colorc{4, 16, 10} \colorc{5, 6, 9}
\colorc{6, 10, 11} \colorc{8, 12, 17} \colorc{11, 12, 13} \colorc{13, 14, 15} \colorc{14, 16, 17}
\,,\nn \\ 
 C_{44} &= 
\colorc{1, 6, 10} \colorc{2, 10, 5} \colorc{3, 7, 13} \colorc{4, 8, 12} \colorc{5, 16, 14}
\colorc{6, 11, 9} \colorc{8, 14, 15} \colorc{9, 17, 15} \colorc{11, 12, 13} \colorc{16, 7, 17}
\,,\nn \\ 
 C_{45} &= 
\colorc{1, 13, 6} \colorc{2, 5, 13} \colorc{3, 9, 14} \colorc{4, 16, 8} \colorc{7, 6, 10}
\colorc{7, 11, 14} \colorc{10, 12, 16} \colorc{11, 8, 15} \colorc{12, 9, 17} \colorc{15, 5, 17}
\,,\nn \\ 
 C_{46} &= 
\colorc{1, 11, 6} \colorc{2, 7, 12} \colorc{3, 5, 9} \colorc{4, 10, 13} \colorc{5, 13, 8}
\colorc{6, 16, 14} \colorc{8, 16, 17} \colorc{10, 14, 15} \colorc{11, 9, 12} \colorc{15, 7, 17}
\,,\nn \\ 
 C_{47} &= 
\colorc{1, 6, 5} \colorc{2, 8, 7} \colorc{3, 12, 13} \colorc{4, 9, 10} \colorc{5, 7, 11}
\colorc{6, 14, 15} \colorc{11, 13, 9} \colorc{14, 10, 16} \colorc{15, 12, 17} \colorc{16, 8, 17}
\,,\nn \\ 
 C_{48} &= 
\colorc{1, 9, 15} \colorc{2, 5, 13} \colorc{3, 14, 6} \colorc{4, 10, 16} \colorc{5, 6, 17}
\colorc{7, 12, 10} \colorc{8, 14, 16} \colorc{9, 11, 8} \colorc{15, 7, 13} \colorc{17, 12, 11}
\,,\nn \\ 
 C_{49} &= 
\colorc{1, 16, 11} \colorc{2, 6, 5} \colorc{3, 12, 17} \colorc{4, 10, 9} \colorc{6, 17, 8}
\colorc{7, 9, 14} \colorc{8, 13, 10} \colorc{11, 13, 15} \colorc{14, 12, 15} \colorc{16, 5, 7}
\,,\nn \\ 
 C_{50} &= 
\colorc{1, 12, 5} \colorc{2, 7, 6} \colorc{3, 9, 8} \colorc{4, 10, 11} \colorc{5, 13, 6}
\colorc{7, 8, 14} \colorc{12, 15, 11} \colorc{13, 16, 17} \colorc{14, 16, 15} \colorc{17, 10, 9}
\,.
\end{align}
Group indices $a_1,a_2,a_3, a_4$ correspond to external color, and are
not summed over. Internal indices $a_5$ through $a_{17}$ are to be
summed over. Thus the suppressed tensor structure of each color factor
is $C_i \equiv C_i^{a_1a_2a_3a_4}$.  It turns out that $C_{50}$ vanishes 
identically for all gauge groups (because of constraints from the
Jacobi identity).


\end{document}